\documentclass[superscriptaddress,prx,twocolumn,nofootinbib,citeautoscript,longbibliography,notitlepage]{revtex4-1}

\usepackage{graphicx}
\usepackage{dcolumn}
\usepackage{bm}
\usepackage{color}
\usepackage{amsmath}
\usepackage{tabularx,graphicx}
\usepackage{epstopdf}
\usepackage{latexsym}
\usepackage{amssymb}
\usepackage{amsmath}
\usepackage{color, colortbl}
\usepackage{psfrag}
\usepackage{bbm}
\usepackage{booktabs}
\usepackage{bm}
\usepackage{titlesec}
\usepackage{dsfont}
\usepackage{feynmp}
\usepackage{slashed}
\usepackage{multirow}
\usepackage[tight]{subfigure}
\usepackage{comment}
\usepackage{float}
\usepackage{diagbox} 
\usepackage{makecell}
\usepackage{array}
\usepackage[papersize={8.5in,11in}]{geometry}
\usepackage{color}
\definecolor{darkblue}{rgb}{0.,0.,0.4}
\definecolor{darkred}{rgb}{0.5,0.,0.}
\definecolor{BlueViolet}{RGB}{138,43,226}
\definecolor{SkyBlue}{RGB}{30,144,255}
\definecolor{DarkGreen}{RGB}{0,100,0}
\definecolor{Green}{RGB}{0, 255, 0}
\usepackage[pdftex,colorlinks=true,linkcolor=darkblue,citecolor=blue,urlcolor=darkred]{hyperref}

\renewcommand{\epsilon}{\varepsilon}

\allowdisplaybreaks[3]

 \UseRawInputEncoding

\usepackage{cases}

\allowdisplaybreaks[4]

\begin{document}



\title{Interaction-driven competing instabilities in two-dimensional checkerboard quadratic band crossing semimetals}

\author{Jia-Hao Wen}
\affiliation{Department of Physics, Tianjin University, Tianjin 300072, P.R. China}
\affiliation{Laboratory of Quantum Information,
University of Science and Technology of China, Hefei, 230026, P.R. China}
\author{Wen-Hao Bian}
\affiliation{School of Physics, Nanjing University, Nanjing, Jiangsu 210093, P.R. China}
\author{Jing Wang}
\altaffiliation{Corresponding author: jing$\textunderscore$wang@tju.edu.cn}
\affiliation{Department of Physics, Tianjin University, Tianjin 300072, P.R. China}
\affiliation{Tianjin Key Laboratory of Low Dimensional Materials Physics and Preparing Technology,
Tianjin University, Tianjin 300072, P.R. China}

\date{\today}


\begin{abstract}

Two-dimensional quadratic band crossing point (QBCP) semimetals
serve as a fertile platform for interaction-driven quantum criticality,
nevertheless, the role of the complete fermion-fermion interactions remains
largely unexplored. To address this, we systematically investigate the low-energy
physics of a spin-1/2 checkerboard-lattice QBCP semimetal by considering all 16
marginally short-range fermion-fermion interactions. Using the momentum-shell
renormalization group, we derive the coupled flow equations, which reveal that
the low-energy physics is governed by a series of fixed points. Depending on the
initial conditions, the system may exhibit a single stable fixed point, several
discrete fixed points, or three continuous groups of fixed points parameterized
by a few variables. By introducing source terms and evaluating susceptibilities,
we identify the dominant and subdominant instabilities. The leading phases include
nematic spin-nematic (NSN), quantum anomalous Hall (QAH), chiral superconducting
($\mathrm{SC}_1$), and quantum spin Hall (QSH) orders. Moreover, several features
distinguish our findings from previous simplified models. Off-diagonal interactions
generate terms that promote orders absent in such models, the continuous groups allow
systematic tuning of the leading instability, and trajectory-dependent instabilities emerge
where the full RG flow, rather than the fixed point alone, determines the dominant state.
These results offer a comprehensive theoretical understanding of low-energy quantum
critical behaviors and phase instabilities in two-dimensional QBCP semimetals and
related electron systems.

\end{abstract}


\maketitle

%
%
%
%
%
%
%
%
%




\section{Introduction}

Semimetal materials constitute a central research frontier in modern condensed matter physics~\cite{Novoselov2025Nat,Castro2009RMP,Fu2007PRL,Roy2009PRB,Moore2010Nat,Hasan2010RMP,Qi2011RMP,Tang2018Sci,
Shen2012Book,Bernevig2013Book,Armitage2018RMP,Roy2018PRX}. Recent decades have witnessed substantial
advances in the study of low-dimensional Dirac and Weyl semimetals and related materials~\cite{Novoselov2025Nat,Castro2009RMP,Fu2007PRL,Roy2009PRB,Moore2010Nat,Hasan2010RMP,
Qi2011RMP,Tang2018Sci,Wang2012PRB,Young2012PRL,Steinberg2014PRL,Liu2014Nat,Liu2014Sci,Xiong2015Sci,
Burkov2011PRL,Yang2011PRB,Wan2011PRB,Huang2015PRX,Xu2015Sci,Xu2015Nat,Lv2015Nat,Weng2015PRX,
Korshunov2014PRB,Hung2016PRB,Nandkishore2013PRB,Potirniche2014PRB,Nandkishore2017PRB,Roy2016PRB}.
These materials are characterized by discrete Dirac points, around which gapless quasiparticle excitations
with linear energy-momentum dispersion emerge along two or three spatial directions~\cite{Castro2009RMP,Fu2007PRL,Hasan2010RMP,Qi2011RMP,Roy2009PRB,Korshunov2014PRB,Potirniche2014PRB,Roy2016PRB}.
More recently, attention has progressively evolved from systems with linear dispersion toward
those exhibiting quadratic dispersion. A particularly active area concerns two-dimensional (2D)
electron systems that host quadratic band crossing points (QBCPs), at which the upper and lower
energy bands touch quadratically at a single point~\cite{Fradkin2008PRB,Fradkin2009PRL,Ray2020PRB,
Yang2010PRB,Vafek2010PRB,Cvetkovic2012PRB,Vafek2014PRB,Wang2017PRB,Zhu2016PRL,DZDW2020PRB,Herbut2012PRB,
Roy2026PRB,Mandal2019CMP,Shahl2021PRB,Mandal2018PRB,Lin2018PRB,Luttinger1956PR,Murakami2004PRB,
Janssen2015PRB,Boettcher2016PRB,Janssen2017PRB,Boettcher2017PRB,Savary2014PRX,Savary2017PRB,Lai,
Goswami2017PRB,Szabo2021PRB,Roy2019PRB,Ray2018PRB,Wang2020PRB,Chong2008PRB,Wang2023EPJB,Fu-Wang2024AP,
Fang2026PRB,Dora2014PRB}. Such QBCP states can be realized in multiple lattice and structural models, including kagom\'{e} lattices~\cite{Fradkin2009PRL,Ray2020PRB,Huse2003PRB}, checkerboard lattices~\cite{Vafek2014PRB,Fradkin2008PRB}, Lieb lattices~\cite{Tsai2015NJP}, and collinear spin-density-wave phases~\cite{Chern2012PRL}.

A key distinction between 2D QBCP semimetals and conventional 2D Dirac/Weyl semimetals lies in their
density of states near the Fermi level. Dirac/Weyl systems possess a vanishing density of states at
the band-crossing point, whereas 2D QBCP systems have a finite density of states at the crossing point~\cite{Fradkin2009PRL,Cvetkovic2012PRB,Vafek2014PRB}. This unique feature gives rise to
gapless fermionic excitations around the QBCP and causes interactions such as fermion-fermion
couplings to play an important role in determining the low-energy physical behavior of the system~\cite{Fradkin2009PRL,Vafek2014PRB,Cvetkovic2012PRB,Vafek2010PRB,Yang2010PRB,Venderbos2016PRB,
Wu2016PRL,Wang2017PRB,Zhu2016PRL,DZDW2020PRB,Zeng-Zhu2018NPJ}. As a direct consequence of this finite density of states, the system
is highly susceptible to interaction-driven instabilities.
Indeed, two-dimensional QBCP systems are thermodynamically unstable against electron-electron interactions,
so that even weak coupling can readily trigger symmetry-breaking phase transitions~\cite{Fradkin2009PRL,Vafek2014PRB,Cvetkovic2012PRB,Vafek2010PRB,Yang2010PRB,Venderbos2016PRB,
Wu2016PRL,Wang2017PRB,Zhu2016PRL,DZDW2020PRB,Zeng-Zhu2018NPJ}. Recent studies have shown that repulsive electron-electron interactions
in checkerboard lattices and two-valley bilayer graphene with intrinsic QBCPs can drive the semimetallic phase
into two distinct gapped states~\cite{Vafek2014PRB,Fradkin2008PRB,Vafek2010PRB,Yang2010PRB}.

Despite these advances, earlier theoretical approaches to four-fermion interactions in QBCP systems
remain insufficient. In particular, prevailing models typically retain four independent interaction channels
~\cite{Fradkin2009PRL,Cvetkovic2012PRB,Wang2017PRB,DZDW2020PRB},
thereby neglecting the cross-channel contributions that arise from the interplay among distinct
fermionic interactions~\cite{Zhu2016PRL,Baum2026}. This simplification is a technical approximation,
and may not capture the effective interactions generated by fermion exchange among different
scattering channels~\cite{Altland2006Book,Baum2026}. Since the coupling among these channels generates several cross channels rooted
in correlation effects, these contributions play a crucial role in modifying
the electronic properties of QBCP systems. Specifically, they can reshape the band structure and the
density-of-states distribution, and can even induce novel symmetry-breaking transitions.
Consequently, a complete understanding of the low-energy physics of 2D QBCP semimetals requires going beyond
the four-channel approximation and incorporating the full set of interaction channels.

Motivated by the unresolved issues outlined above,
we systematically investigate how the interplay among distinct fermion-fermion interactions shapes
the low-energy properties of two-dimensional checkerboard quadratic band crossing point (QBCP) semimetals.
To this end, we construct a theoretical model that incorporates all sixteen independent fermionic
interaction channels, covering all feasible fermionic coupling configurations. We then adopt the
momentum-shell renormalization group (RG) approach~\cite{Shankar1994RMP,Wilson1975RMP,Polchinski9210046}
to derive the energy-dependent RG flow equations for these couplings, from which several interesting
low-energy behaviors are extracted.

At the outset, our RG analysis shows that the fermion-fermion interactions, which are marginal at the tree level,
flow toward divergence at a certain critical scale, with the low-energy physics captured by emergent
fixed points [as defined in Eq.~(\ref{Eq_definition_FP})]. Depending on the initial parameters, there
exist three distinct scenarios. In the isotropic scenario, all sixteen couplings flow to a single stable fixed point, FP$_0$~(\ref{Eq_FP0}). Moving to the anisotropic scenario, we find seven stable fixed points,
FP$_1$-FP$_7$~(\ref{Eq_FP1})-(\ref{Eq_FP7}), which are paired by a discrete off-diagonal index-exchange
symmetry. By contrast, in the general scenario, the fixed points organize into three continuous
fixed-point groups, FPG$_1$-FPG$_3$~(\ref{Eq_FPG1})-(\ref{Eq_FPG3}), characterized by the group
parameters $\alpha$, $\beta$, and $\gamma$.

To identify which symmetry-broken phase dominates at each fixed point, we next introduce source
terms for competing candidate phases (see Table~\ref{table:xiangbian} for details) and evaluate the corresponding
susceptibilities, thereby determining the leading and subleading instabilities~\cite{Halboth2000RPB,Metzner2000PRL,Maiti2010PRB,Chubukov2016PRX,Vafek2014PRB,
Vafek2010PRB,Cvetkovic2012PRB}. Evaluating and comparing susceptibilities near each fixed point,
we find $\mathrm{SC}_1$ leading at $\mathrm{FP}_0$ (NSN subleading)
and $\mathrm{NSN}(\mathrm{site})\text{-}z$ leading at $\mathrm{FP}_1$-$\mathrm{FP}_4$
and $\mathrm{FP}_6$ ($\mathrm{SC}_1$ or QAH subleading). In comparison, near $\mathrm{FP}_5$
and $\mathrm{FP}_7$ both chiral $\mathrm{SC}_1$ and QAH have relatively high probabilities of becoming the leading instability.
Within three fixed-point groups (see Fig.~\ref{fig:GZFPS123}), increasing $\alpha$ drives FPG$_1$
from single-dominant NSN(site)-$x$ to a competing regime where QAH, QSH, NSN, and $\mathrm{SC}_1$ coexist and compete.
For FPG$_2$, $\beta$ has negligible influence and NSN(site)-$z$ remains dominant throughout.
For FPG$_3$, increasing $\gamma$ drives a continuous transition from NSN(site)-$z$ to a
near-degenerate QAH/chiral SC$_1$ regime.

Moreover, three features distinguish our results from earlier studies based on only the four diagonal interactions~\cite{Fradkin2009PRL,Wang2017PRB,DZDW2020PRB,Vafek2014PRB}. Off-diagonal interactions
introduce additional linear and quadratic terms into the susceptibility exponents, promoting chiral
SC and NSN phases that cannot dominate under diagonal-only interactions. The continuous fixed-point
groups permit systematic tuning of the leading instability via the group parameters. Trajectory-dependent
instabilities emerge at FP$_5$ and FP$_7$, where the full RG trajectory, rather than the fixed point alone,
determines the leading instability. Together, these results establish that all sixteen fermion-fermion
interaction channels are essential for understanding the low-energy quantum-critical behavior of 2D QBCP semimetals.

The remainder of this paper is organized as follows. In Sec.~\ref{Sec_model}, we introduce the microscopic
model for two-dimensional QBCP semimetals and construct the low-energy effective theory that incorporates the full
set of fermion-fermion interactions. Then, section~\ref{Sec_RG_analysis} presents the RG analysis and
derives the coupled flow equations for all interaction parameters. Next, we within Sec.~\ref{Sec_classification_FPs} analyze the parameter-dependent RG trajectories and classify all physically relevant fixed points. Furthermore,
in Sec.~\ref{Sec_phase_transition}, we systematically investigate interaction-driven phase instabilities and compare our results with earlier works. At last, section~\ref{Sec_summary} provides a brief summary for our findings.

\section{Microscopic model and effective theory}\label{Sec_model}

We initiate our theoretical investigation of the low-energy quantum criticality in two-dimensional (2D) quadratic band crossing point (QBCP) semimetals by presenting the microscopic model and establishing the effective theory.

\subsection{Microscopic model}

The non-interacting part of the Hamiltonian, which describes spin-1/2 fermions hopping on a checkerboard lattice,
is expressed as~\cite{Fradkin2008PRB,Fradkin2009PRL,Huse2003PRB,Vafek2014PRB},
\begin{eqnarray}
H_0=\sum_{\sigma
=\uparrow\downarrow}
\sum_{\left| \mathbf{k} \right|<\Lambda}\Psi^\dag_{\mathbf{k}\sigma}\mathcal{H}_0(\mathbf{k})\Psi_{\mathbf{k}\sigma},\label{1}
\end{eqnarray}
with $\Lambda$ denoting the momentum cutoff determined by the lattice spacing.
Here, the fermionic spinor $\Psi_{\mathbf{k}\sigma}\equiv (c_{a\sigma},c_{b\sigma})^{\mathrm{T}}$ incorporates
sublattice indices $a,b$ and spin index $\sigma$. The corresponding Hamiltonian density is given by~\cite{Fradkin2008PRB,Fradkin2009PRL,Huse2003PRB},
\begin{eqnarray}
\mathcal{H}_0(\mathbf{k}) = t_I \mathbf{k}^2 \sigma_{0} + 2t_x k_xk_y \sigma_{1}
+ t_z (k_x^2 - k_y^2) \sigma_{3},\label{2}
\end{eqnarray}
where the material-specific parameters $t_I$, $t_x$, and $t_z$ encode the band anisotropy.
The Pauli matrices $\sigma_{1,2,3}$ and identity $\sigma_0$ operate on the sublattice degrees of freedom.

Upon diagonalizing the Hamiltonian density in Eq.~(\ref{2}), we obtain the two quadratic dispersion branches~\cite{Fradkin2008PRB,Fradkin2009PRL,Huse2003PRB,Vafek2014PRB},
\begin{eqnarray}
E_{\mathbf{k}}^\pm=\frac{\mathbf{k}^2\left[\lambda\pm\sqrt{\cos^2\eta\cos^22\theta_{\mathbf{k}}
+\sin^2\eta\sin^22\theta_{\mathbf{k}}}\right]}{\sqrt{2}m},\label{Eq_dispersions}
\end{eqnarray}
with the accompanying definitions
\begin{eqnarray}
m&\equiv&\frac{1}{\sqrt{2}\tilde{t}},\,\,\,\,\lambda\equiv\frac{t_I}{\tilde{t}},\,\,\,\,
\sin\eta\equiv\frac{t_x}{\tilde{t}},\nonumber\\
\cos\eta&\equiv&\frac{t_z}{\tilde{t}},\,\,\sin\theta_{\mathbf{k}}\equiv\frac{k_y}{|\mathbf{k}|},
\,\,\cos\theta_{\mathbf{k}}\equiv\frac{k_x}{|\mathbf{k}|},
\end{eqnarray}
where we have defined $\tilde{t}\equiv \sqrt{t_x^2+t_z^2}$, and $\theta_{\mathbf{k}}$ represents the
angular coordinate of the momentum $\mathbf{k}$ in the 2D plane.
As evident from the spectrum~(\ref{Eq_dispersions}), the upper and lower bands undergo a quadratic touching
at the zone center $\mathbf{k}=0$, provided the condition $|t_I| < \min(|t_x|, |t_z|)$ is met.
Besides, the symmetry content of the model is dictated by the relative magnitudes of above band parameters.
Specifically, a finite $t_I$ explicitly breaks particle-hole (PH) symmetry, while rotational symmetry is present exclusively in the isotropic limit $t_x=t_z$ and broken otherwise~\cite{Fradkin2008PRB,Fradkin2009PRL,Vafek2014PRB}.

Without loss of generality, we within this work put our focus on the PH-symmetric and rotationally-invariant
scenario, thereby adopting the parameterization $t_I = 0$ and $t_x = t_z \equiv t$. It is of particular importance to highlight that the free model~(\ref{2}) would serve as the unperturbed starting point for our subsequent RG analysis and be associated with the non-interacting fixed point of our theory.

\subsection{Fermion-fermion interactions and effective action}

To proceed, let us consider the fermion-fermion interactions that play an important role
in pinning down the low-energy physics of 2D QBCP semimetals~\cite{Fradkin2009PRL,Vafek2014PRB,Vafek2010PRB,
Yang2010PRB,Wang2017PRB,Zhu2016PRL,DZDW2020PRB,Fu-Wang2024AP,Cvetkovic2012PRB,Zhai2021NPB}.

To characterize the role of low-energy excitations near the QBCP, early theoretical works commonly employed
simplified short-range fermion interactions of the type $(\Psi^\dagger\sigma_i\Psi)^2$~\cite{Fradkin2009PRL,Vafek2014PRB,Wang2017PRB}.
However, such a simplified treatment typically retains only diagonal scattering channels, thereby
potentially neglecting cross-channel couplings and interference effects
that emerge among different sublattice and spin degrees of freedom. As a consequence,
they may play a crucial role in the quantum critical regime.
To remedy this and to provide a more complete description, we introduce the most general form of the marginally
fermion-fermion interaction, given by~\cite{Fradkin2008PRB,Fradkin2009PRL,Ray2020PRB,Yang2010PRB,Vafek2010PRB,Cvetkovic2012PRB,
Vafek2014PRB,Wang2017PRB,DZDW2020PRB,Herbut2012PRB,Roy2026PRB,Shahl2021PRB,Janssen2015PRB,Boettcher2016PRB,Janssen2017PRB,
Boettcher2017PRB,Savary2014PRX,Savary2017PRB,Goswami2017PRB,Szabo2021PRB,Roy2019PRB,Ray2018PRB,Wang2020PRB,
Fu-Wang2024AP,Fang2026PRB,Zhang2025PRB,Zhu2016PRL,Mandal2019CMP,Mandal2018PRB},
\begin{eqnarray}
S_{\rm int}
&=& \sum_{i,j=0}^3 \lambda_{ij} \int_{-\infty}^{+\infty}\frac{d\omega_{1}d\omega_{2}d\omega_{3}}{(2\pi)^{3}}
\int^\Lambda\frac{d^{2}\mathbf{k}_{1}d^{2}\mathbf{k}_{2}d^{2}\mathbf{k}_{3}}{(2\pi)^{6}}\nonumber\\
&&
\times\Psi^{\dagger}(\omega_{1},\mathbf{k}_{1})\sigma_{i}\Psi(\omega_{2},\mathbf{k}_{2})
\Psi^{\dagger}(\omega_{3},\mathbf{k}_{3})\sigma_{j}\nonumber\\
&&\times\Psi(\omega_{1}+\omega_{2}-\omega_{3},\mathbf{k}_{1}+\mathbf{k}_{2}-\mathbf{k}_{3}),\label{Eq:Sint}
\end{eqnarray}
where the positive constants $\lambda_{ij}$ with $i,j=0,1,2,3$ represent the repulsive interaction
strengths in the various sublattice-scattering channels distinguished by the Pauli matrices $\sigma_i$ and $\sigma_j$.

With the free and interacting sectors defined in Eqs.~(\ref{2}) and~(\ref{Eq:Sint}), respectively, we
are left with the full effective action as follows,
\begin{eqnarray}
S_{\text{eff}}
&=&\!\!\!\int\!\frac{d\omega d^2 \mathbf{k}}{(2\pi)^3} \Psi^\dagger(\omega,\mathbf{k})
\!\left( -i\omega + \mathcal{H}_0 \right)\Psi(\omega,\mathbf{k})+ S_{\rm int}.\label{Eq:Seff}
\end{eqnarray}
The noninteracting (bare) fermion propagator is readily extracted from the quadratic (free) term of
the above action, and is given by
\begin{eqnarray}
G_0(i\omega,\mathbf{k})=\frac{1}{-i\omega +\mathcal{H}_0(\mathbf{k})}.
\end{eqnarray}

In the context of 2D QBCP semimetals, the unique quadratic dispersion and finite density of states at the band-touching point render these fermionic couplings in the effective action~(\ref{Eq:Seff}) marginal at the tree level. However, upon incorporating one-loop quantum fluctuations, they can acquire essential corrections and become relevant operators. This accordingly gives rise to nontrivial
behaviors in the low-energy regime, which will be addressed in the forthcoming sections.

\section{RG analysis and flow equations}\label{Sec_RG_analysis}

To probe the low-energy quantum critical fluctuations in two-dimensional (2D) QBCP semimetals, we begin with the
effective theory~(\ref{Eq:Seff}) and then proceed to derive its renormalization group (RG) flow equations within this section.

Following the standard Wilsonian momentum-shell renormalization group (RG) approach~\cite{Shankar1994RMP,Wilson1975RMP,Polchinski9210046}, we endeavor to derive the energy scale-dependent
evolutions of interaction parameters in the effective action~(\ref{Eq:Seff}). These evolutions are of particular
importance to dictate the low-energy behavior of the QBCP system. To this end, at first, one needs to explicitly integrate
out the high-energy fermionic modes whose momenta lie in the thin momentum shell
$b\Lambda < k < \Lambda$, where $\Lambda$ denotes the ultraviolet cutoff and $b = e^{-l}$ with $l>0$ serving
as the flow parameter that tracks the reduction of the energy scale~\cite{Vafek2014PRB,Cvetkovic2012PRB,Wang2017PRB,Maiti2010PRB,Altland2006Book,Wang2011PRB,Wang2013PRB,
Huh2008PRB,Kim2008PRB,She2010PRB,Roy2016PRB}.

Then, in accordance with the Wilsonian RG framework~\cite{Shankar1994RMP,Wilson1975RMP,Polchinski9210046}, the free part of the low-energy effective action~(\ref{Eq:Seff}) is taken as the Gaussian fixed point, which is kept invariant under the RG flow.
This consequently gives rise to the RG rescaling transformations for momentum, energy, and fermionic fields as follows
~\cite{Vafek2014PRB,Huh2008PRB,She2010PRB,Wang2011PRB,Kim2008PRB,Shankar1994RMP,She2015PRB}
\begin{eqnarray}
\omega&\rightarrow& \omega e^{-2l},\label{eq:omega_trans}\\
k_{x}&\rightarrow& k_{x}' e^{-l},\\
k_{y}&\rightarrow& k_{y}' e^{-l},\\
\Psi(i\omega,\mathbf{k})&\rightarrow& \Psi'(i\omega',\mathbf{k}')e^{\frac{1}{2}\int dl(6-\eta_f)}\label{eq:Psi_trans},
\end{eqnarray}
which are adopted to connect the step-to-step RG procedures~\cite{Shankar1994RMP,Wilson1975RMP,Polchinski9210046}.
Hereby, $\eta_f$ serves as the anomalous dimension associated with the fermionic fields, which
vanishes at the one-loop order due to the peculiar properties of 2D QBCP semimetals~\cite{Fradkin2009PRL,Vafek2014PRB}.

Further, we include higher-order quantum fluctuations beyond the tree level.
Specifically, we perform a systematic analytic evaluation of the one-loop fermion-fermion corrections to all fermion-fermion
interactions in Eq.~(\ref{Eq:Sint}). The explicit expressions for these corrections are collected in Appendix~\ref{Appendix_1L_corrections}.
Combining these results with the scaling transformations~(\ref{eq:omega_trans})-(\ref{eq:Psi_trans}), we obtain the full set of
coupled RG flow equations, which read
\begin{widetext}
\begin{eqnarray}
\frac{d\lambda_{00}}{dl}
&=&-(\lambda_{00}\lambda_{11}+3\lambda_{01}\lambda_{10}+4\lambda_{02}\lambda_{20}+3\lambda_{03}\lambda_{30}
+\lambda_{00}\lambda_{33}),\label{Eq_lambda00}\\
\frac{d\lambda_{01}}{dl}
&=&\lambda_{00}\lambda_{01}-\lambda_{00}\lambda_{10}-2\lambda_{01}\lambda_{11}-\lambda_{02}\lambda_{12}
-\lambda_{03}\lambda_{13}-4\lambda_{02}\lambda_{21}-\lambda_{01}\lambda_{22}-2\lambda_{03}\lambda_{31}
-\lambda_{01}\lambda_{33},\\
\frac{d\lambda_{02}}{dl}
&=&2\lambda_{00}\lambda_{02}-2\lambda_{02}\lambda_{11}-2\lambda_{01}\lambda_{12}-2\lambda_{02}\lambda_{22}
-2\lambda_{02}\lambda_{33}-2\lambda_{03}\lambda_{32},\\
\frac{d\lambda_{03}}{dl}
&=&
\lambda_{00}\lambda_{03}-\lambda_{03}\lambda_{11}-2\lambda_{01}\lambda_{13}-\lambda_{03}\lambda_{22}
-4\lambda_{02}\lambda_{23}-\lambda_{00}\lambda_{30}-\lambda_{01}\lambda_{31}-\lambda_{02}\lambda_{32}-2\lambda_{03}\lambda_{33},\\
\frac{d\lambda_{10}}{dl}
&=&
-(3\lambda_{10}\lambda_{11}+4\lambda_{12}\lambda_{20}+\lambda_{20}\lambda_{21}+\lambda_{00}\lambda_{01}
+2\lambda_{13}\lambda_{30}+\lambda_{30}\lambda_{31}),\\
\frac{d\lambda_{11}}{dl}
&=&
\frac{1}{2}(2\lambda_{00}\lambda_{11}-\lambda_{00}\lambda_{00}-\lambda_{01}\lambda_{01}-\lambda_{02}\lambda_{02}-\lambda_{03}\lambda_{03}
-\lambda_{10}\lambda_{10}-3\lambda_{11}\lambda_{11}-\lambda_{12}\lambda_{12}-\lambda_{13}\lambda_{13}\nonumber\\&&
-\lambda_{20}\lambda_{20}-\lambda_{21}\lambda_{21}-\lambda_{22}\lambda_{22}-\lambda_{23}\lambda_{23}
-\lambda_{30}\lambda_{30}-\lambda_{31}\lambda_{31}-\lambda_{32}\lambda_{32}-\lambda_{33}\lambda_{33}\nonumber\\&&
-8\lambda_{12}\lambda_{21}-2\lambda_{11}\lambda_{22}-2\lambda_{13}\lambda_{31}
-4\lambda_{23}\lambda_{32}-4\lambda_{11}\lambda_{33}+4\lambda_{22}\lambda_{33}),\\
\frac{d\lambda_{12}}{dl}
&=&
2\lambda_{00}\lambda_{12}-4\lambda_{11}\lambda_{12}-2\lambda_{12}\lambda_{22}+2\lambda_{23}\lambda_{31}
-\lambda_{13}\lambda_{32}-3\lambda_{12}\lambda_{33}-2\lambda_{21}\lambda_{33},\\
\frac{d\lambda_{13}}{dl}
&=&
\lambda_{00}\lambda_{13}-3\lambda_{11}\lambda_{13}-\lambda_{13}\lambda_{22}-4\lambda_{12}\lambda_{23}
-2\lambda_{22}\lambda_{31}+2\lambda_{21}\lambda_{32}-\lambda_{13}\lambda_{33},\\
\frac{d\lambda_{20}}{dl}
&=&
-(2\lambda_{10}\lambda_{21}+4\lambda_{20}\lambda_{22}+2\lambda_{23}\lambda_{30}),\\
\frac{d\lambda_{21}}{dl}
&=&
\lambda_{00}\lambda_{21}-\lambda_{11}\lambda_{21}-5\lambda_{21}\lambda_{22}-\lambda_{23}\lambda_{31}
+2\lambda_{13}\lambda_{32}-2\lambda_{12}\lambda_{33}-2\lambda_{21}\lambda_{33},\\
\frac{d\lambda_{22}}{dl}
&=&2\lambda_{00}\lambda_{22}-\lambda_{12}\lambda_{21}-3\lambda_{11}\lambda_{22}-2\lambda_{22}\lambda_{22}
-2\lambda_{13}\lambda_{31}-\lambda_{23}\lambda_{32}+2\lambda_{11}\lambda_{33}-3\lambda_{22}\lambda_{33},\\
\frac{d\lambda_{23}}{dl}
&=&
\lambda_{00}\lambda_{23}-\lambda_{13}\lambda_{21}-2\lambda_{11}\lambda_{23}-5\lambda_{22}\lambda_{23}
+2\lambda_{12}\lambda_{31}-2\lambda_{11}\lambda_{32}-\lambda_{23}\lambda_{33},\\
\frac{d\lambda_{30}}{dl}
&=&
-(\lambda_{00}\lambda_{03}+\lambda_{20}\lambda_{23}+\lambda_{10}\lambda_{13}+2\lambda_{10}\lambda_{31}
+4\lambda_{20}\lambda_{32}+3\lambda_{30}\lambda_{33}),\\
\frac{d\lambda_{31}}{dl}
&=&
2\lambda_{12}\lambda_{23}-2\lambda_{13}\lambda_{22}-\lambda_{11}\lambda_{31}-\lambda_{22}\lambda_{31}
+\lambda_{00}\lambda_{31}-4\lambda_{21}\lambda_{32}-3\lambda_{31}\lambda_{33},\\
\frac{d\lambda_{32}}{dl}
&=&
2\lambda_{13}\lambda_{21}-\lambda_{12}\lambda_{31}+2\lambda_{00}\lambda_{32}-2\lambda_{22}\lambda_{32}
-2\lambda_{11}\lambda_{23}-3\lambda_{11}\lambda_{32}-4\lambda_{32}\lambda_{33},\\
\frac{d\lambda_{33}}{dl}
&=&
\frac{1}{2}(2\lambda_{00}\lambda_{33}-\lambda_{00}\lambda_{00}-\lambda_{01}\lambda_{01}-\lambda_{02}\lambda_{02}-\lambda_{03}\lambda_{03}
-\lambda_{10}\lambda_{10}-\lambda_{11}\lambda_{11}-\lambda_{12}\lambda_{12}-\lambda_{13}\lambda_{13}\nonumber\\&&
-\lambda_{20}\lambda_{20}-\lambda_{21}\lambda_{21}-\lambda_{22}\lambda_{22}-\lambda_{23}\lambda_{23}
-\lambda_{30}\lambda_{30}-\lambda_{31}\lambda_{31}-\lambda_{32}\lambda_{32}-3\lambda_{33}\lambda_{33}\nonumber\\&&
-4\lambda_{12}\lambda_{21}+4\lambda_{11}\lambda_{22}-2\lambda_{13}\lambda_{31}-8\lambda_{23}\lambda_{32}
-4\lambda_{11}\lambda_{33}-2\lambda_{22}\lambda_{33}).\label{Eq_lambda33}
\end{eqnarray}
\end{widetext}
Due to the relevant one-loop corrections stemmed from the vanishing density of states near the QBCP,
the coupled RG flow equations exhibit strong interdependence and nontrivial competition among distinct kinds
of fermion-fermion interactions, producing scale-dependent effective couplings that dictates the low-energy
collective behavior of the system. In order to decode the underlying low-energy physics of 2D QBCP systems,
we proceed to seek and classify the potential fixed points of these interaction parameters in Sec.~\ref{Sec_classification_FPs},
and in Sec.~\ref{Sec_phase_transition}, we systematically investigate the correlated instabilities around
the fixed points and determine associated phase transitions.


\section{Relative fixed points}\label{Sec_classification_FPs}

We now turn to a systematic analysis of the coupled RG flow equations~(\ref{Eq_lambda00})-(\ref{Eq_lambda33}),
with the aim of elucidating the low-energy behavior of fermion-fermion interactions that control the physical
structure of the 2D QBCP system.

Since the RG trajectories depend sensitively on the initial coupling
values (i.e., $\lambda_{ij}(l=0)$ and dubbed $\lambda_{ij}(0)$ from now on for brevity), we adopt a
stepwise strategy, which progressively relaxes the constraints on the initial conditions and clusters
into three scenarios: (A) the isotropic scenario, wherein all sixteen couplings share a common value at $l=0$,
and (B) the anisotropic scenario, in which a single coupling is initially dominant over all others, as
well as (C) the general scenario, with all couplings unbiasedly chosen from a broad range.

This hierarchical progression, moving from the simplest idealized limit to the most complex realistic setting,
enables us to identify both the universal features of the RG flow and those that are sensitive to the initial conditions.
In particular, this clarifies the relative roles of universal fixed points versus bare-parameter sensitivity
in interaction-driven instabilities and phase transitions.

\begin{figure}[htpb]
\subfigure[]
{\includegraphics[width=2.98in]{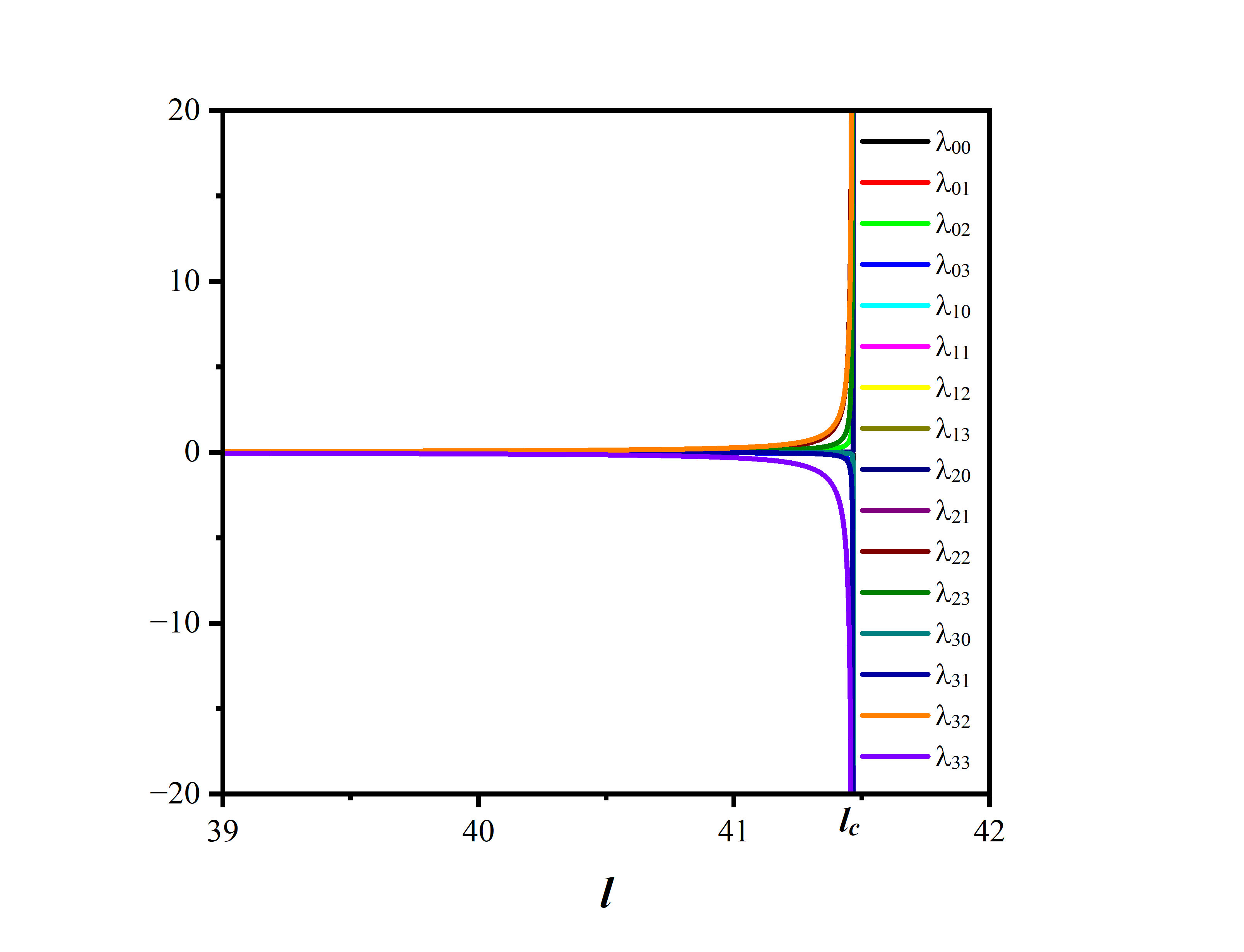}}\\
\vspace{-0.35cm}
\hspace{0.1cm}
\subfigure[]
{\includegraphics[width=3.05in]{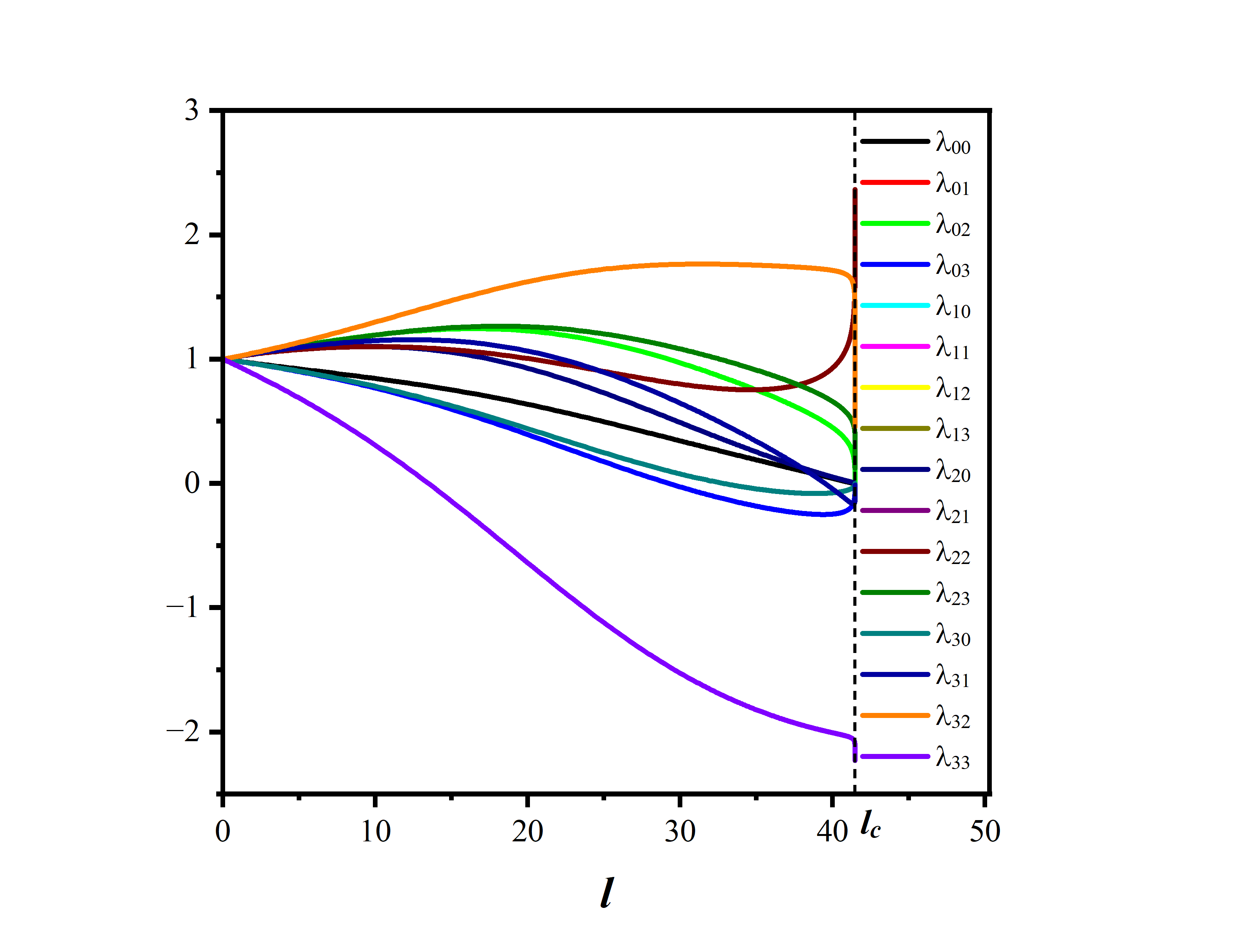}}
\vspace{-0.2cm}
\caption{(Color online) Energy-dependent evolution of all sixteen fermion-fermion couplings
with initial value $\lambda_0=10^{-2}$. (a) Unrescaled couplings diverge at a common critical
scale $l_c\simeq 41.5$. (b) Rescaled couplings remain finite and flow to the unique fixed point FP$_0$.}
\label{fig:evolution}
\end{figure}

\subsection{Isotropic scenario}\label{Sec_FP_Limit}

At first, we consider the isotropic scenario in which $\lambda_{ij}(0)=\lambda_0$ for all $i,j$.
As the RG flow proceeds toward lower energies, these couplings evolve along distinct quantitative trajectories
but eventually diverge at a common critical scale $l_c$ shown in Fig.~\ref{fig:evolution}(a).

To regularize the flow and systematically identify the fixed-point structure, it is helpful to
introduce a reference scale~\cite{Vafek2014PRB,Fu-Wang2024AP}
\begin{eqnarray}
\lambda_{\rm scale}(l)\equiv
\sqrt{\frac{1}{16}\sum_{i,j=0}^3\lambda_{ij}^2(l)},
\end{eqnarray}
and rescale each coupling according to
\begin{eqnarray}
\lambda_{ij}(l)\to \frac{\lambda_{ij}(l)}{\lambda_{\rm scale}(l)}.
\end{eqnarray}
Hereafter, unless explicitly noted otherwise, all $\lambda_{ij}$ denote these rescaled quantities.

As depicted in Fig.~\ref{fig:evolution}(b), the scale $l_c$ marks the critical
point~\cite{Shankar1994RMP,Sachdev2011Book}, and the fixed point is defined as
\begin{eqnarray}
\mathrm{FP}\equiv\left.\{\lambda_{ij}\}\right|_{l=l_c}.\label{Eq_definition_FP}
\end{eqnarray}
Such FPs are fundamentally connected to quantum critical phenomena~\cite{Vojta2003RPP,Sachdev2011Book,Halboth2000RPB,Metzner2000PRL,Maiti2010PRB,Vafek2014PRB,Chubukov2016PRX,
Roy2018PRX,Nandkishore2012NP,Cvetkovic2012PRB} and are instrumental in governing the low-energy behavior
of 2D QBCP systems. In the rest of this section, we aim to search for all possible FPs of
the RG equations~(\ref{Eq_lambda00})-(\ref{Eq_lambda33}) and postpone the
analysis of their physical implications to Sec.~\ref{Sec_phase_transition}.

Specifically, we find that in the isotropic scenario, the rescaled RG flows are essentially insensitive to the
specific value of $\lambda_0>0$ (i.e., the initial conditions) as depicted in
Fig~\ref{fig:evolution}(b). All trajectories go towards a unique attractive FP, designated by FP$_0$,
which takes the form of
\begin{eqnarray}
\mathrm{FP_0}\approx
\left(
  \begin{array}{cccc}
    0 & 0 & 0 & 0 \\
    0 & -2.2342& 0.4072 & -0.0090 \\
    0 & 0.1052 & 2.3796 & 0.1052 \\
    0 & 0.0090 & 0.4072 & -2.2343 \\
  \end{array}
\right).\label{Eq_FP0}
\end{eqnarray}
Although the critical scale $l_c$ depends on $\lambda_0$, the FP$_0$ remains invariant. Consequently, in this scenario,
the low-energy physics is determined by a universal fixed point, independent of the overall strength of
the initial interactions.

\subsection{Anisotropic scenario}\label{Sec_FP_Special}

Subsequently, we turn to the anisotropic scenario, in which one specific interaction parameter
is initially bigger than all the others. Specifically, we set one coupling $\lambda_{ij}(0)$ to
a value in the range $10^{-5}$ to $10^{-2}$, while keeping the remaining fifteen parameters at the baseline value $10^{-5}$. By systematically varying the dominant coupling, we can trace how the fixed-point structure
of fermion-fermion interactions evolves as a function of the anisotropic initial condition.

\begin{figure}[htpb]
\subfigure[]{
\includegraphics[width=3.05in]{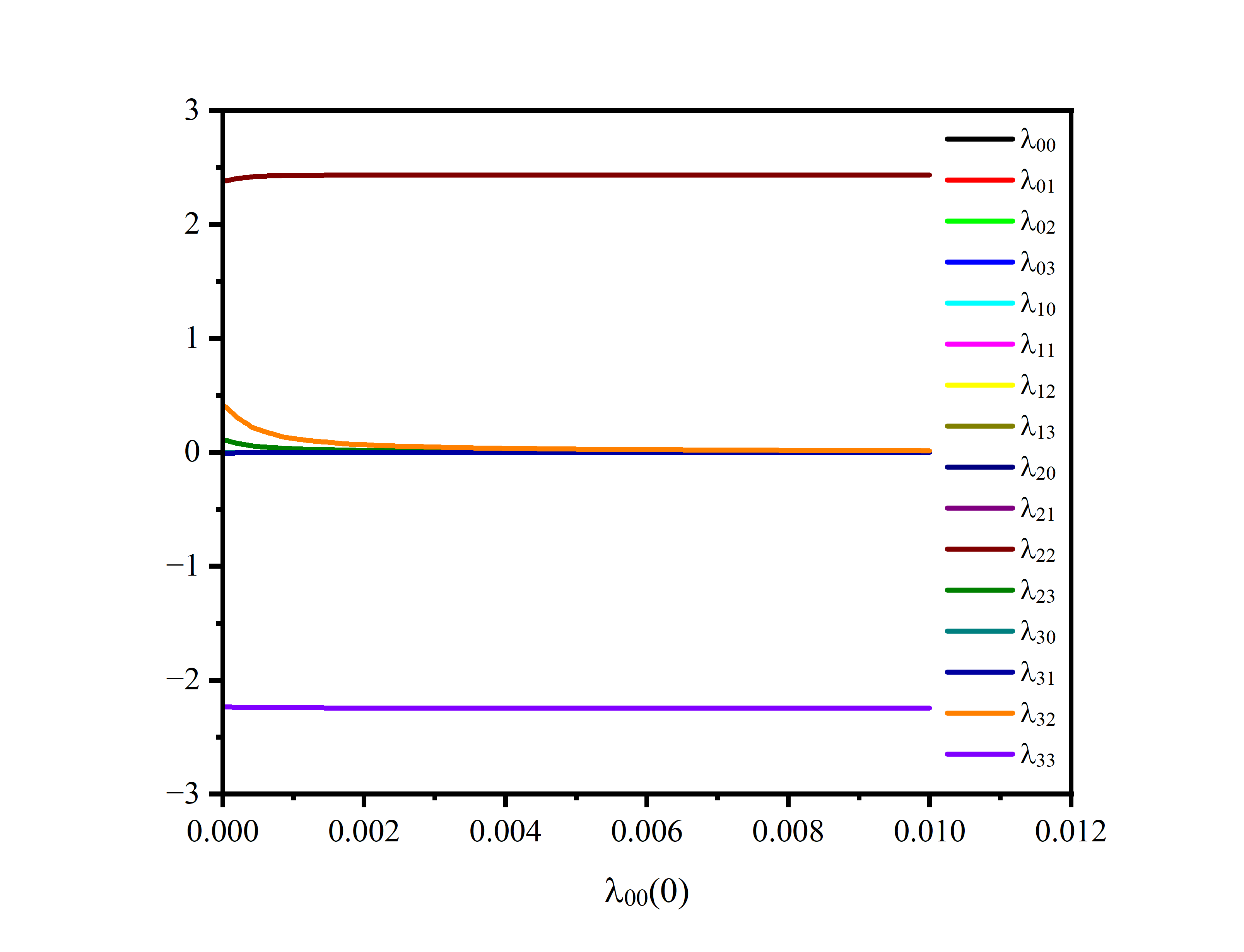}}\\
\vspace{-0.40cm}
\subfigure[]{
\includegraphics[width=3.05in]{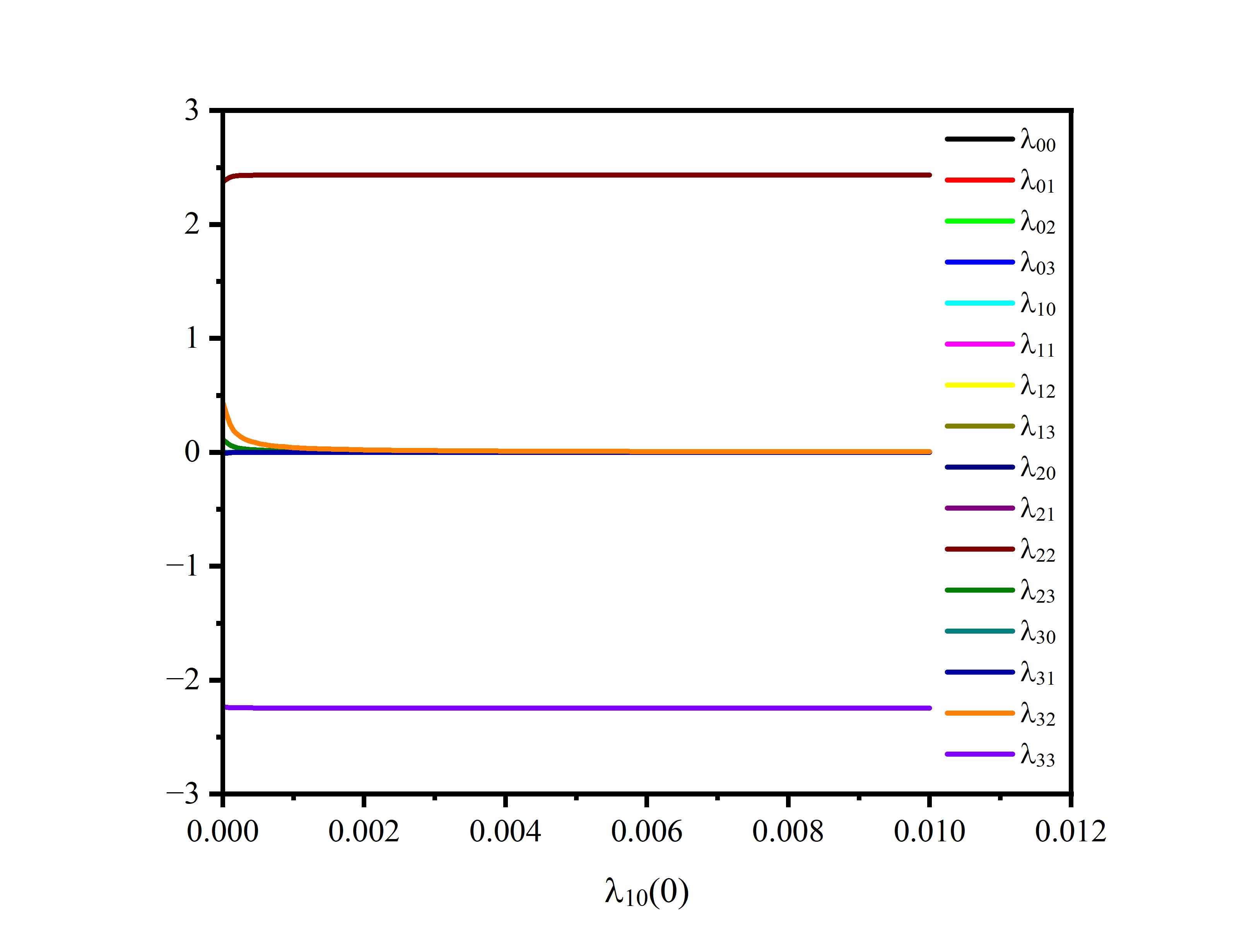}}
\vspace{-0.2cm}
\caption{(Color online) Evolution of critical values at $l_c$ with (a) $\lambda_{00}(0)$ and (b) $\lambda_{10}(0)$ varied.
All other initial couplings are $10^{-5}$. For large dominant couplings, all rescaled couplings converge
to the same FP$_1$, demonstrating its universality.}
\label{fig:FP1}
\end{figure}

The numerical results indicate that the emergent FPs fall into several distinct categories, depending
on which interaction dominates initially. For clarity, we group the results as follows.

When any of the eight couplings $\lambda_{00}$, $\lambda_{01}$, $\lambda_{02}$, $\lambda_{03}$, $\lambda_{10}$, $\lambda_{20}$, $\lambda_{22}$, or $\lambda_{30}$ dominates, the system is attracted by the same FP, which we denote as FP$_1$ and reads,
\begin{eqnarray}
\mathrm{FP_1}\approx\left(
  \begin{array}{cccc}
    0 & 0 & 0 & 0 \\
    0 & -2.2441 & 0 & 0 \\
    0 & 0 & 2.4347 & 0 \\
    0 & 0 & 0  &-2.2441\\
  \end{array}
\right).\label{Eq_FP1}
\end{eqnarray}
In Fig.~\ref{fig:FP1}, we present representative plots of the RG evolution as a function of $\lambda_{00}(0)$ and $\lambda_{10}(0)$. These plots confirm that the system consistently flows to FP$_1$ provided the dominant coupling is sufficiently large compared to the baseline value of $10^{-5}$ (the basic results are insensitive to the concrete values of dominant coupling).

\begin{figure*}[htpb]
\centering
\subfigure[FP$_2$]
{\includegraphics[width=2.5in]{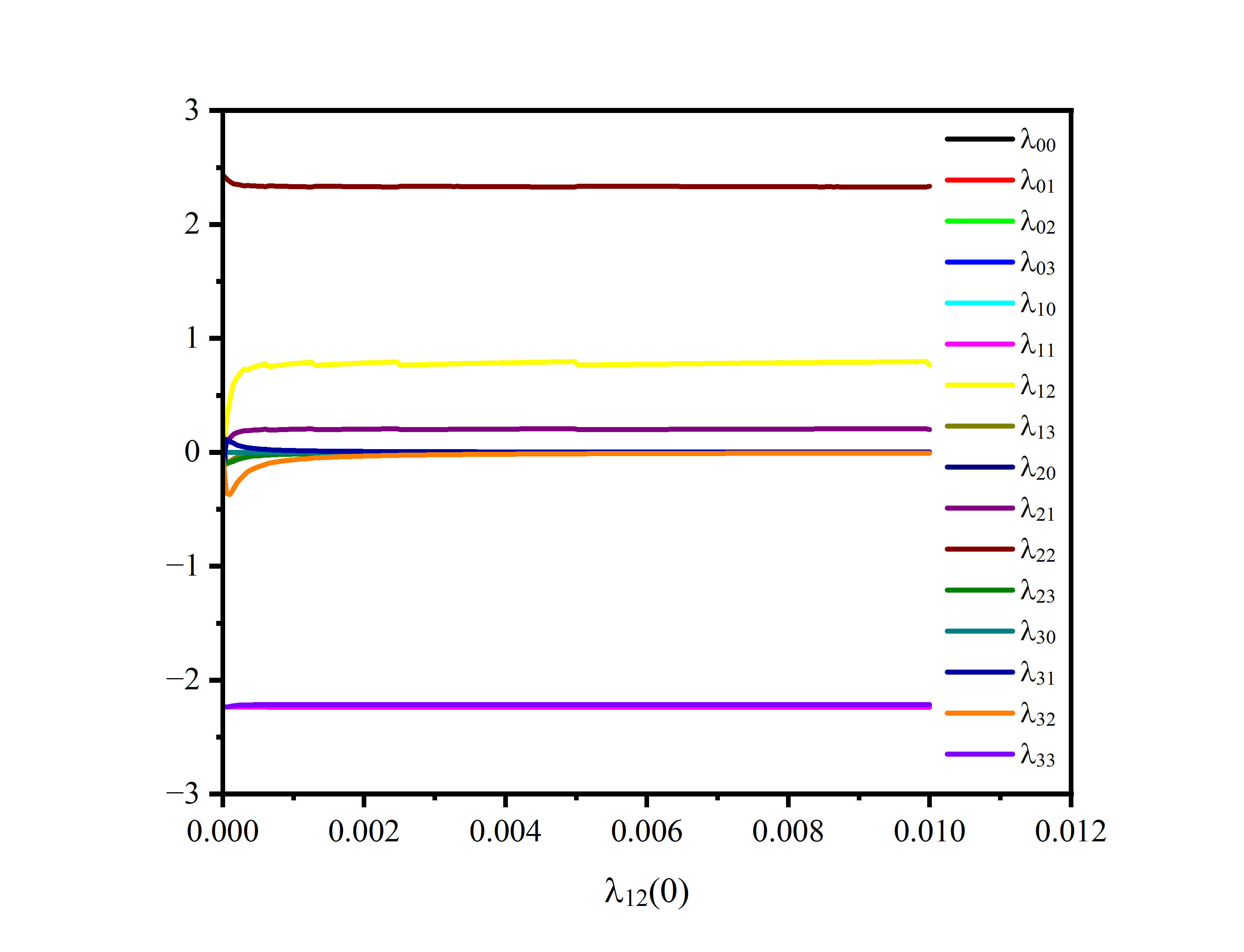}}
\hspace{-.95cm}
\subfigure[FP$_7$]
{\includegraphics[width=2.5in]{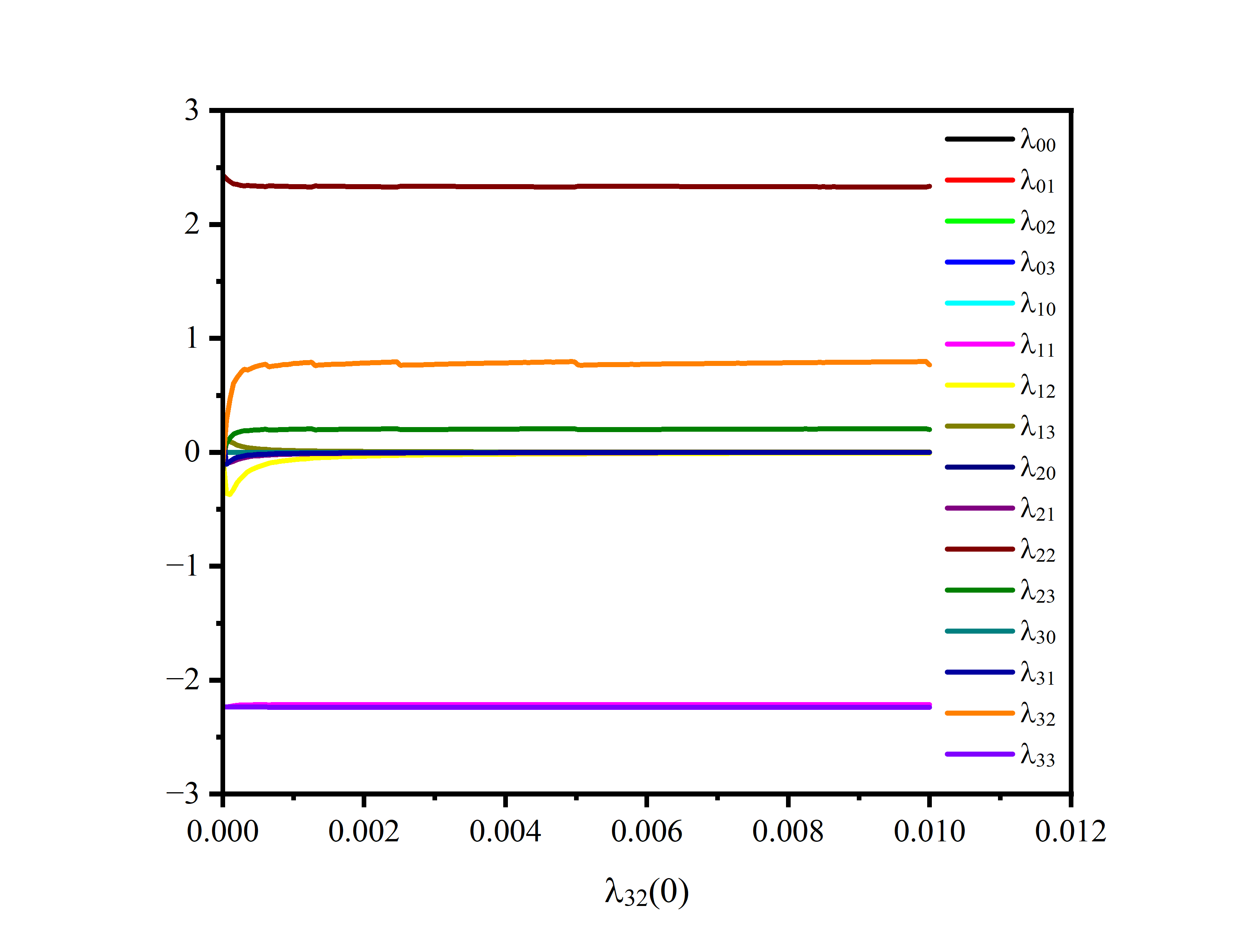}}
\hspace{-0.95cm}
\subfigure[FP$_3$]
{\includegraphics[width=2.5in]{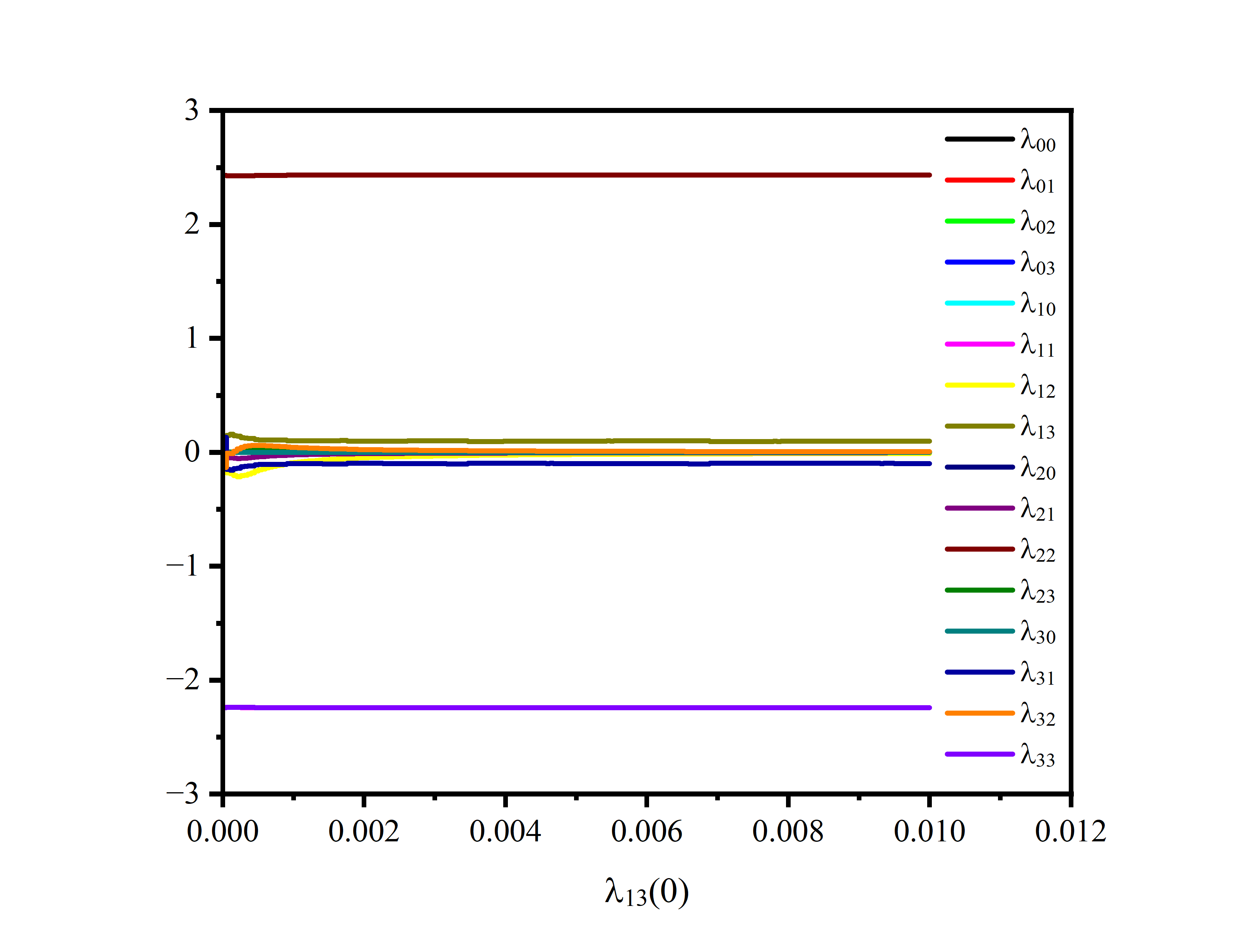}}\\
\vspace{-0.15cm}
\subfigure[FP$_6$]
{\includegraphics[width=2.5in]{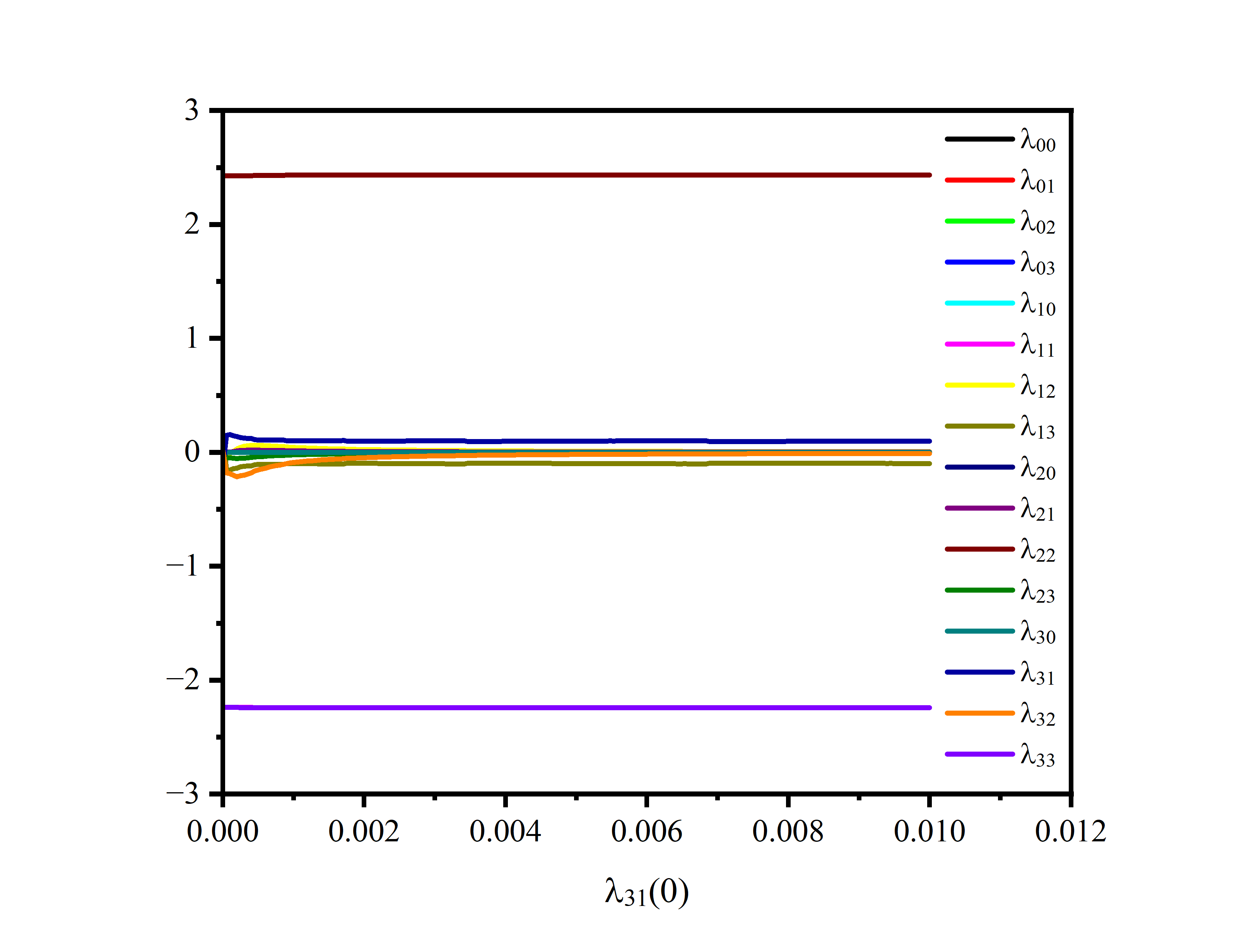}}
\hspace{-0.95cm}
\subfigure[FP$_4$]
{\includegraphics[width=2.5in]{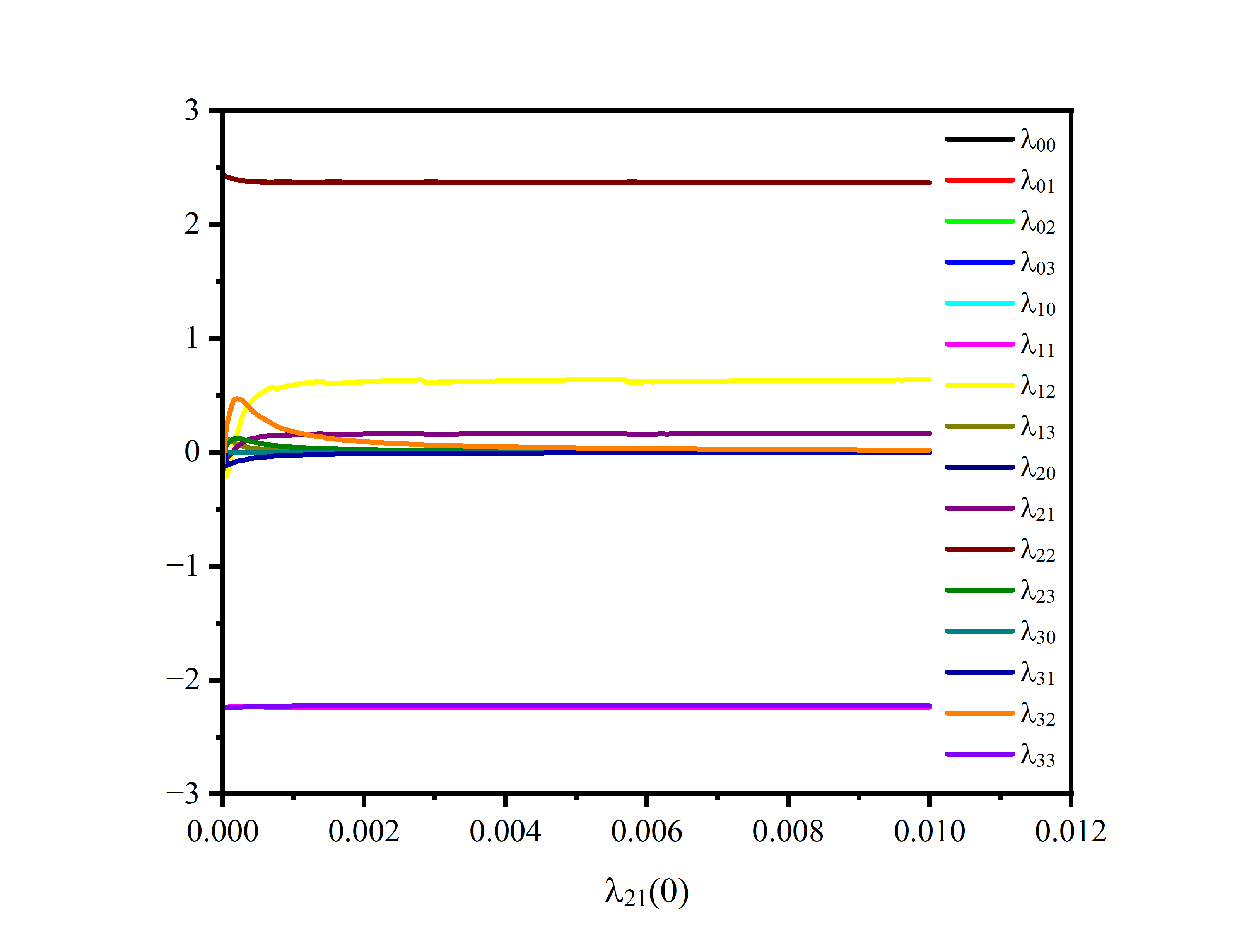}}
\hspace{-0.95cm}
\subfigure[FP$_5$]
{\includegraphics[width=2.5in]{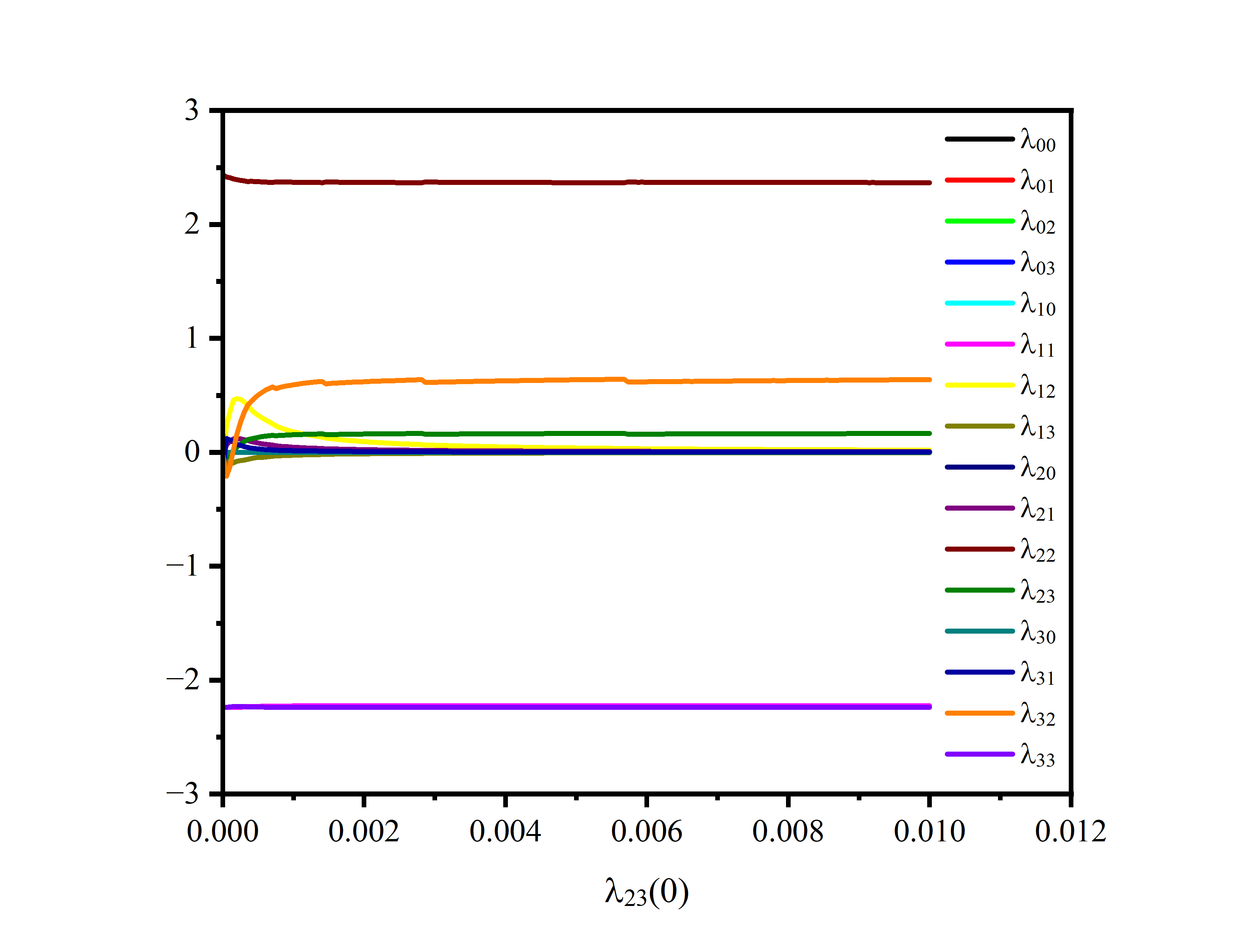}}\\
\vspace{-0.15cm}
\subfigure[Unstable]
{\includegraphics[width=2.5in]{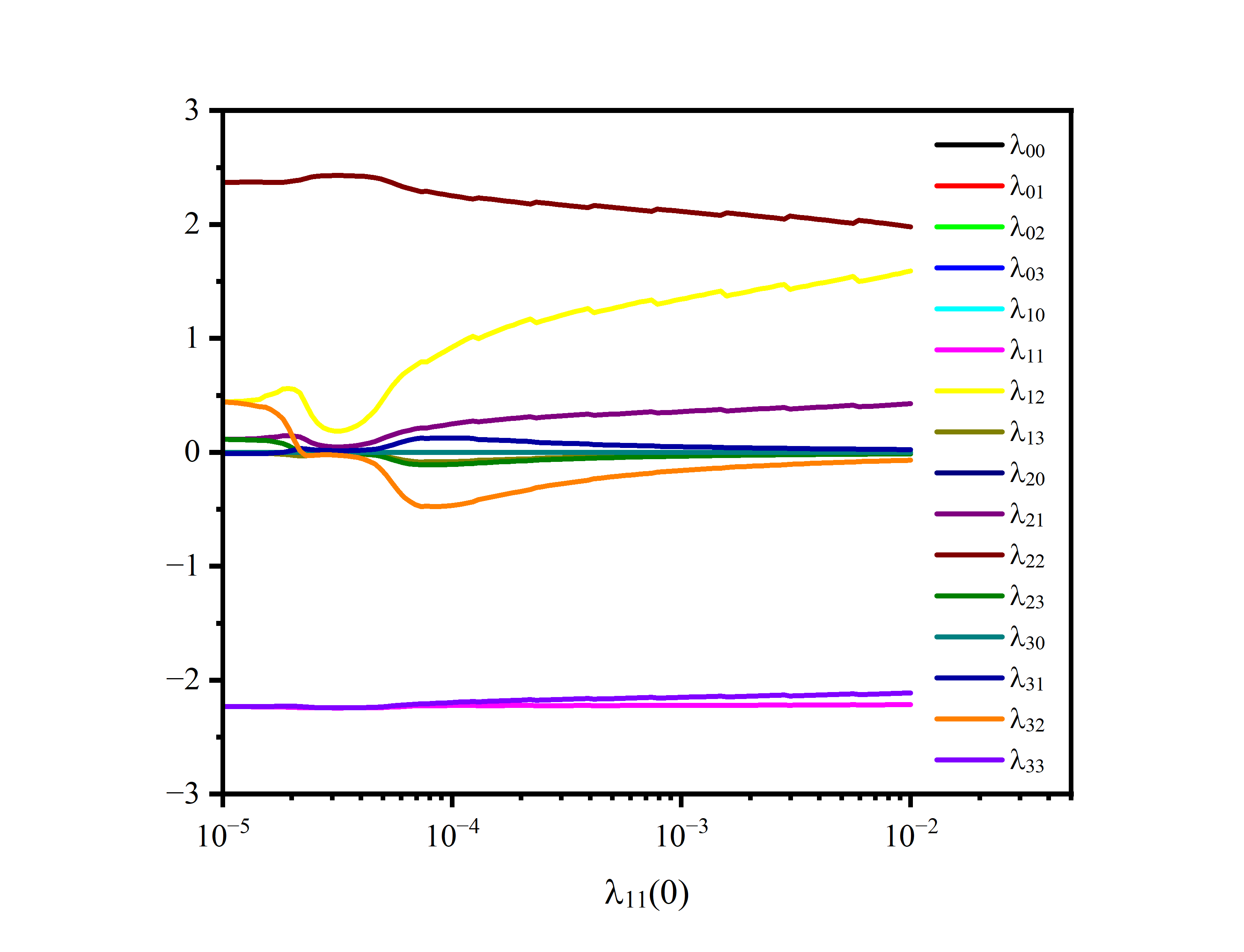}}
\hspace{0.85cm}
\subfigure[Unstable]
{\includegraphics[width=2.5in]{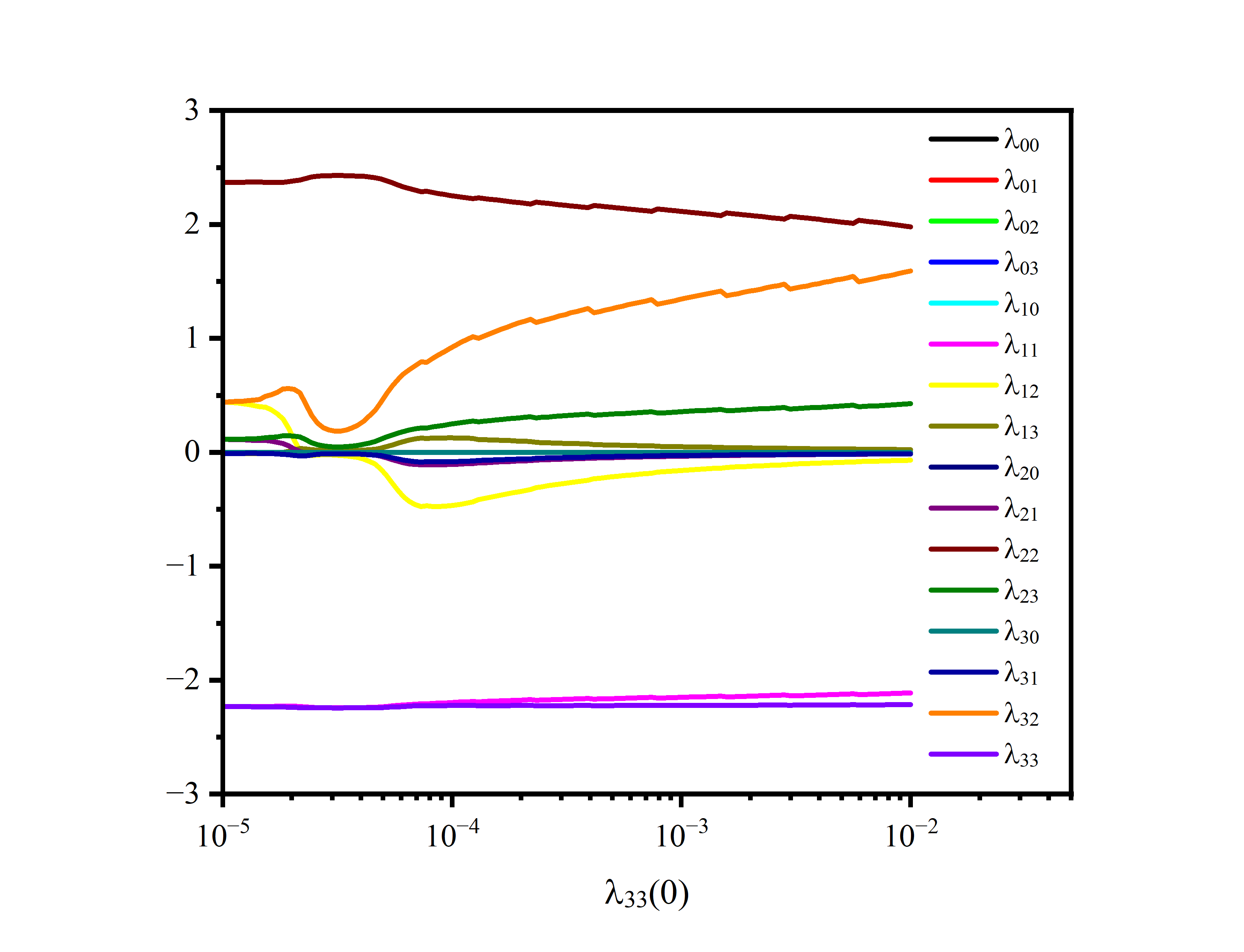}}
\vspace{-0.0cm}
\caption{(Color online) Evolution of FP$_{2}$¨CFP$_{7}$ and the unstable cases.
(a)¨C(f) Fixed-point couplings versus the dominant initial coupling: (a) FP$_2$ ($\lambda_{12}$),
(b) FP$_7$ ($\lambda_{32}$), (c) FP$_3$ ($\lambda_{13}$), (d) FP$_6$ ($\lambda_{31}$), (e) FP$_4$
($\lambda_{21}$), (f) FP$_5$ ($\lambda_{23}$). (g),(h) Unstable flows for varied $\lambda_{11}(0)$
and $\lambda_{33}(0)$, respectively. All other initial couplings are $10^{-5}$.}
\label{fig:FPs}
\end{figure*}

With respect to the remaining six couplings, we find that each dominant parameter drives the system to a distinct FP.
For convenience, they are labelled as FP$_2$ through FP$_7$ and expressed as,
\begin{eqnarray}
\mathrm{FP_2}
&\approx&
\left(
  \begin{array}{cccc}
    0 & 0 & 0 & 0 \\
    0 & -2.2375 & 0.7653 & 0 \\
    0 & 0.1986 & 2.3516 & 0 \\
    0 & 0 & 0 &-2.2156\\
  \end{array}
\right),\label{Eq_FP2}\\
\mathrm{FP_3}
&\approx&
\left(
  \begin{array}{cccc}
    0 & 0 & 0 & 0 \\
    0 & -2.2441 & 0 & 0.0981 \\
    0 & 0 & 2.4347 & 0 \\
    0 & -0.0981 & 0 &-2.2441\\
  \end{array}
\right),\label{Eq_FP3}\\
\mathrm{FP_4}
&\approx&
\left(
  \begin{array}{cccc}
    0 & 0 & 0 & 0 \\
    0 & -2.2395 & 0.6385 & 0 \\
    0 & 0.1653 & 2.3668 & 0 \\
    0 & 0 & 0 &-2.2244\\
  \end{array}
\right),\label{Eq_FP4}\\
\mathrm{FP_5}
&\approx&
\left(
  \begin{array}{cccc}
    0 & 0 & 0 & 0 \\
    0 & -2.2244 & 0 & 0 \\
    0 & 0 & 2.3668 & 0.1653 \\
    0 & 0 & 0.6385 &-2.2395\\
  \end{array}
\right),\label{Eq_FP5}\\
\mathrm{FP_6}
&\approx&
\left(
  \begin{array}{cccc}
    0 & 0 & 0 & 0 \\
    0 & -2.2441 & 0 & -0.0981 \\
    0 & 0 & 2.4347 & 0 \\
    0 & 0.0981 & 0 &-2.2441\\
  \end{array}
\right),\label{Eq_FP6}\\
\mathrm{FP_7}
&\approx&
\left(
  \begin{array}{cccc}
    0 & 0 & 0 & 0 \\
    0 & -2.2156 & 0 & 0 \\
    0 & 0 & 2.2516 & 0.1986 \\
    0 & 0 & 0.7653 &-2.2375\\
  \end{array}
\right),\label{Eq_FP7}
\end{eqnarray}
corresponding to the dominance of $\lambda_{12}$, $\lambda_{13}$, $\lambda_{21}$, $\lambda_{23}$, $\lambda_{31}$, and $\lambda_{32}$, respectively. Fig.~\ref{fig:FPs} quantitatively shows how these FPs evolve with the initial value of dominant interaction.
Notably, as illustrated in Fig.~\ref{fig:FPs}(a)-(f), these FPs fall into three related pairs,
namely FP$_2$ $\leftrightarrow$ FP$_7$, FP$_3$ $\leftrightarrow$ FP$_6$, and FP$_4$ $\leftrightarrow$ FP$_5$,
which are connected by exchanging the corresponding off-diagonal indices $\lambda_{12}\leftrightarrow\lambda_{32}$, $\lambda_{13}\leftrightarrow\lambda_{31}$, $\lambda_{21}\leftrightarrow\lambda_{23}$, respectively.
This indicates an underlying discrete symmetry of FPs in the interaction space.

Interestingly, for the cases where $\lambda_{11}(0)$ or $\lambda_{33}(0)$ is dominant, the rescaled fermion-fermion interactions
cannot flow towards a stable FP as shown in Fig.~\ref{fig:FPs}(g)-(h). As a consequence,
we do not define FPs for these two cases, suggesting that the corresponding channels cause
the system exhibit runaway flow.

\subsection{General scenario}\label{Sec_FP_General}

Furthermore, we address the most general and physically realistic scenario, where all sixteen
initial interaction parameters are unbiasedly chosen.

To this end, we sample each $\lambda_{ij}(0)$ with $i,j=0,1,2,3$
from a log-uniform distribution over the broad interval $[10^{-5}, 10^{-2}]$
(to avoid overlap with the isotropic scenario, we exclude this case hereby), generating a large ensemble
of random initial conditions for subsequent RG analysis.
With the help of systematically statistical analysis of the resulting FPs, we find that the overwhelming majority
can be classified into three parameter-dependent families, which we refer to as fixed-point groups (FPGs),
\begin{eqnarray}
\mathrm{FPG_1}
&\approx&
\left(
  \begin{array}{cccc}
    0 & 0 & 0 & 0 \\
    0 & -2.2 & 0 & \alpha \\
    0 & 0 & 2.4 & 0 \\
    0 & -\alpha & 0 & -2.2 \\
  \end{array}
\right),\label{Eq_FPG1}\\
\mathrm{FPG_2}
&\approx&
\left(
  \begin{array}{cccc}
    0 & 0 & 0 & 0 \\
    0 & -2.2 & 3.86\beta & 0 \\
    0 & \beta & 2.4 & 0 \\
    0 & 0 & 0 & -2.2 \\
  \end{array}
\right),\label{Eq_FPG2}\\
\mathrm{FPG_3}
&\approx&
\left(
  \begin{array}{cccc}
    0 & 0 & 0 & 0 \\
    0 & -2.2 & 0 & 0 \\
    0 & 0 & 2.4 & \gamma \\
    0 & 0 & 3.86\gamma & -2.2 \\
  \end{array}
\right).\label{Eq_FPG3}
\end{eqnarray}
Here, the parameters $\alpha$, $\beta$, and $\gamma$ are determined by the specific initial conditions,
which are approximately confined by $[-0.25,0.25]$, $[0,0.2]$, and $[0,0.2]$, respectively. It is of
particular importance to highlight that all FPs previously found in the anisotropic scenario~\ref{Sec_FP_Special} can
be recovered as limiting cases of these three FPGs. Specifically, FP$_1$ coincides with the $\alpha=0$
limit of FPG$_1$. Moreover, for appropriately chosen values of the parameters $(\alpha,\beta,\gamma)$,
FP$_{3,6}$ reduce to members of FPG$_1$, FP$_{2,4}$ are recovered by FPG$_2$, and FP$_{5,7}$ fall within
FPG$_3$, respectively.  This thereby furnishes a
bridge between the anisotropic scenario and the general scenario,
demonstrating that the latter continuously connects to the former as the initial conditions
are progressively constrained.


To recapitulate, both FPs or FPGs are of direct physical relevance and associated with certain physical
implications. The unrescaled fermion-fermion interaction strengths, as shown in Fig.~\ref{fig:evolution}, diverge
at the critical scale $l_c$, signaling the onset of strong-coupling phenomena. Near a FP, the system becomes unstable toward some symmetry-breaking and phase transitions. The specific FP reached by the RG flow thus determines the dominant low-energy instability,
which will be studied in the following section~\ref{Sec_phase_transition}.

\begin{figure*}[htbp]
\centering
\subfigure[]{
\includegraphics[width=3.5in]{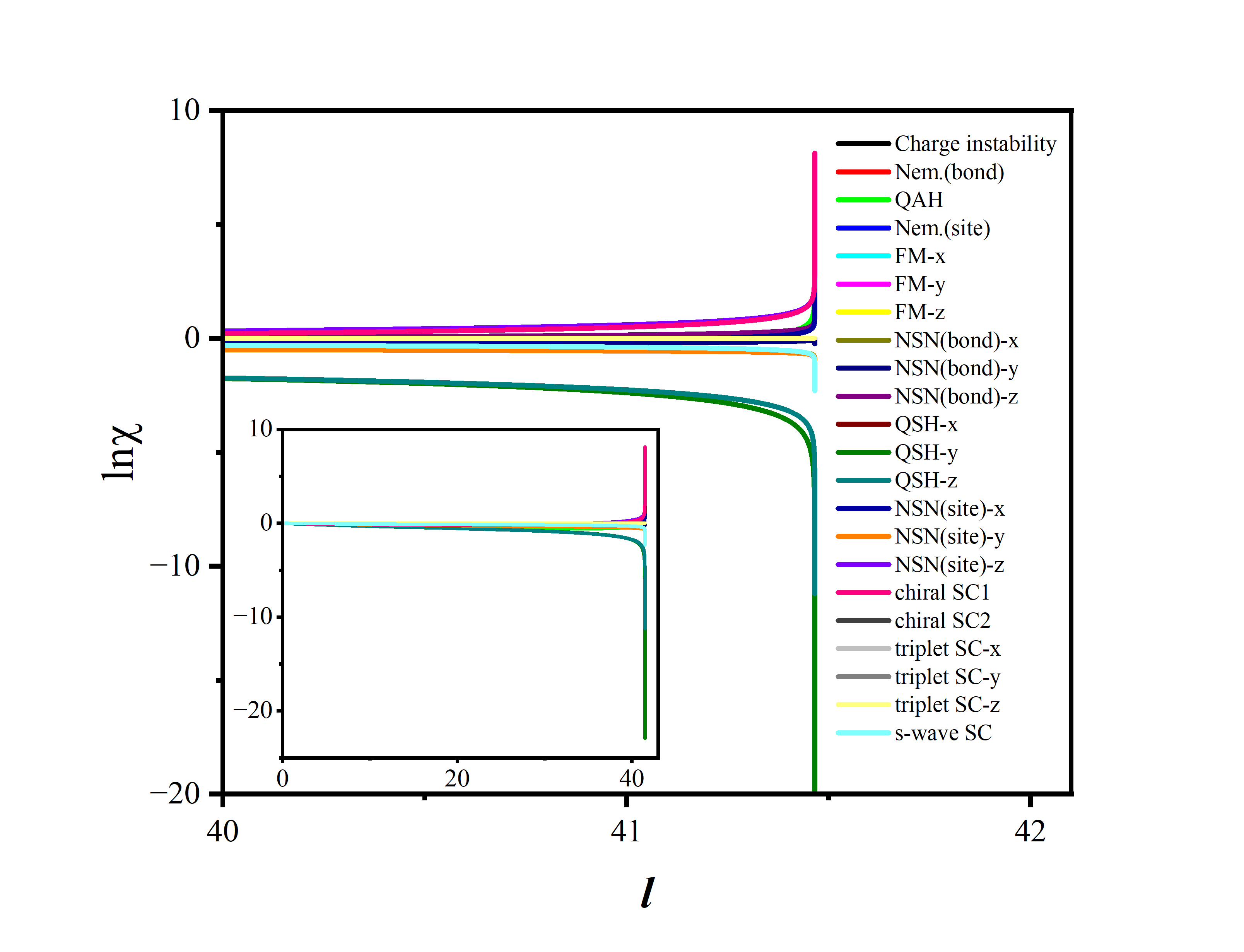}
}\hspace{-0.9cm}
\subfigure[]{
\includegraphics[width=3.5in]{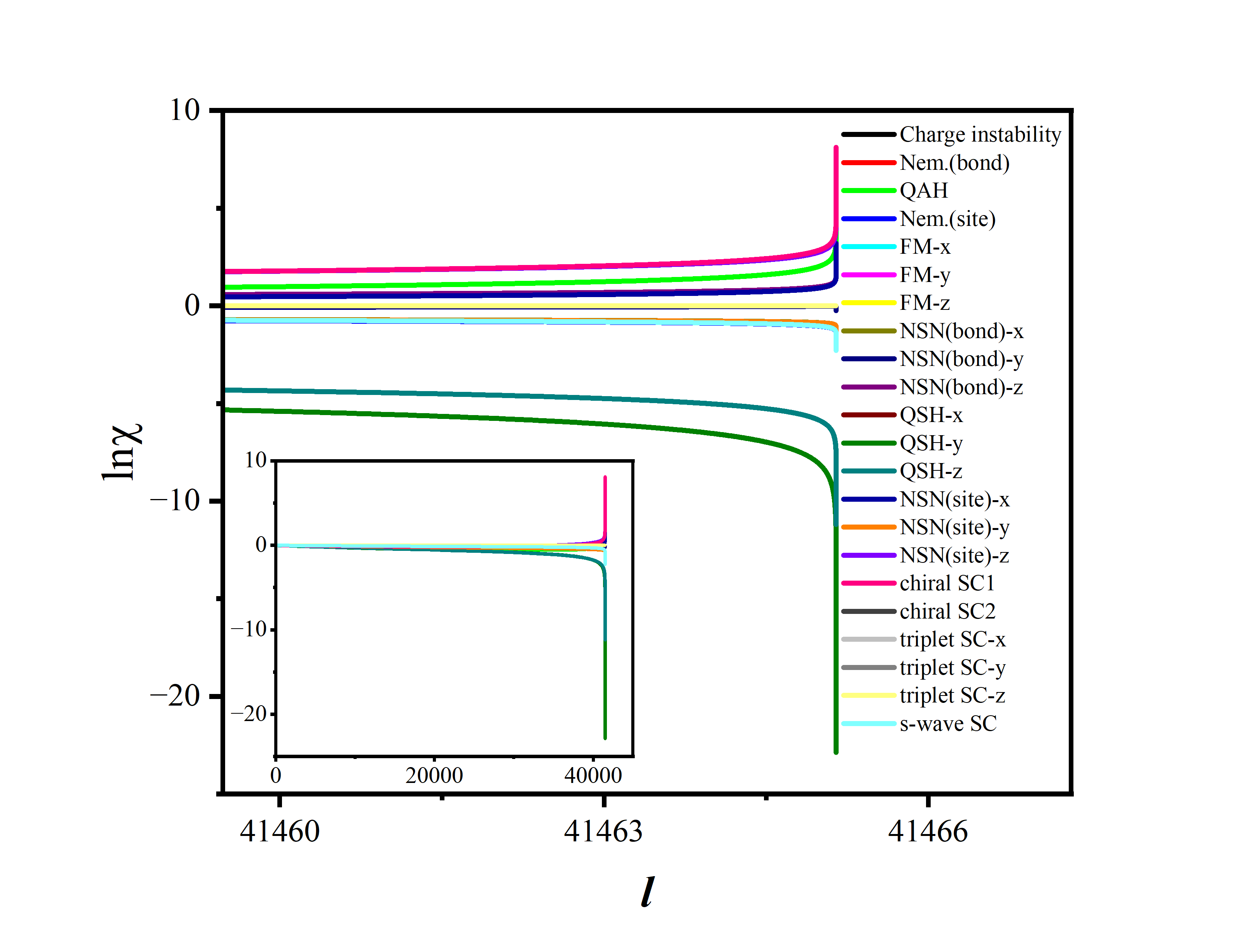}
}
\\
\vspace{-0.3cm}
\subfigure[]{
\includegraphics[width=3.5in]{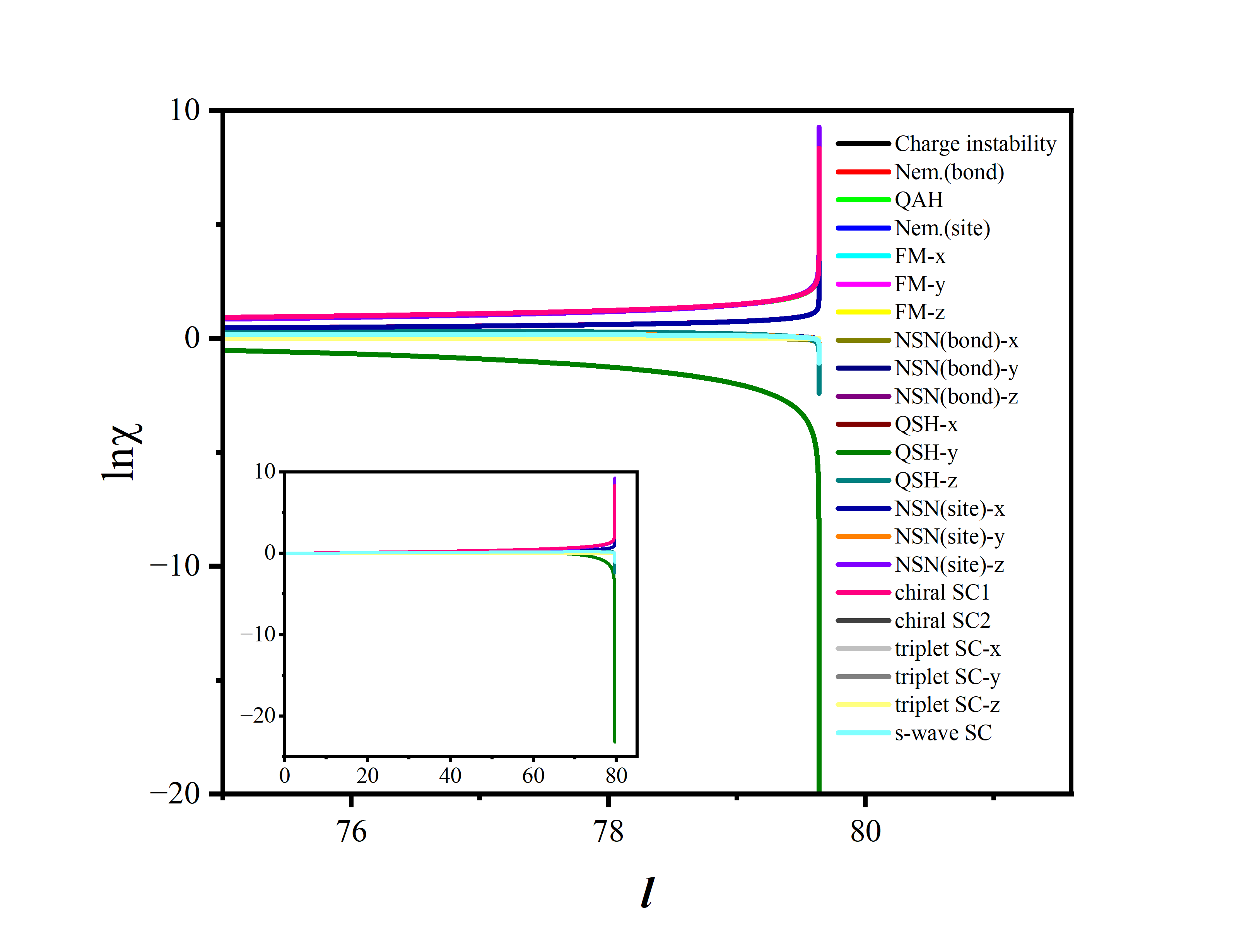}
}
\hspace{-1.1cm}
\subfigure[]{
\includegraphics[width=3.5in]{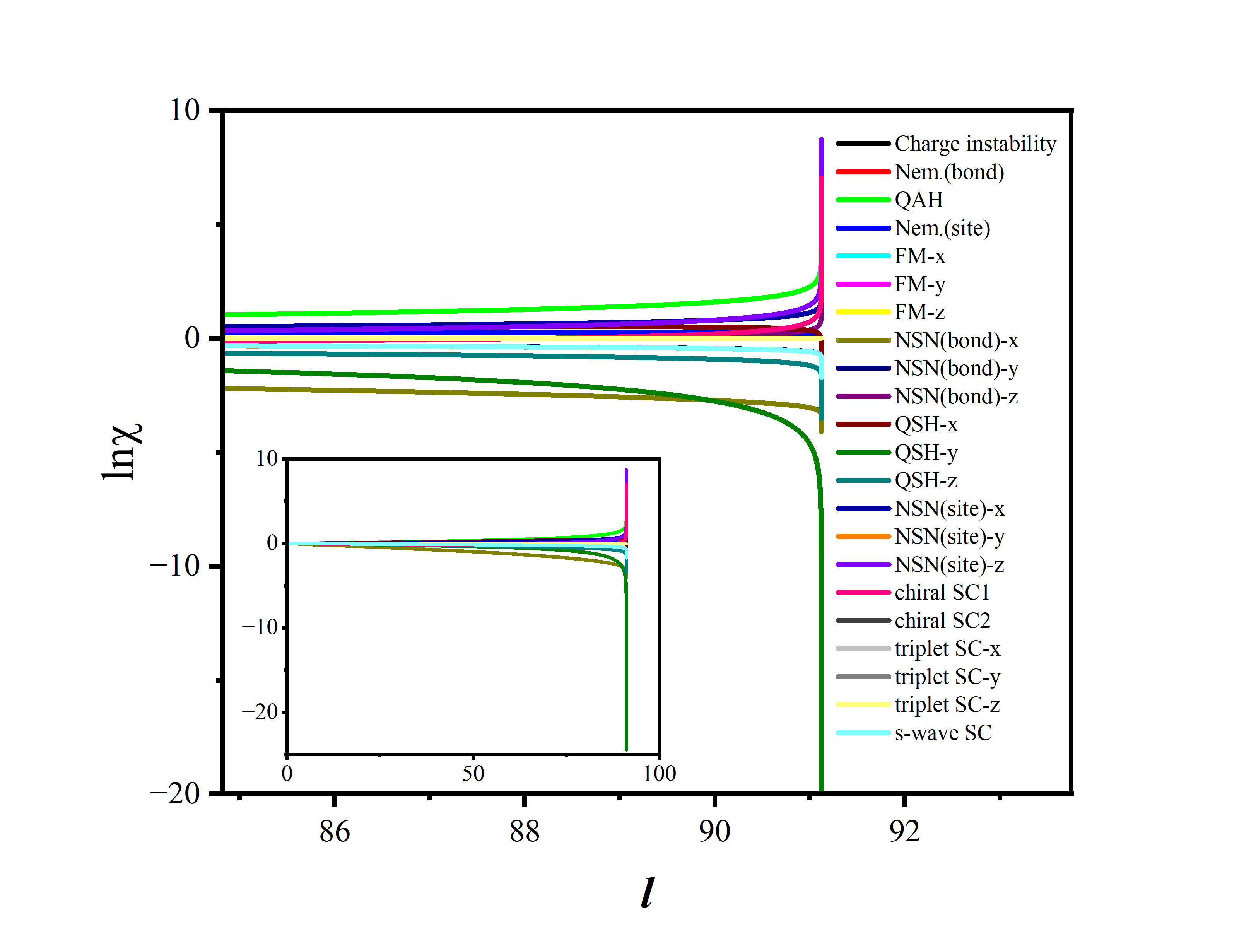}
}\vspace{-0.1cm}
\caption{(Color online) Energy-dependent susceptibilities of candidate instabilities for
(a) $\rm FP_0$ with $\lambda_0=10^{-2}$, (b) $\rm FP_0$ with $\lambda_{0}=10^{-5}$,
(c) $\rm FP_1$ with $\lambda_{00}= 10^{-2}$ and the rest being $10^{-5}$, and (d) $\rm FP_1$ with
$\lambda_{10}=10^{-2}$ and the rest being $10^{-5}$.}
\label{fig_FP01}
\end{figure*}

\section{Phase transitions driven by fermion-fermion interactions}\label{Sec_phase_transition}

Principally, near each FP, an instability is induced and linked to the onset of a symmetry-broken phase~\cite{Cvetkovic2012PRB,Vafek2014PRB,Wang2017PRB,Maiti2010PRB,Altland2006Book,Vojta2003RPP,
Metzner2000PRL,Halboth2000RPB,Eberlein2014PRB,Chubukov2012ARCMP,Nandkishore2012NP,Chubukov2016PRX,Roy2018PRX,Wang2020NPB,Zhang2025PRB}.
In what follows, we systematically investigate the leading instabilities associated with all FPs identified
in previous section~\ref{Sec_classification_FPs}. Our results show that these FPs yield distinct instabilities and phase
transitions. In particular, the competition among these phases is critically controlled by the initial conditions. This systematic
classification is essential for a comprehensive understanding of interaction-driven phase transitions in 2D QBCP semimetals.

\subsection{Source terms and susceptibility formalism}\label{Sec_source_suscept}

With the FPs established, we now turn to the low-energy physics in their vicinities. To this end, we introduce a source term composed of fermionic bilinear operators, which serves as a systematic tool for quantitatively analyzing potential instabilities and phase-transition candidates. The source term is given by~\cite{Roy2018PRX,Vafek2014PRB,Fradkin2008PRB,Fu-Wang2024AP,Szabo2021PRB}
\begin{eqnarray}
S_{\rm sou}
&=&
\int\mathrm{d}\tau\int\mathrm{d^2}\bm{x}\sum_{i=1}^4
(\Delta_{i}^{c}\Psi^{\dag}\mathcal{M}_{i}^{c}\Psi+\vec{\Delta}_{i}^{s}\cdot\Psi^{\dag}\vec{\mathcal{M}}_{i}^{s}\Psi)\nonumber\\
&&+\int\mathrm{d}\tau\int\mathrm{d^2}\bm{x}\sum_{i=1}^4(\Delta_{i}^{\mathrm{PP}}
\Psi^{\dag}\mathcal{M}_{i}^{\mathrm{PP}}\Psi+\rm H.c.),\label{equation:waiyuanxiang}
\end{eqnarray}
in which $\mathcal{M}_i^{c/s}$ represent the fermionic bilinears in the particle-hole channel, with the superscripts $c$ and $s$ referring to the charge and spin sectors, respectively, while $\mathcal{M}_i^{\rm PP}$ denotes the bilinear in the particle-particle (pairing) channel. The fields $\Delta_i$ play a dual role~\cite{Vafek2014PRB,Vafek2010PRB,Szabo2021PRB}. On one hand, they serve as source fields that probe the system's response. On the other, they can be considered as order parameters characterizing the corresponding symmetry-broken phases.

\begin{table}[htbp]
\centering
\caption{Candidate instabilities and phases near the fixed points induced by fermion-fermion  interactions~\cite{Vafek2014PRB,Vafek2010PRB,Cvetkovic2012PRB}. For brevity, we adopt the following abbreviations: SC for superconductivity, FM for ferromagnetism, QAH for the quantum anomalous Hall state, QSH for the quantum spin Hall
state, Nem for nematic order, and NSN for nematic-spin-nematic order.
}\label{table:xiangbian}
\vspace{0.3cm}
\renewcommand{\arraystretch}{1.5} 
\begin{tabular}{>{\centering\arraybackslash}p{1.7cm}|>{\centering\arraybackslash}p{3.2cm}|>{\centering\arraybackslash}p{2.6cm}}  
\hline\hline
\makecell{Order \\ parameters} & \makecell{Vertex matrixes \\ of fermionic bilinears} & \makecell{Potential phases \\ } \\
\hline
$\Delta_1^c$ & $\mathcal{M}_1^c=\tau_0\otimes\sigma_0$ & Charge instability\\
\hline
$\Delta_2^c$ & $\mathcal{M}_2^c=\tau_0\otimes\sigma_1$ & Nem.(bond)\\
\hline
$\Delta_3^c$ & $\mathcal{M}_3^c=\tau_0\otimes\sigma_2$ & QAH\\
\hline
$\Delta_4^c$ & $\mathcal{M}_4^c=\tau_0\otimes\sigma_3$ & Nem.(site)\\
\hline
$\vec{\Delta}_1^s$ & $\vec{\mathcal{M}}_1^s=\vec{\tau}\otimes\sigma_0$ & Ferromagnet\\
\hline
$\vec{\Delta}_2^s$ & $\vec{\mathcal{M}}_2^s=\vec{\tau}\otimes\sigma_1$ & NSN(bond)\\
\hline
$\vec{\Delta}_3^s$ & $\vec{\mathcal{M}}_3^s=\vec{\tau}\otimes\sigma_2$ & QSH\\
\hline
$\vec{\Delta}_4^s$ & $\vec{\mathcal{M}}_4^s=\vec{\tau}\otimes\sigma_3$ & NSN(site)\\
\hline
$\Delta_1^{\mathrm{PP}}$ & $\mathcal{M}_1^{\mathrm{PP}}=\tau_2\otimes\sigma_0$ & chiral SC1\\
\hline
$\Delta_2^{\mathrm{PP}}$ & $\mathcal{M}_2^{\mathrm{PP}}=\tau_2\otimes\sigma_1$ & chiral SC2\\
\hline
$\Delta_3^{\mathrm{PP}}$ & $\mathcal{M}_3^{\mathrm{PP}}=\tau_{0,1,3}\otimes\sigma_2$ & triplet SC\\
\hline
$\Delta_4^{\mathrm{PP}}$ & $\mathcal{M}_4^{\mathrm{PP}}=\tau_2\otimes\sigma_3$ & s-wave SC\\
\hline\hline
\end{tabular}
\end{table}

A comprehensive list of candidate phases induced by FPs and their associated bilinear matrices is provided in Table~\ref{table:xiangbian}. To characterize the response of the system to each source field, we designate the
corresponding susceptibility for each candidate channel via the second derivative
of the free energy with respect to the corresponding source field~\cite{Vafek2014PRB,Vafek2010PRB,Szabo2021PRB},
\begin{equation}
\chi^{c,s,\mathrm{PP}}_i \equiv -\frac{\partial^2 f}{\partial \Delta^{c,s,\mathrm{PP}}_i
\partial \Delta^{c,s,\mathrm{PP}}_i} \bigg|_{\Delta^{c,s,\mathrm{PP}}_i=0},\label{Eq_chi_i}
\end{equation}
where $f$ is the free energy density. The susceptibility $\chi_i$ thus measures the linear response of the system to the source field $\Delta_i$ and serves as a diagnostic for the associated instability~\cite{Vafek2014PRB,Vafek2010PRB,Metzner2000PRL,Halboth2000RPB,
Nandkishore2012NP,Chubukov2016PRX,Roy2018PRX}.

\subsection{Leading instabilities and competing orders}

The formula~(\ref{Eq_chi_i}) establishes a direct link between the susceptibility and the physical fluctuations
of the corresponding order parameter. A divergence of $\chi_i$ signals the onset of long-range order in that channel.
By paralleling the same RG procedure as employed in Sec.~\ref{Sec_RG_analysis}, we derive the flow equations for
all source terms in Table~\ref{table:xiangbian} (see Appendix~\ref{Appendix_source_terms} for details).
Combined with the RG flows of the interaction parameters~(\ref{Eq_lambda00})-(\ref{Eq_lambda33}) and the formula of susceptibilities~(\ref{Eq_chi_i}), this yields the energy dependence of each susceptibility near the FPs.
The resulting scale-dependent susceptibility $\chi_i(l)$ allows us to identify the leading instability,
which is the channel exhibiting the strongest divergence of its susceptibility at the critical scale $l_c$~\cite{Vafek2014PRB,Vafek2010PRB,Metzner2000PRL,Halboth2000RPB,
Nandkishore2012NP,Chubukov2016PRX,Roy2018PRX}. Following the definitions addressed in Sec.~\ref{Sec_classification_FPs},
we are going to consider three distinct scenarios of FPs one by one in the following.

\subsubsection{Isotropic scenario}

We first consider the isotropic scenario, wherein all sixteen fermion-fermion
interaction parameters share a common initial value. Under this condition, the RG flow
goes towards to a unique fixed point, namely, FP$_0$ as given in Eq.~(\ref{Eq_FP0}).

After performing the numerical analysis presented in Sec.~\ref{Sec_source_suscept}, Fig.~\ref{fig_FP01}(a)-(b)
display the main results of the susceptibilities of all candidate phases as approaching the FP$_0$.
We can find that the chiral SC$_1$ channel exhibits the most rapid divergence and the NSN(site)-$z$ channel is subordinate.
Importantly, the basic results are insensitive to the initial values of fermion-fermion interactions ranging from
$10^{-2}$ to $10^{-5}$. In other words, the initial interaction strength affects only the critical scale $l_c$ at
which the divergence occurs, but not the hierarchy of the competing instabilities.

\begin{figure*}[htpb]
\centering
\subfigure[FP$_2$]{
\includegraphics[width=2.5in]{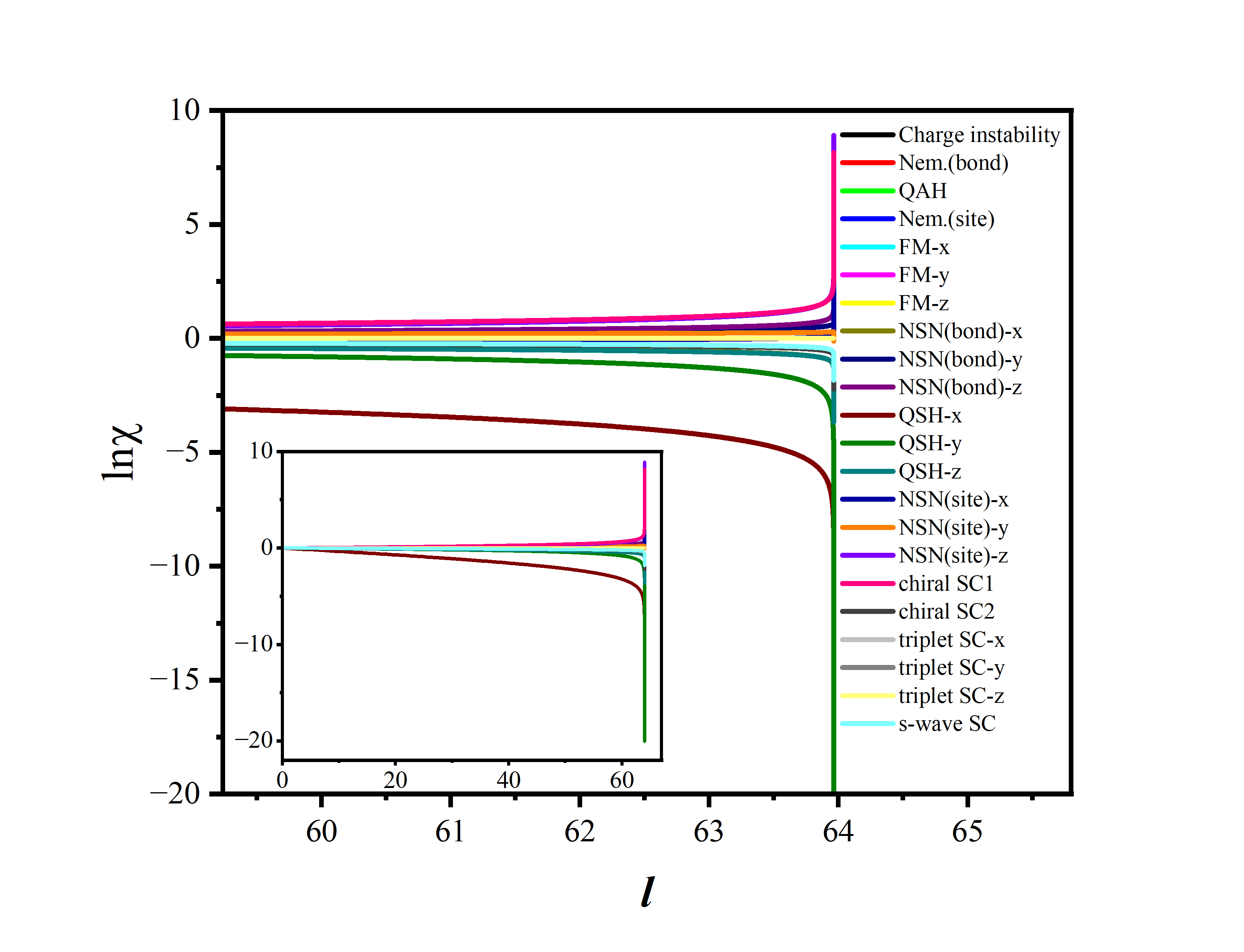}
}\hspace{-1.18cm}
\subfigure[FP$_3$]{
\includegraphics[width=2.5in]{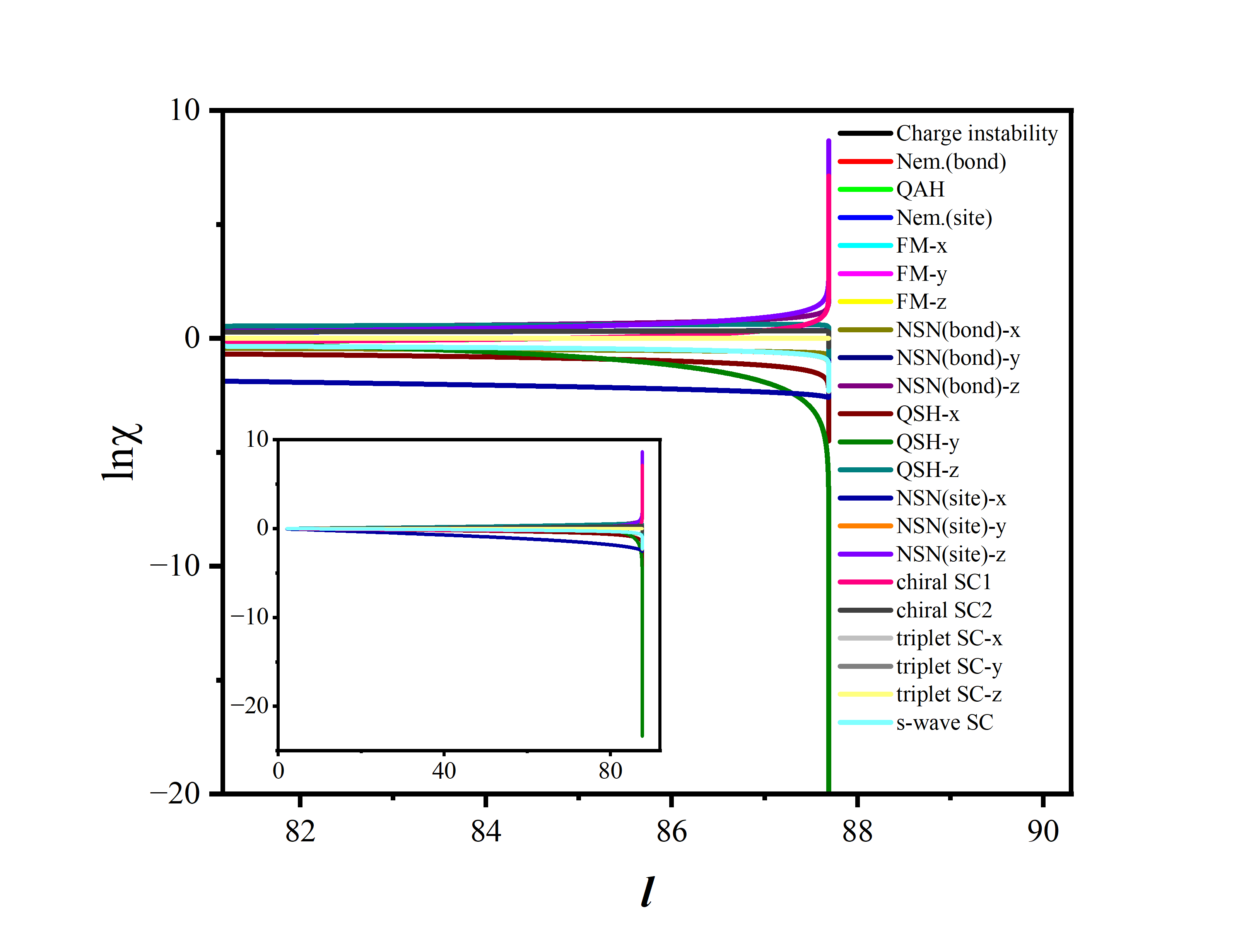}
}\hspace{-1.18cm}
\subfigure[FP$_4$]{
\includegraphics[width=2.5in]{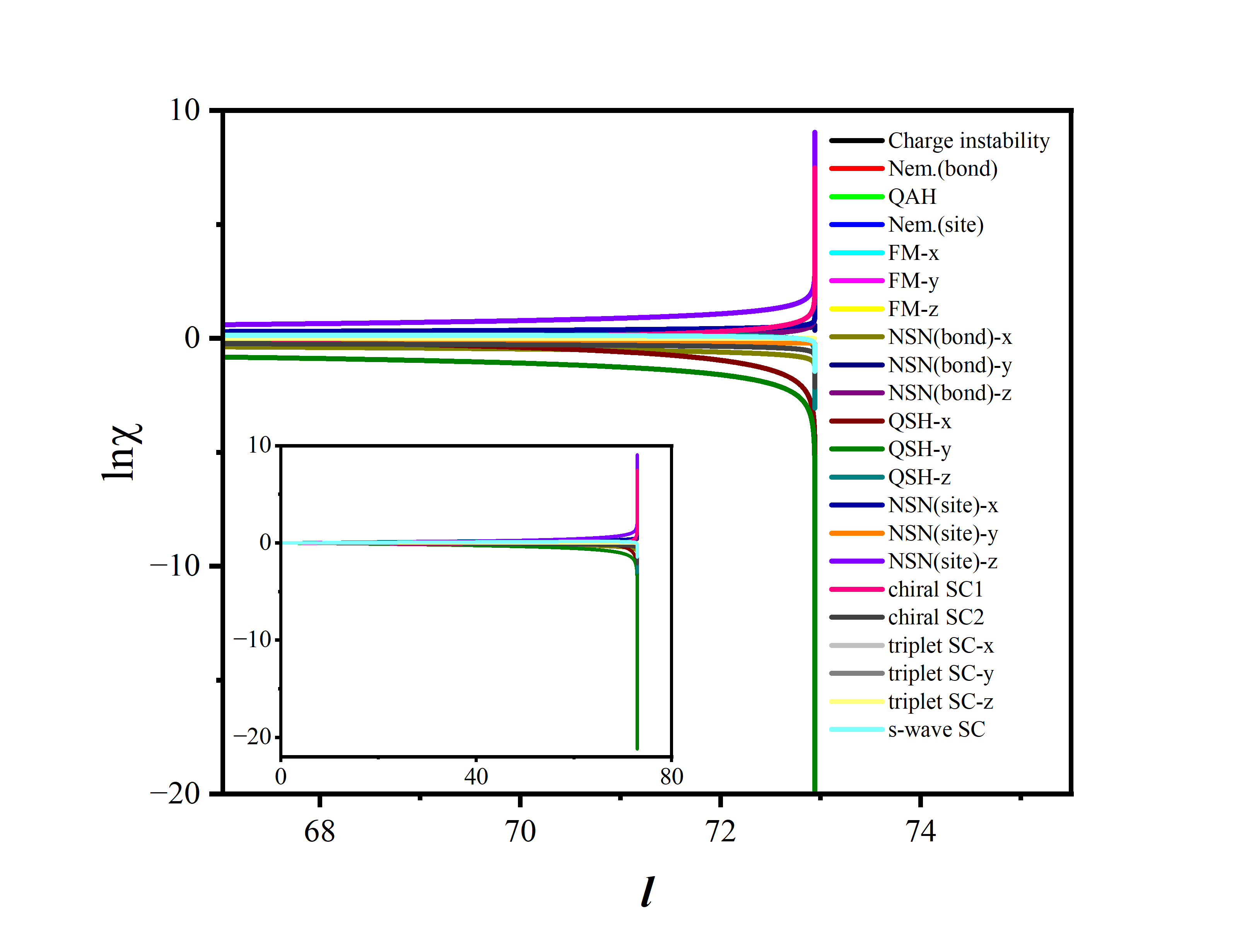}
}\\
\vspace{-0.2cm}
\subfigure[FP$_6$]{
\includegraphics[width=2.5in]{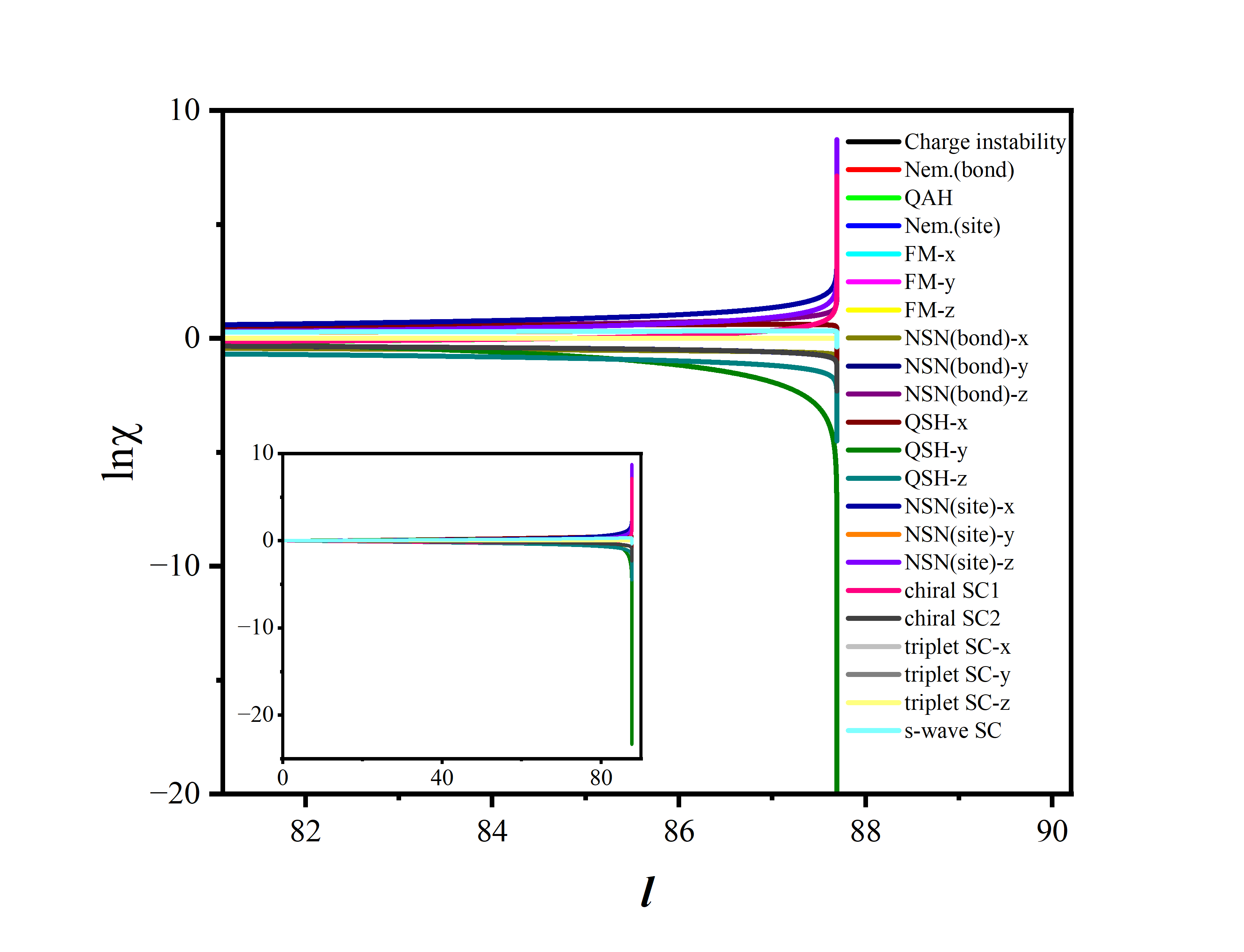}
}\hspace{-1.18cm}
\subfigure[]{
\includegraphics[width=2.5in]{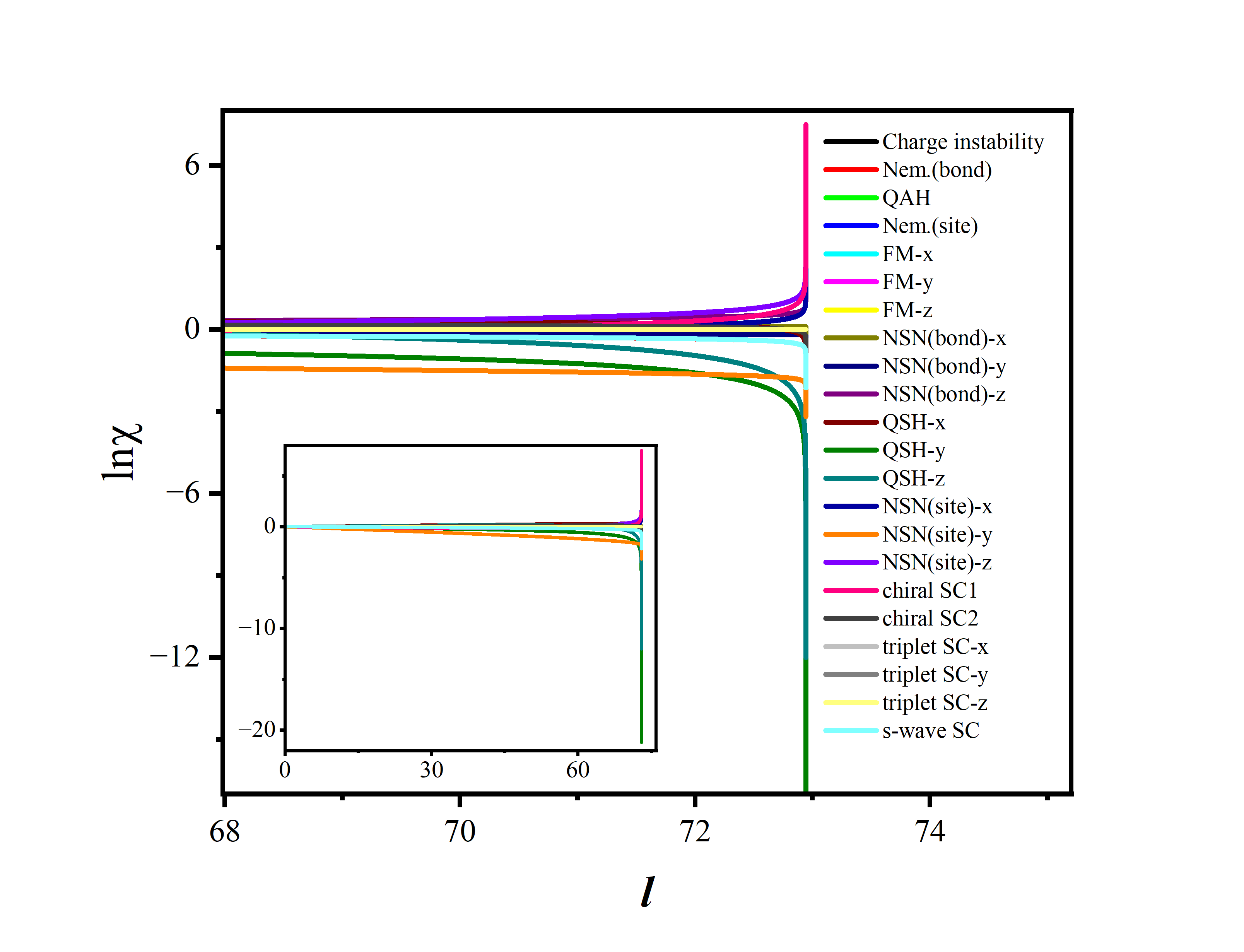}
}\hspace{-1.18cm}
\subfigure[]{
\includegraphics[width=2.5in]{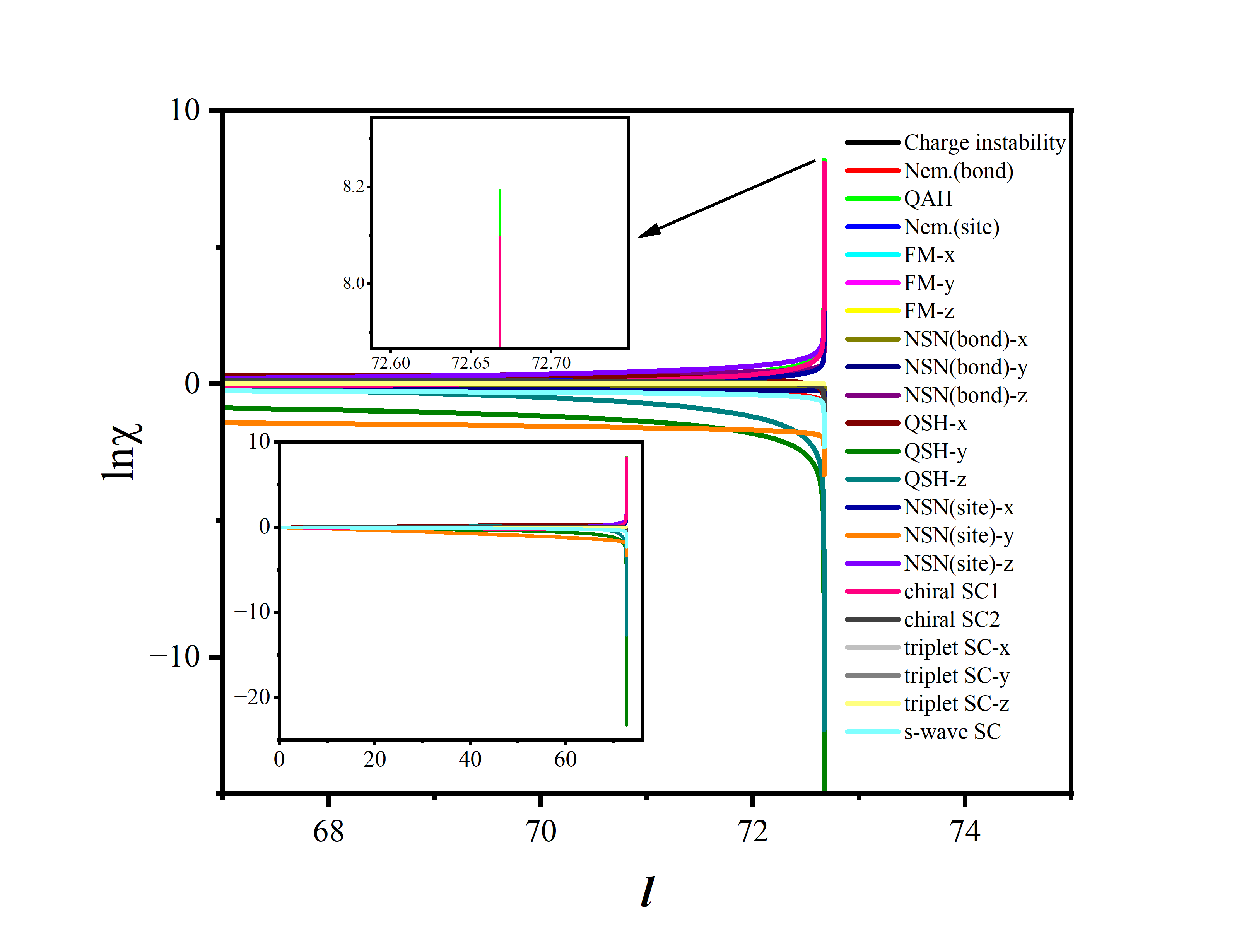}
}\\
\vspace{-0.2cm}
\subfigure[]{
\includegraphics[width=2.5in]{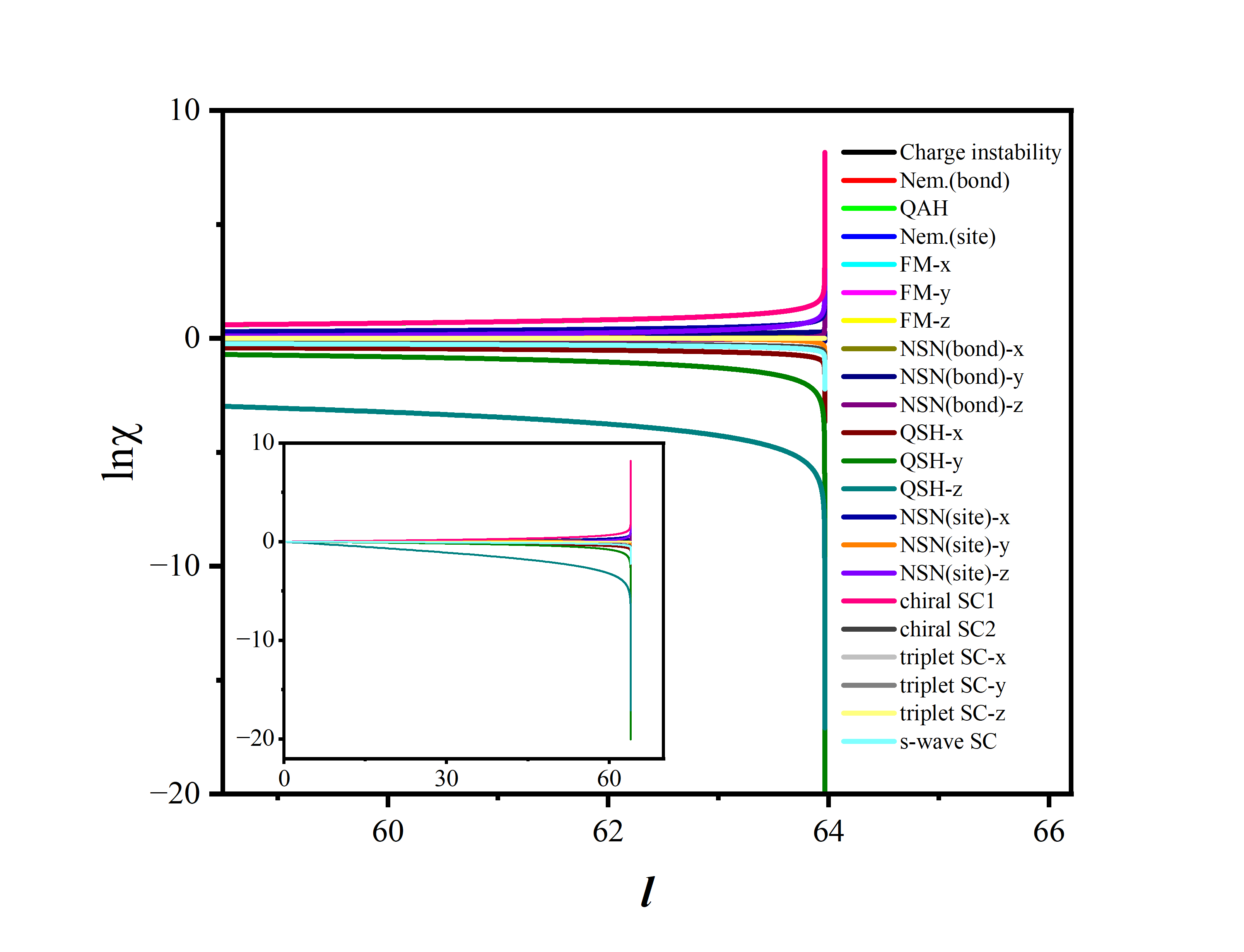}
}\hspace{-0.1cm}
\subfigure[]{
\includegraphics[width=2.5in]{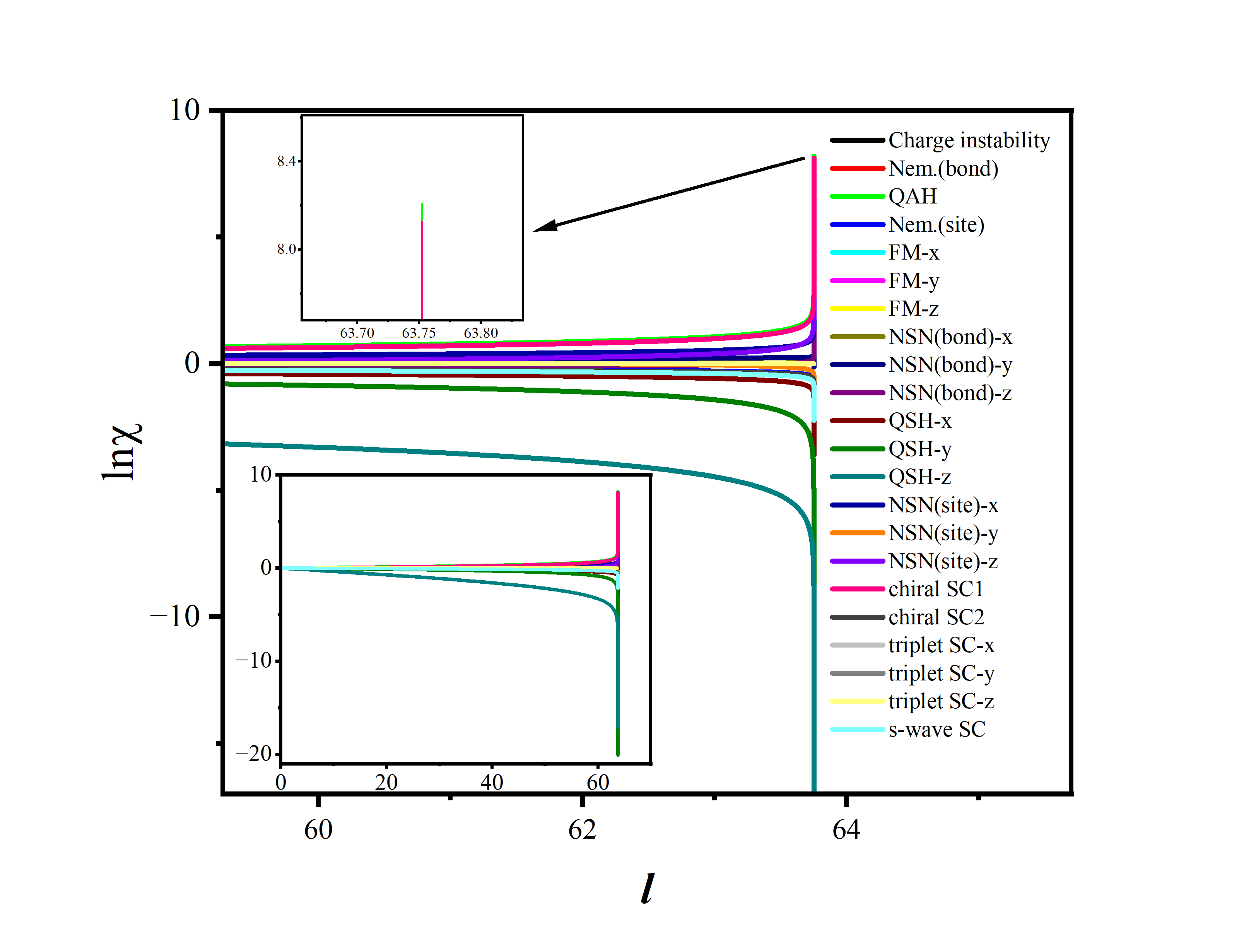}
}\vspace{-0.1cm}
\caption{(Color online) Energy dependence of the susceptibilities for the candidate instabilities.
Panels (a)¨C(f) correspond to $\mathrm{FP}_{2}$, $\mathrm{FP}_{3}$, $\mathrm{FP}_{4}$, $\mathrm{FP}_{6}$,
$\mathrm{FP}_{5}$, and $\mathrm{FP}_{7}$, respectively, with one dominant coupling initialized at $10^{-2}$
and all remaining couplings at $10^{-5}$. Panel (g) shows $\mathrm{FP}_{5}$ with $\lambda_{23}=10^{-2}$,
$\lambda_{10}=10^{-3}$, and all other couplings set to $10^{-5}$. Panel (h) shows $\mathrm{FP}_{7}$ with
$\lambda_{32}=10^{-2}$,
$\lambda_{10}=10^{-3}$, and all other couplings set to $10^{-5}$.}
\label{fig_FP27}
\end{figure*}

\subsubsection{Anisotropic scenario}

Subsequently, we turn to the anisotropic scenario, in which one interaction parameter is
initialized significantly larger than the other fifteen fermion-fermion couplings.
This causes the system to flow to one of the seven FPs, namely, FP$_1$-FP$_7$ as specified in
Eqs.~(\ref{Eq_FP1})-(\ref{Eq_FP7}). Paralleling the similar numerical analysis, we present
energy-dependent susceptibilities as approaching each FP
in Fig.~\ref{fig_FP01}(c)-(d) and Fig.~\ref{fig_FP27}. A systematic comparison of these susceptibilities
allows us to identify the leading and subleading instabilities for each fixed point, which are
summarized in Table~\ref{table:stability}.

\begin{table}[htbp]
\centering
\caption{Leading and subleading instabilities near the distinct FPs (FP$_1$-FP$_7$).}\label{table:stability}
\vspace{0.3cm}
\renewcommand{\arraystretch}{1.2}
\begin{tabular}{>{\centering\arraybackslash}p{1.1cm}|>{\centering\arraybackslash}p{3.0cm}|>{\centering\arraybackslash}p{3.0cm}}  
\hline\hline
\makecell{FPs} & \makecell{leading phase} & \makecell{subleading phase} \\
\hline
FP1 & NSN(site)-z & chiral\ SC1\ or\ QAH\\
\hline
FP2 & NSN(site)-z & chiral\ SC1 or\ QAH\\
\hline
FP3 & NSN(site)-z &chiral\ SC1 or\ QAH\\
\hline
FP4 & NSN(site)-z &chiral\ SC1 or\ QAH\\
\hline
FP5 & chiral\ SC1\ / QAH &QAH / chiral\ SC1\\
\hline
FP6 & NSN(site)-z &chiral\ SC1 or\ QAH\\
\hline
FP7 & chiral\ SC1\ / QAH &QAH / chiral\ SC1\\
\hline
\hline
\end{tabular}
\end{table}

\begin{figure*}[htbp]
\centering
\hspace{-1.05cm}
\subfigure[]{
\includegraphics[width=3.0in]{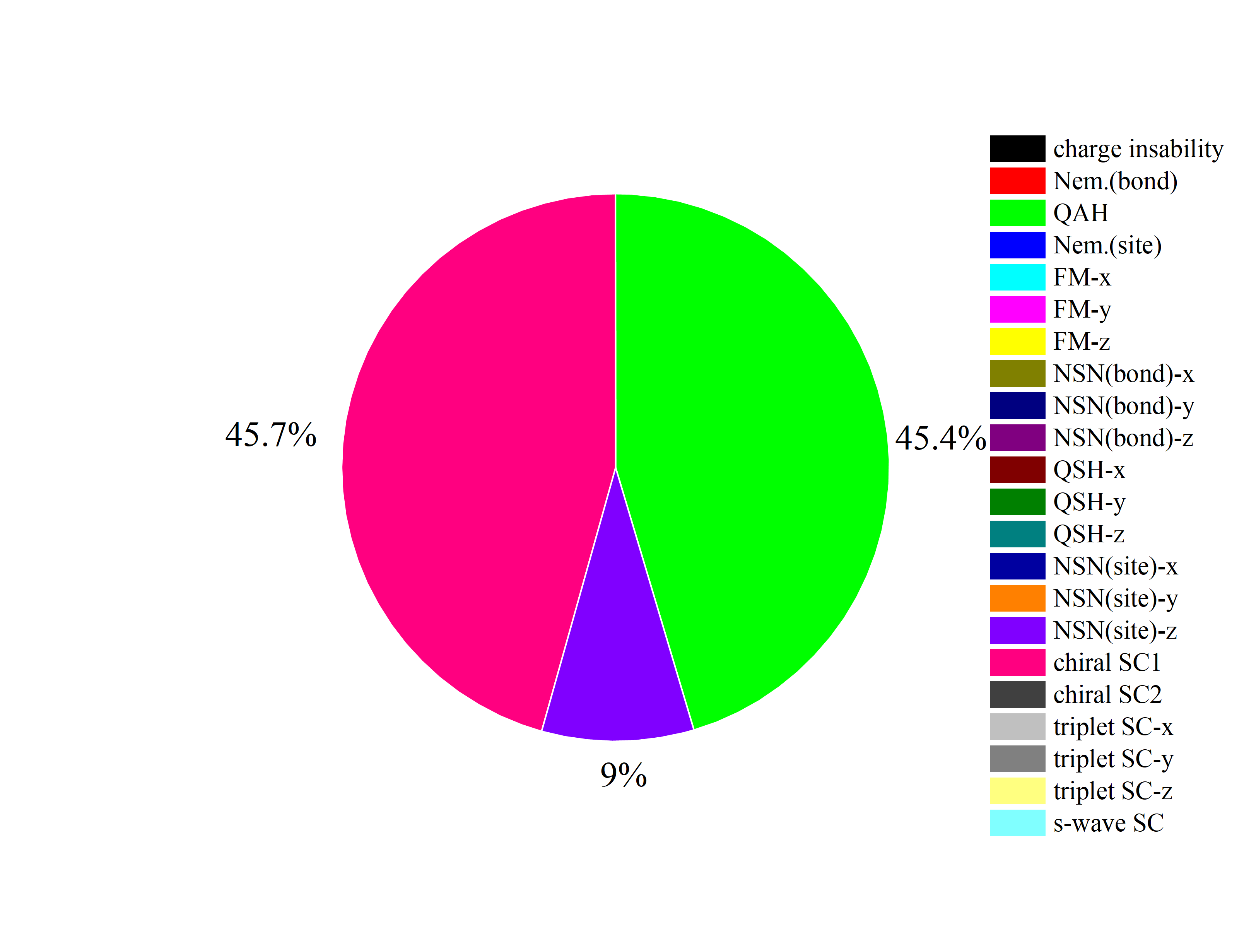}
}\hspace{0.75cm}
\subfigure[]{
\includegraphics[width=3.0in]{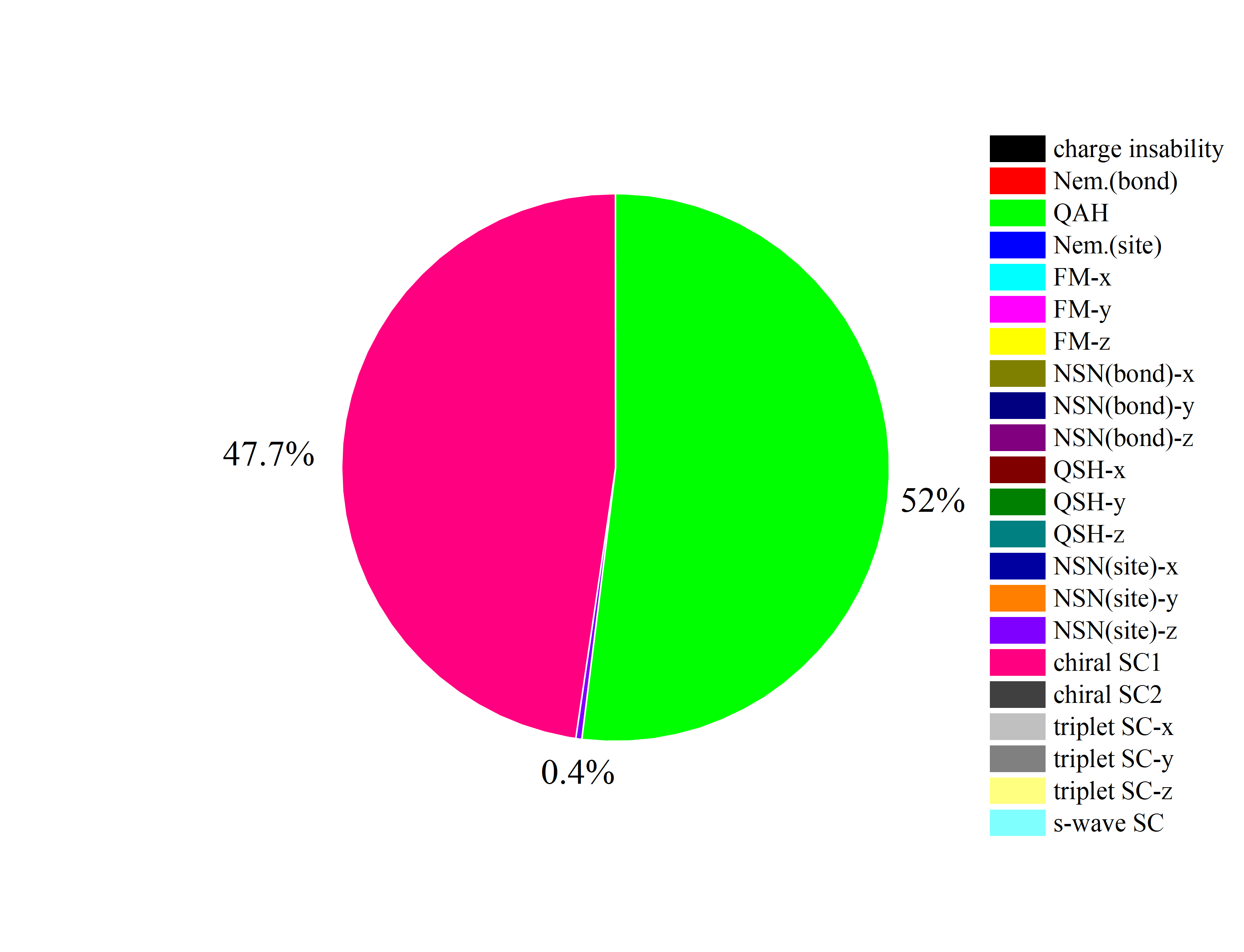}
}\\
\vspace{-0.15cm}
\hspace{-1.05cm}
\subfigure[]{
\includegraphics[width=3.0in]{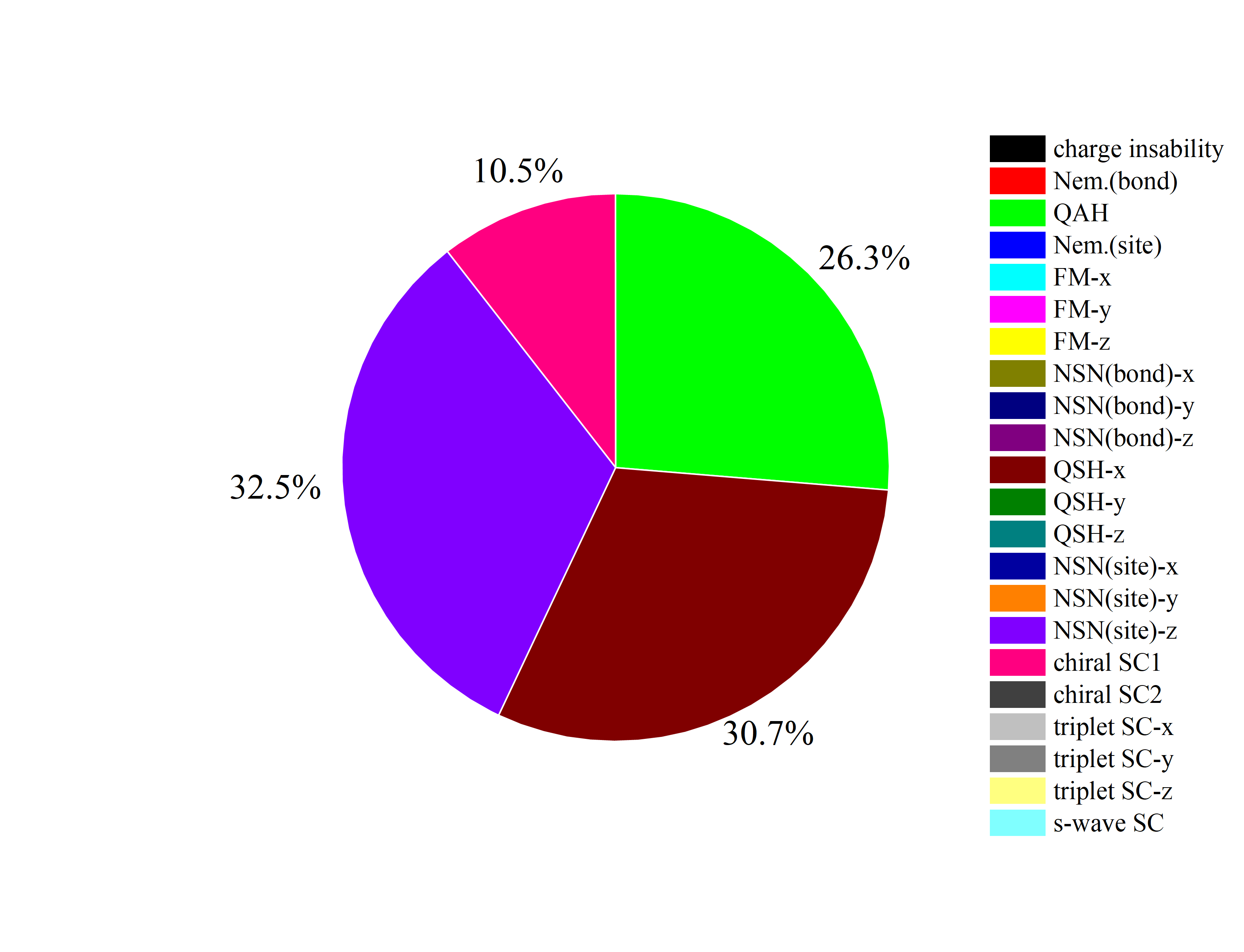}
}\hspace{0.75cm}
\subfigure[]{
\includegraphics[width=3.0in]{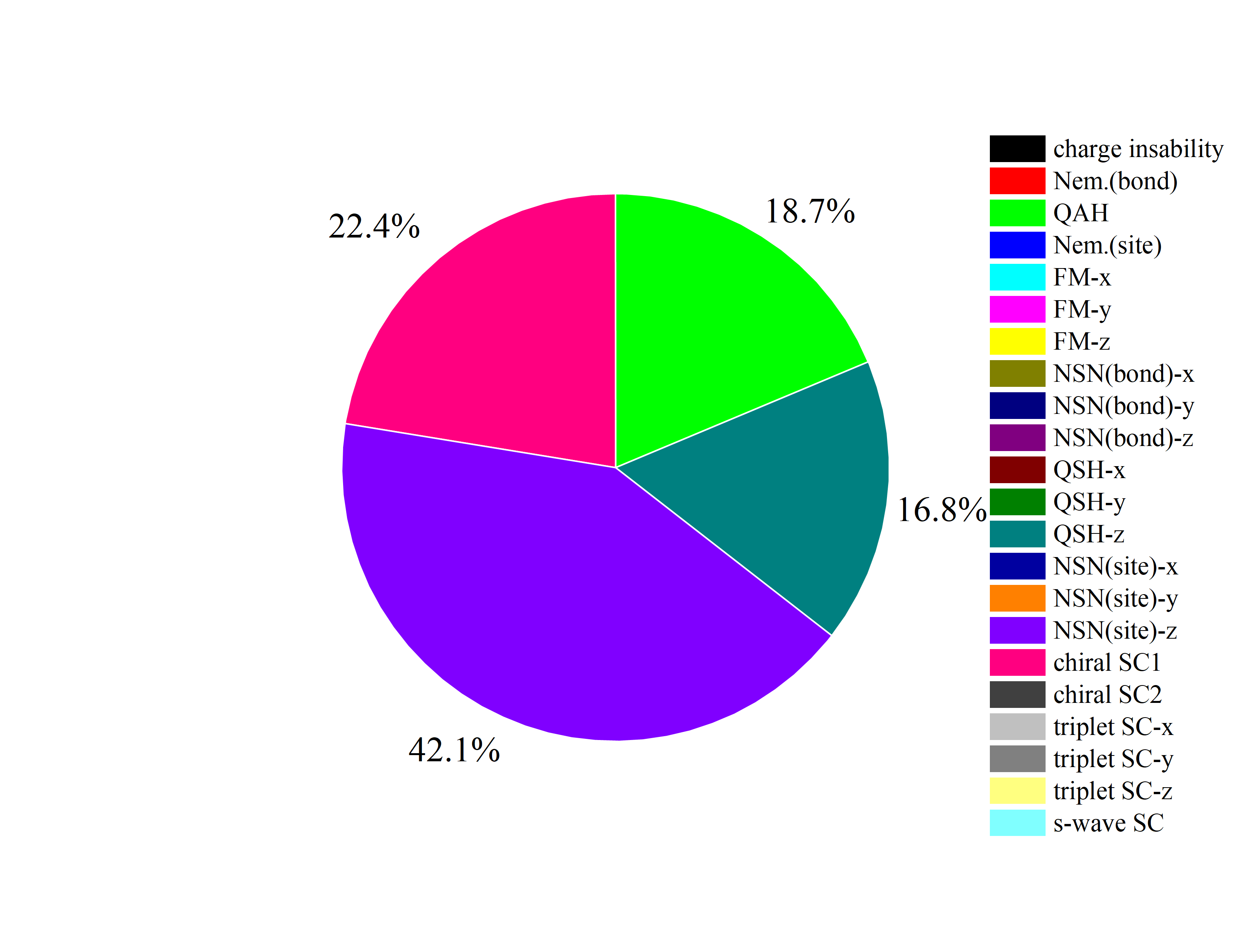}
}
\vspace{-0.1cm}
\caption{(Color online) Statistical probability distributions of the leading and subleading instabilities obtained from random initial conditions. (a) and (b) show the probability that each candidate instability becomes the leading instability near FP$_5$ and FP$_7$, respectively. (c) Probability that each instability becomes the subleading instability at $\alpha \approx -0.25$. (d) Probability that each instability becomes the leading instability at $\alpha \approx 0.25$. The color legend identifies the corresponding instability channels, and the percentages denote the relative probabilities.}
\label{fig:BFP57}
\end{figure*}

Specifically, for FP$_{1,2,3,4,6}$, Figs.~\ref{fig_FP01}(c)-(d) and \ref{fig_FP27}(a)-(d) show that NSN(site)-$z$ is always the leading instability, while the subleading channel alternates between chiral SC$_1$ and QAH depending on the initial conditions. Examining the structures of these FPs~[see Eqs.~(\ref{Eq_FP1})-(\ref{Eq_FP4}), (\ref{Eq_FP6})], we can qualitatively understand this phenomenon. They all share a common feature, namely, a dominant positive $\lambda_{22}(l_c)$ component ranging from 2.3516 to 2.4347, together with negative elements $\lambda_{11}(l_c)$ and $\lambda_{33}(l_c)$. Substituting these values into the flow equation of source terms~(\ref{Eq_B1})-(\ref{Eq_B22}), it becomes evident that the coefficients of $\lambda_{22}$, $\lambda_{11}$, $\lambda_{33}$ in
Eqs.~(\ref{Eq_B3}), (\ref{Eq_B16}), and (\ref{Eq_B17}) are positive, negative, and negative, respectively.
This accordingly is helpful to increase the related instabilities, namely, QAH, NSN(site)-z and chiral SC$_1$,
and make them own large opportunities to be leading and subleading instabilities. To clarify this,
let us make a rough estimation. Neglecting the the off-diagonal elements and substituting the values of the
diagonal elements at these FPs into the Eqs.~(\ref{Eq_B1})-(\ref{Eq_B22}),
we can find that NSN(site)-$z$ linked to Eq.~(\ref{Eq_B16}) acquires the largest exponent,
which approximately explains why it emerges as the leading instability in all these cases.

In addition, we comment on the subleading instability. Taking FP$_1$ as an example (other fixed points are similar), Figs.~\ref{fig_FP01}(c)-(d) show that the subleading instability switches between chiral SC$_1$ and QAH depending on the initial conditions. This behavior originates from the fact that the susceptibility is not determined solely by the critical values at FPs,
rather, it is an integral over the entire RG trajectory from $l_0$ to $l_c$. Different initial conditions produce different
trajectories in the sixteen-dimensional parameter space. Since the subleading exponent is smaller, its integrated value
is more sensitive to intermediate-scale couplings, leading to the observed switching.

We next examine FP$_5$ and FP$_7$,which exhibit a qualitatively different behavior from
the previously discussed FPs. As illustrated in Figs.~\ref{fig_FP27}(e)-(f), the susceptibilities
of the chiral SC$_1$ and QAH channels diverge at nearly the same rate, indicating a near-degeneracy
between these two competing instabilities. The origin of this near-degeneracy can be traced to the
structure of the FPs.  Both FP$_5$~(\ref{Eq_FP5})
and FP$_7$~(\ref{Eq_FP7}) contain off-diagonal elements $\lambda_{23}$ and $\lambda_{32}$.
However, $\lambda_{32}$ is significantly larger and therefore plays a more important role than $\lambda_{23}$
in the source-term equations. Learning from flow equations
of source terms~(\ref{Eq_B3}), (\ref{Eq_B16}), and (\ref{Eq_B17}) that are respectively
associated with QAH, NSN(site)-$z$, and chiral SC$_1$, one can notice
that the coefficient of $\lambda_{32}$ is positive in both Eq.~(\ref{Eq_B3}) and Eq.~(\ref{Eq_B17}),
while in Eq.~(\ref{Eq_B16}) is negative. This enables QAH and chiral SC$_1$ to overtake NSN(site)-$z$ and become
the leading instabilities.

\begin{figure}[htpb]
\centering
\subfigure[]{
\includegraphics[width=3.05in]{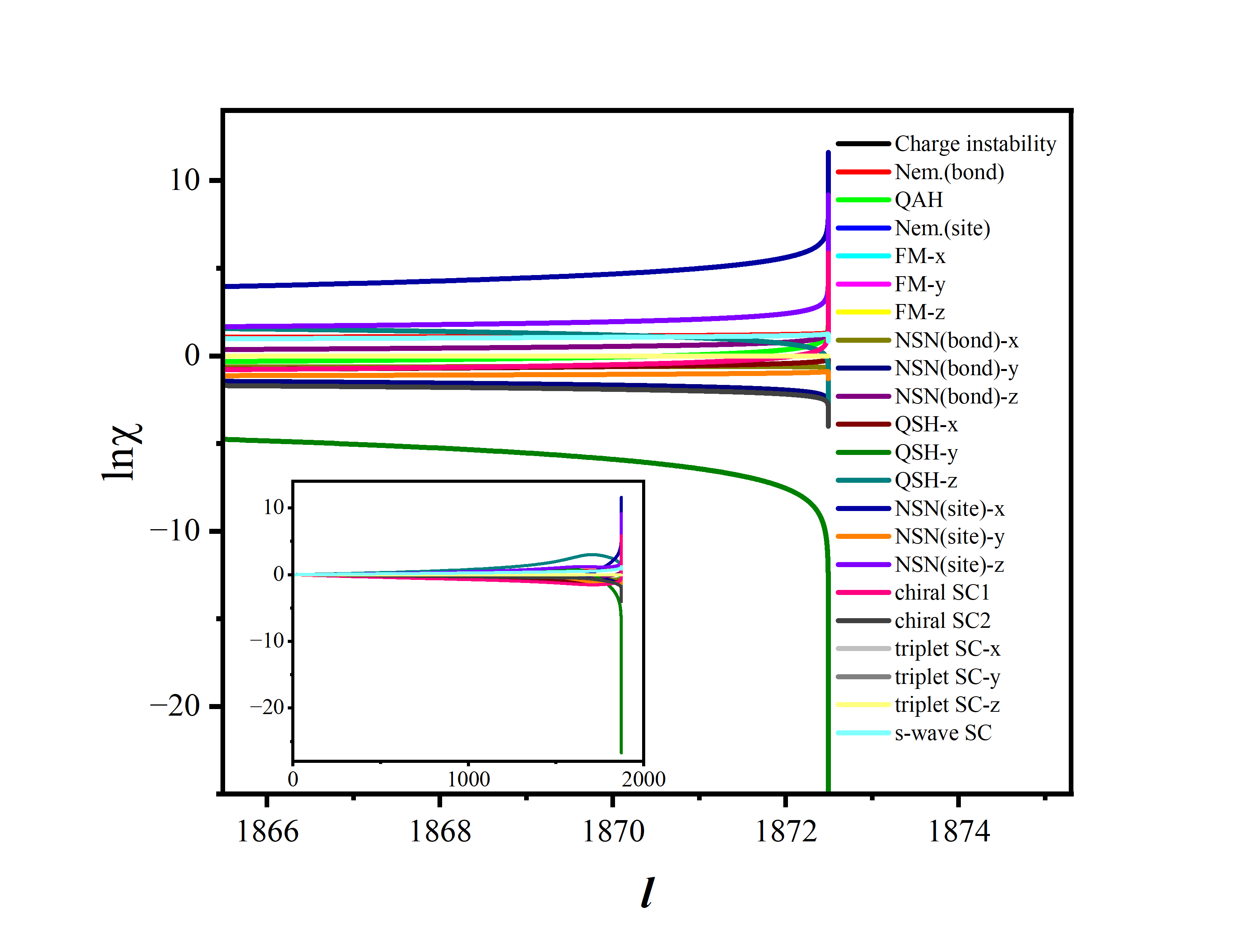}
}\\
\vspace{-0.25cm}
\subfigure[]{
\includegraphics[width=3.05in]{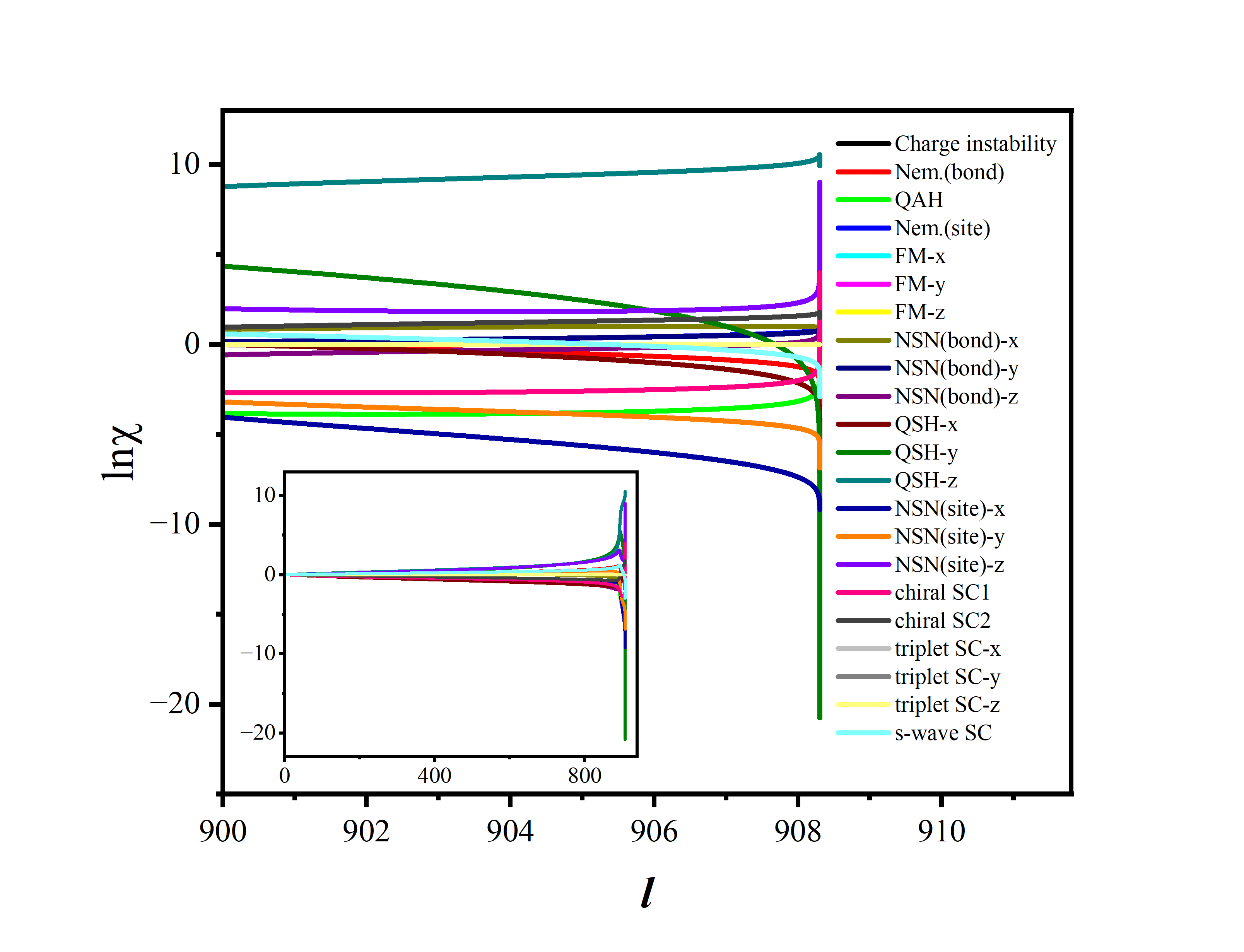}
}\vspace{-0.1cm}
\caption{(Color online) Energy-dependent susceptibilities nearby the FPG$_1$ for the candidate instabilities
for (a) $\alpha \approx -0.25$ and (b) $\alpha \approx 0.25$.} 
\label{fig:FPS1}
\end{figure}

Given the exponents are nearly degenerate, the susceptibility becomes more sensitively on the initial values of the subordinate fifteen parameters. Compared Fig.~\ref{fig_FP27}(e) with Fig.~\ref{fig_FP27}(g) nearby the FP$_5$, increasing $\lambda_{10}(0)$ from $10^{-5}$ to $10^{-3}$ reverses the leading instability from chiral SC$_1$ to QAH [the basic results
are analogous for FP$_7$ as illustrated in Fig.~\ref{fig_FP27}(f) and Fig.~\ref{fig_FP27}(h)]. This sensitivity arises because the accumulated contributions from the other couplings along the RG trajectory can tip the balance between the two channels.
Hence, at FP$_5$ and FP$_7$, the leading instability is not fixed by the FP alone but depends on the full RG trajectory.
Our statistical sampling, shown in Fig.~\ref{fig:BFP57}(a)-(b), confirms this picture that both chiral SC$_1$ and QAH have substantial probabilities (approximately $45\%$-$52\%$) of becoming the leading instability.
Generally, the critical values of FPs primarily determine the leading instability.
However, near FP$_5$ and FP$_7$, the competition between QAH and chiral SC$_1$ becomes so
intense that a secondary factor, namely, the interaction parameters along the RG trajectory also
play an important role. As a result, both QAH and chiral SC$_1$ have substantial probabilities
of emerging as the leading instability.

This highlights the trajectory-dependent nature of the competition in this regime.
Such trajectory-dependent behavior has no analogue in the simplified model that consists of four fermion-fermion interactions~\cite{Fradkin2009PRL,Wang2017PRB,DZDW2020PRB,Vafek2014PRB}, where the FP uniquely dictates
the leading instability. It thus constitutes a new and important feature emerging from the full sixteen-channel coupling space, emphasizing
the necessity of retaining all interaction channels for a complete understanding of the low-energy physics.

\subsubsection{General scenario}

To proceed, we now consider the general scenario, in which all sixteen initial interaction parameters are sampled independently from a log-uniform distribution. In this circumstance, the FPs are found to organize into three continuous fixed-point groups, namely,
FPG$_1$-FPG$_3$ as specified in Eqs.~(\ref{Eq_FPG1})-(\ref{Eq_FPG3}). Within each group, we endeavor to investigate how the leading instability depends on the group parameter.

We begin with FPG$_1$~(\ref{Eq_FPG1}), characterized by the group parameter $\alpha\in[-0.25,0.25]$. This parameter
enters $\lambda_{13}(l_c)$ and $\lambda_{31}(l_c)$ with opposite signs, thereby controlling the asymmetry between
these two couplings. The corresponding susceptibility flows for $\alpha=\pm0.25$ are presented in
Fig.~\ref{fig:FPS1} as approaching FPG$_1$. When $\alpha$ is tuned to the lower end of its range, e.g., $\alpha\approx -0.25$, Fig.~\ref{fig:FPS1}(a) shows that the NSN(site)-$x$ susceptibility diverges most rapidly, thereby emerging as the leading instability.
At this endpoint, the subleading instabilities are distributed among QAH,
QSH-$x$, NSN(site)-$z$, and chiral SC$_1$, with their relative probabilities quantified in Fig.~\ref{fig:BFP57}(c).
This distribution indicates a near-degenerate competition in the subleading instability, where no
single channel decisively dominates over the others.

As $\alpha$ approaches the upper limit $0.25$, the behavior changes qualitatively. The dominant
instability is no longer unique. In sharp contrast to the case of a single leading instability, QAH,
QSH-$z$, NSN(site)-$z$, and chiral SC$_1$ exhibit comparable divergence rates, resulting in a near-degenerate
competition among these four candidate states. Quantitatively, QAH and chiral SC$_1$ each acquire probabilities
of approximately $18\%$¨C$22\%$, NSN(site)-$z$ carries $42.1\%$, and QSH-$z$ also has a non-negligible probability.
Figure~\ref{fig:FPS1}(b) shows an example in which QSH-$z$ becomes the dominant instability at $\alpha\approx0.25$
in FPG$_1$. This contrasts with the lower endpoint, where the leading instability is unique. In particular,
QSH-$x,z$ emerges as a possible leading instability only in this regime, which is absent in the anisotropic scenario.

\begin{figure}[htpb]
\centering
\subfigure[]
{\includegraphics[width=2.6in]{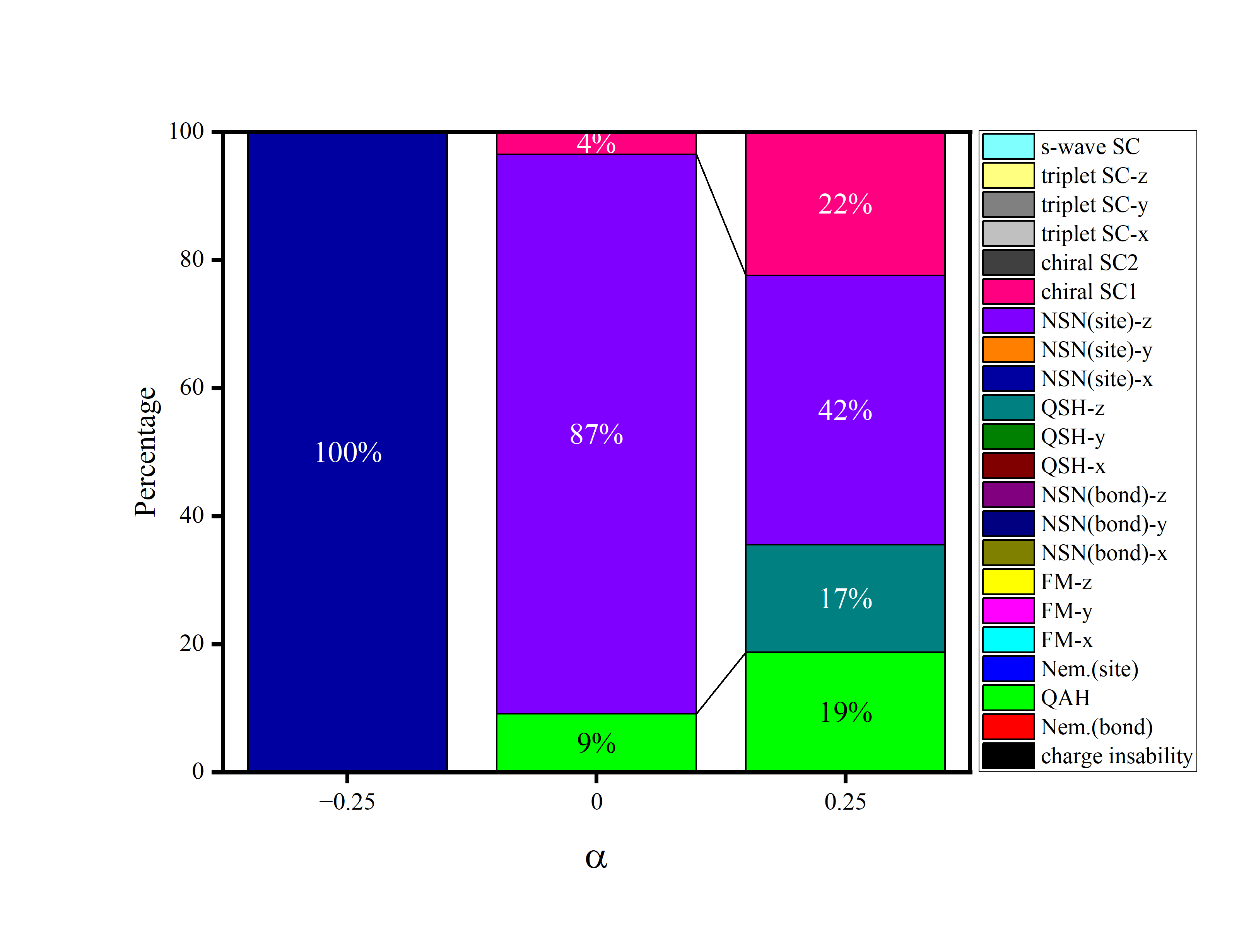}}\\
\vspace{-0.35cm}
\subfigure[]
{\includegraphics[width=2.6in]{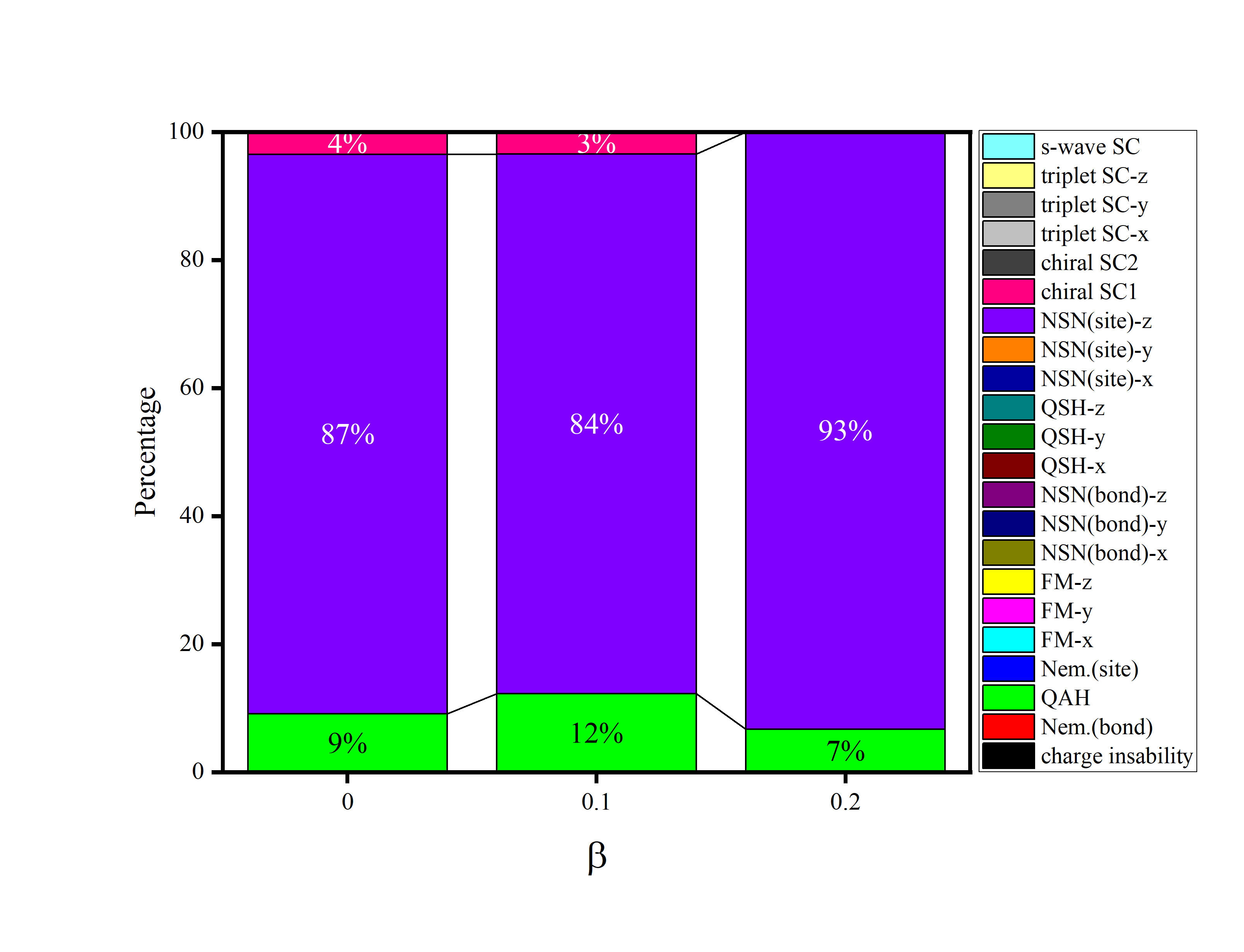}}\\
\vspace{-0.35cm}
\subfigure[]
{\includegraphics[width=2.6in]{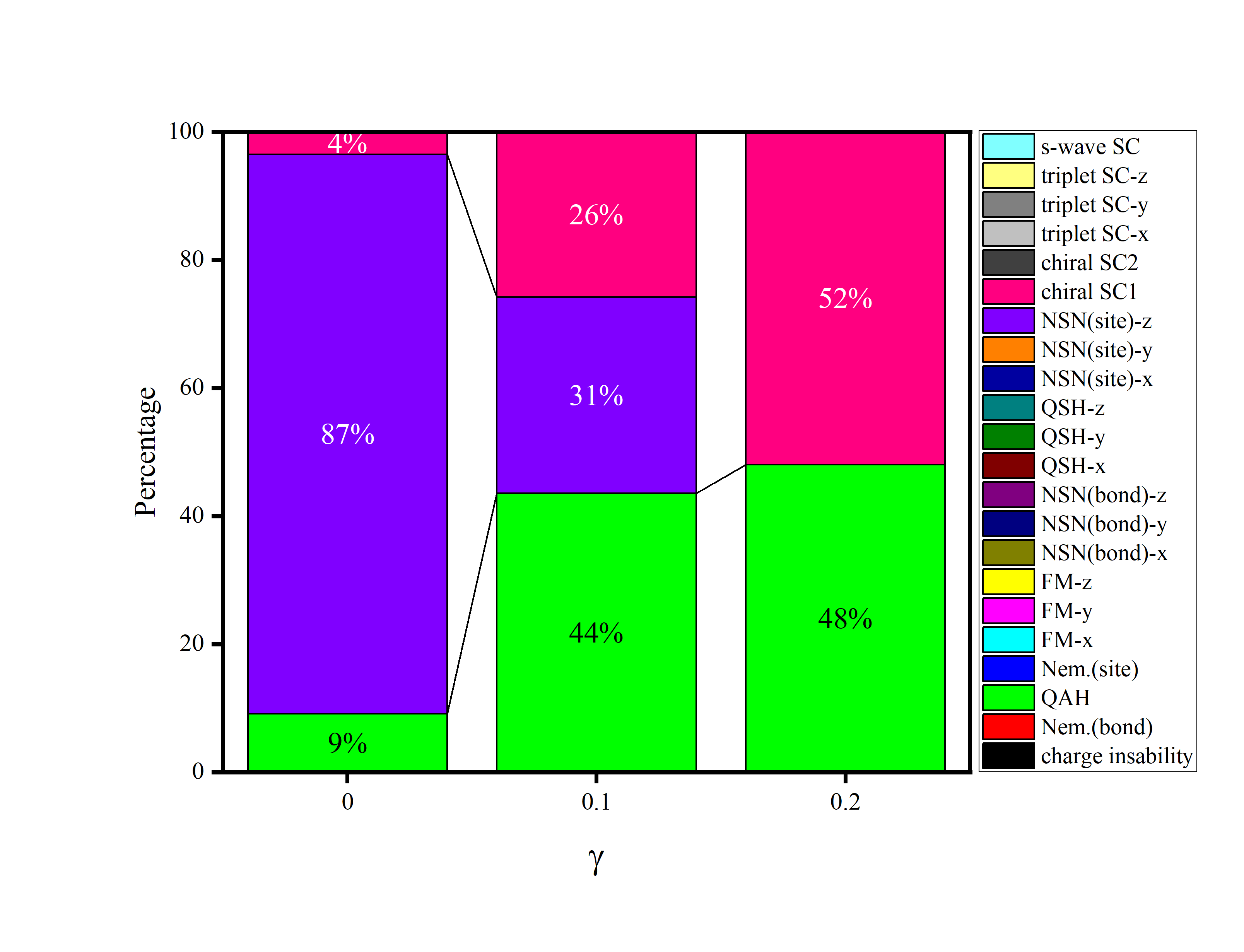}}\\
\vspace{-0.1cm}
\caption{(Color online) Probability evolution of the candidate dominant instabilities as functions of (a) $\alpha$, (b) $\beta$, and (c) $\gamma$. Stacked bars show the percentage of random initial conditions for which each instability channel emerges as the leading instability. The vertical axis gives the probability in percent, and the color legend identifies the corresponding instability channels.}
\label{fig:GZFPS123}
\end{figure}

Beyond the endpoints $\alpha=\pm0.25$, the full dependence of the probabilities for leading instabilities
on $\alpha$ is displayed in Fig.~\ref{fig:GZFPS123}(a). The probability of NSN(site)-$x$ being the leading instability
decreases monotonically from nearly $100\%$ at $\alpha=-0.25$ to essentially $0\%$ at $\alpha=0.25$. In contrast,
the probabilities of QAH and chiral SC$_1$ exhibit a monotonic increase with $\alpha$. Interesting, the probability
for NSN(site)-$z$ first increases, attains a maximum near $\alpha=0$, and then decreases as $\alpha$ grows further.
The QSH-$z$ channel, however, only acquires a non-negligible probability when $\alpha$ is sufficiently close to $0.25$.
Consequently, the leading instability within FPG$_1$ is closely associated with the parameter $\alpha$, transitioning
from NSN(site)-$x$ at negative $\alpha$ to a multi-competitive regime involving QAH, chiral SC$_1$, and QSH-$z$ at
positive $\alpha$.

We then turn to FPG$_2$~(\ref{Eq_FPG2}), which is parameterized by $\beta\in[0,0.2]$. It is manifest that
this group parameter governs the off-diagonal couplings $\lambda_{12}(l_c)$ and $\lambda_{21}(l_c)$.
Paralleling the numerical analysis of FPG$_1$, we present the probability that each candidate phase
emerges as the leading instability as $\beta$ is varied in Fig.~\ref{fig:GZFPS123}(b).

Learning from Fig.~\ref{fig:GZFPS123}(b), we notice that three candidate states, i.e.,
NSN(site)-$z$, QAH, and chiral SC$_1$, compete for the leading instability, while the remaining channels
have no opportunity to become dominant. On the one hand, the NSN(site)-$z$ channel maintains a probability
of approximately $80\%$ or higher throughout the entire range of $\beta$, unambiguously dominating over all
other competing states. On the other hand, the probabilities of QAH and chiral SC$_1$ remain below $12\%$,
showing only weak, non-monotonic variations with $\beta$. This indicates that within FPG$_2$, the leading
instability is essentially insensitive to $\beta$, and the system is most likely to adopt the NSN(site)-$z$ ordered state.

At last, we consider the FPG$_3$~(\ref{Eq_FPG3}). This fixed-point group is characterized by the
parameter $\gamma\in[0,0.2]$, which appears in the off-diagonal elements
$\lambda_{23}(l_c)$ and $\lambda_{32}(l_c)$.  Analogous numerical calculations
yield the probabilities of leading instabilities as a function
of the group parameter as illustrated in Fig.~\ref{fig:GZFPS123}(c).

Reading from Fig.~\ref{fig:GZFPS123}(c), we find that this group displays the most dramatic parametric
dependence among the three FPGs.

Fig.~\ref{fig:GZFPS123}(c) reveals a clear crossover in FPG$_3$ as $\gamma$ varies.
For $\gamma \leq 0.05$, the off-diagonal couplings that are proportional to $\gamma$ are negligible,
so the FPG$_3$ is essentially diagonal, resembling FP$_1$~(\ref{Eq_FP1}).
Consequently, NSN(site)-$z$ emerges as the overwhelming leading instability, with a probability
very close to $100\%$. As $\gamma$ increases beyond 0.05, however, the growing off-diagonal couplings
drive a monotonic decrease in the NSN(site)-$z$ probability, while chiral SC$_1$ and QAH rise
steadily and almost symmetrically. It is instructive to comment further on the near-symmetric increase of the chiral SC$_1$ and QAH probabilities in Fig.~\ref{fig:GZFPS123}(c). This originates from the fact that the coefficients of $\lambda_{11}$, $\lambda_{22}$,
$\lambda_{33}$, and $\lambda_{32}$ in the source-term equations~(\ref{Eq_B3}) and~(\ref{Eq_B17})
are equivalent and play the essential role in determining the susceptibilities. Since these dominant
contributions are identical, it follows that the relative susceptibilities of chiral SC$_1$ and QAH
are governed by the remaining minor perturbations.

As $\gamma$ grows, the off-diagonal terms (in particular, $\lambda_{32}$) in FPG$_3$
simultaneously enhance both chiral SC$_1$ and QAH, while suppress NSN(site)-$z$.
By $\gamma = 0.2$, the NSN(site)-$z$ probability is nearly zero,
and chiral SC$_1$ and QAH each reach approximately $50\%$. Throughout the variation of $\gamma$, no
other states become competitive enough to challenge the leading instability.
To recapitulate, Fig.~\ref{fig:GZFPS123}(c) quantitatively illustrates how
the off-diagonal coupling $\gamma$ drives the system continuously from a NSN(site)-$z$
state to a regime of competing superconducting and topological insulating orders.

\subsection{Brief comparison with previous works}

Before closing this section, it is necessary to briefly compare our primary results with previous
studies~\cite{Fradkin2009PRL,Wang2017PRB,DZDW2020PRB,Vafek2014PRB}
based on a simplified interaction model, and highlight the qualitatively new features that emerge from our model.

Particularly, it is instructive to compare our key results with those of the simplified model studied in Ref.~\cite{Vafek2014PRB}, which retains only the four diagonal interactions $\lambda_{ii}$ and neglects all off-diagonal couplings. In that model, the RG equations reduce to a closed set for $\lambda_{00}$, $\lambda_{11}$, $\lambda_{22}$, and $\lambda_{33}$ appearing in our model, and the source-term susceptibilities are linear combinations of only these four parameters. Consequently, the fixed-point structure is discrete and limited to two parameter ratios, with no continuous groups analogous to FPG$_1$-FPG$_3$~\cite{Fradkin2009PRL,Wang2017PRB,DZDW2020PRB,Vafek2014PRB}. Besides, they find that the leading instabilities are restricted to QAH and QSH phases. The absence of off-diagonal fermion-fermion couplings eliminates the enhancement mechanisms for chiral SC and NSN channels that are present in our model.

In sharp contrast, all full sixteen kinds of fermion-fermion interactions in our model introduce several qualitatively
new features. At first, the off-diagonal fermion-fermion interactions generate additional linear and quadratic terms in the susceptibility exponents, which can promote chiral SC and NSN phases to become leading instabilities. In addition, the continuous FPG$_1$-FPG$_3$ allow for a systematic tuning of the leading instability via the group parameters $\alpha$, $\beta$, and $\gamma$, as quantitatively demonstrated in Fig.~\ref{fig:GZFPS123}. Furthermore, we uncover the trajectory-dependent leading instabilities that have no analogue in the simplified model. Specifically, at FP$_5$ and FP$_7$, the chiral SC$_1$ and QAH channels exhibit nearly degenerate exponents, making the leading instability sensitive to the full RG trajectory rather than solely to the FP itself. This phenomenon arises from the competition among all sixteen kinds of fermionic interactions.


\section{Summary}\label{Sec_summary}

In summary, we have systematically investigated the interaction-driven competing instabilities
in 2D QBCP semimetals within the spin-1/2 checkerboard-lattice model, using the Wilsonian RG approach~\cite{Shankar1994RMP,Wilson1975RMP,Polchinski9210046}. To capture the full range of
physical ingredients, we start from the general set of sixteen marginal fermion-fermion couplings
and treat all marginally relevant interactions on an equal footing. Evaluating the one-loop corrections
to all channels yields the complete set of sixteen coupled RG flow equations~(\ref{Eq_lambda00})-(\ref{Eq_lambda33}),
which we then use to systematically classify the resulting fixed points and determine the dominant
and subleading symmetry-broken phases that develop in their vicinity. This unbiased treatment
reveals several qualitatively new features as summarized below.

At first, we study the fixed-point structure in Sec.~\ref{Sec_classification_FPs}, which is divided
into three scenarios based on the initial conditions. In the isotropic scenario, all sixteen couplings
flow to a single fixed point, FP$_0$~(\ref{Eq_FP0}), which is completely independent of the bare
interaction strength. In the anisotropic scenario, where one coupling initially dominates, there
exist seven stable fixed points, FP$_1$-FP$_7$~(\ref{Eq_FP1})-(\ref{Eq_FP7}), whereas dominance
of the $\lambda_{11}$ or $\lambda_{33}$ channel leads to an unstable runaway flow, as shown in
Figs.~\ref{fig:FPs}(g)-(h). These stable fixed points are paired by a discrete off-diagonal
index-exchange symmetry. By contrast, in the general scenario, with all couplings sampled
independently from a broad distribution, the fixed points organize into
three fixed-point groups, FPG$_1$-FPG$_3$~(\ref{Eq_FPG1})-(\ref{Eq_FPG3}),
characterized by the group parameters $\alpha$, $\beta$, and $\gamma$.

Since symmetry breaking can generally induce a variety of instabilities~\cite{Vafek2014PRB,Vafek2010PRB,
Metzner2000PRL,Halboth2000RPB,Nandkishore2012NP,Chubukov2016PRX,Roy2018PRX}, we next identify the leading
and subleading instabilities in the vicinity of all FPs in Sec.~\ref{Sec_phase_transition}. At FP$_0$, the
leading instability is chiral SC$_1$, with NSN order subleading. Near FP$_{1,2,3,4}$ and FP$_6$, the
NSN(site)-$z$ phase is the leading instability, while the subleading instability is either chiral SC$_1$
or QAH. By contrast, both the chiral SC$_1$ and QAH phases have a relatively high probability of becoming
the leading instability near FP$_5$ and FP$_7$. Table~\ref{table:stability} summarizes the key results
for FP$_{1}$-FP$_{7}$.

In addition, we examine the evolution of the leading instabilities across the three fixed-point groups,
as illustrated in Fig.~\ref{fig:GZFPS123}. For FPG$_1$, as the group parameter $\alpha$ increases from
$-0.25$ to $0.25$, the system undergoes a crossover from a single dominant state, NSN(site)-$x$, to a
multi-competitive regime in which QAH, QSH, NSN, and chiral SC compete for dominance. For FPG$_2$, the
parameter $\beta$ has no significant influence on the leading instability, and NSN(site)-$z$ remains dominant.
By contrast, as the parameter $\gamma$ for FPG$_3$ increases from $0$ to $0.2$, the leading
instability gradually shifts from NSN(site)-$z$ to a near-degenerate chiral SC$_1$/QAH state,
which requires further perturbations to distinguish between the two.

Compared with previous studies~\cite{Fradkin2009PRL,Wang2017PRB,DZDW2020PRB,Vafek2014PRB},
several features of our results have no counterparts.
Initially, the off-diagonal fermion-fermion interactions generate additional linear and quadratic terms
in the susceptibility exponents. These terms can promote chiral SC and NSN phases to become leading instabilities,
which cannot become dominant when only diagonal fermion-fermion interactions are present. Besides, the continuous
fixed-point groups FPG$_1$-FPG$_3$ allow for a systematic tuning of the leading instability via the group
parameters $\alpha$, $\beta$, and $\gamma$ as quantitatively demonstrated in Fig.~\ref{fig:GZFPS123}.
Moreover, we uncover trajectory-dependent leading instabilities at FP$_5$ and FP$_7$, where the chiral
SC$_1$ and QAH channels are nearly degenerate and the dominant channel is determined by the full RG trajectory
rather than by the FP alone. These findings together indicate that the competition among all sixteen
fermion-fermion interaction channels is indispensable for determining the low-energy quantum-critical
behavior of 2D QBCP semimetals.

Our comprehensive analysis provides a useful clue for interpreting interaction-driven symmetry breaking
in checkerboard-lattice materials and related quadratic-touching systems. In particular, it highlights the
essential role played by the full set of sixteen fermion-fermion interactions in shaping the low-energy
physics of these systems. We hope that this work will stimulate further theoretical and experimental efforts
to explore the rich interplay between quantum fluctuations and competing orders in 2D QBCP semimetals and
similar systems.

\section*{ACKNOWLEDGEMENTS}

We thank Mr. Yi-Kun Fang for the helpful discussions.
J.W. is supported  by Tianjin Natural Science Foundation Project (25JCYBJC01640) and
was partially supported by the National Natural
Science Foundation of China under Grant No. 11504360.

\begin{figure*}[htbp]
\subfigure[]{
\includegraphics[width=1.0in]{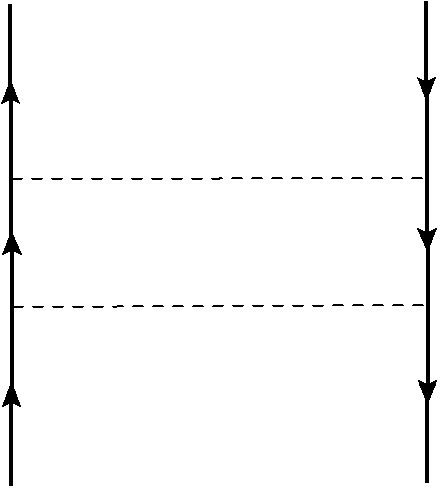}
} \hspace{0.2cm}
\subfigure[]{
\includegraphics[width=1.0in]{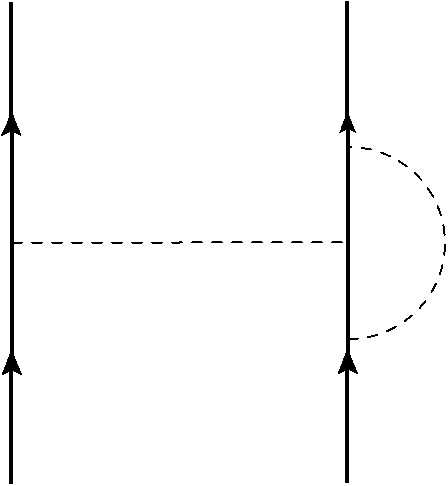}
} \hspace{0.2cm}
\subfigure[]{
\includegraphics[width=1.0in]{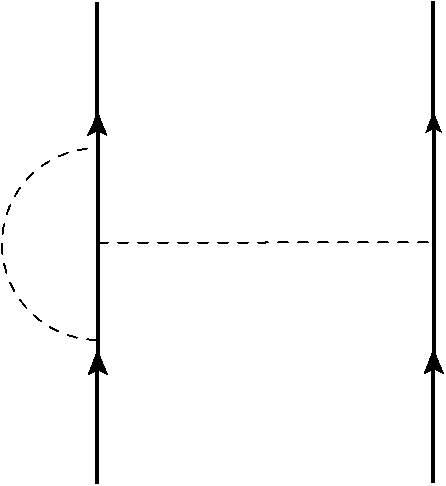}
} \hspace{0.2cm}
\subfigure[]{
\includegraphics[width=1.0in]{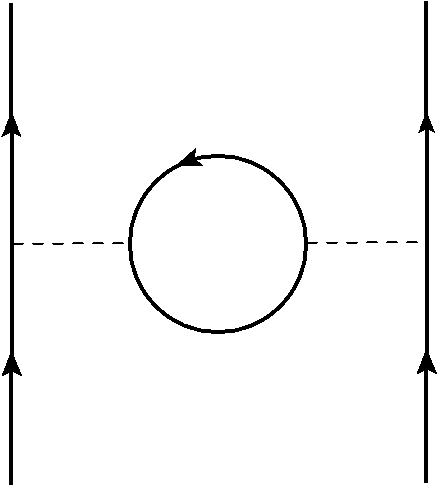}
} \hspace{0.2cm}
\subfigure[]{
\includegraphics[width=1.0in]{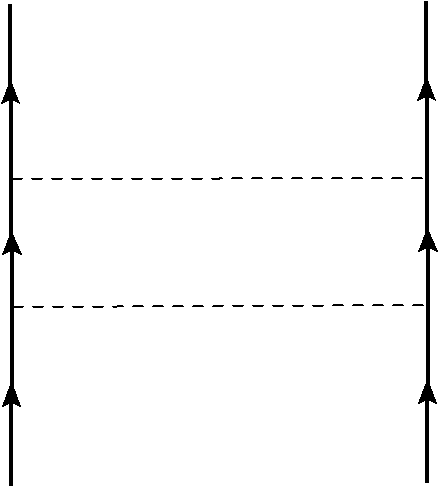}
}
\vspace{-0.1cm}
\caption{One-loop Feynman diagrams for the renormalization of the fermion-fermion interactions. Panels (a)¨C(e) correspond to the distinct one-loop correction channels. The solid and dashed lines specify the fermionic propagator and the fermion-fermion interaction, respectively.}
\label{figure:one-loop}
\end{figure*}

\appendix

\section{One-loop corrections to the fermion-fermion interactions}\label{Appendix_1L_corrections}

Starting from the effective action~(\ref{Eq:Seff}), the one-loop diagrams contributing to the renormalization of
the fermion-fermion interaction are shown in Fig.~\ref{figure:one-loop}. To render the calculation more tractable,
we implement a conventional rescaling procedure that rescales all momenta and frequencies by a reference scale
$\Lambda_0$, which is naturally set by the inverse lattice spacing. The corresponding scaling transformations
are given by $k \rightarrow k/\Lambda_0$ and $\omega \rightarrow \omega/\Lambda_0$
~\cite{Vafek2014PRB,Wang2017PRB, Wang2011PRB,She2010PRB,Huh2008PRB}.
After a series of lengthy but standard calculations~\cite{Vafek2014PRB,Wang2017PRB,Wang2017PRB}, we obtain
the following one-loop corrections to the fermion-fermion interaction parameters:
\begin{widetext}
\begin{eqnarray}
\delta S_{\lambda_{00}}
&=&-4|t|\pi(\lambda_{00}\lambda_{11}+3\lambda_{01}\lambda_{10}+4\lambda_{02}\lambda_{20}+3\lambda_{03}\lambda_{30}
+\lambda_{00}\lambda_{33})l\mathcal{A}_{00},\label{Ap_F00}\\
\delta S_{\lambda_{01}}
&=&4|t|\pi
(\lambda_{00}\lambda_{01}-\lambda_{00}\lambda_{10}-2\lambda_{01}\lambda_{11}-\lambda_{02}\lambda_{12}
-\lambda_{03}\lambda_{13}-4\lambda_{02}\lambda_{21}-\lambda_{01}\lambda_{22}-2\lambda_{03}\lambda_{31}
-\lambda_{01}\lambda_{33})l\mathcal{A}_{01},\\
\delta S_{\lambda_{02}}
&=&4|t|\pi(2\lambda_{00}\lambda_{02}-2\lambda_{02}\lambda_{11}-2\lambda_{01}\lambda_{12}-2\lambda_{02}\lambda_{22}
-2\lambda_{02}\lambda_{33}-2\lambda_{03}\lambda_{32})l\mathcal{A}_{02},\\
\delta S_{\lambda_{03}}
&=&4|t|\pi
(\lambda_{00}\lambda_{03}-\lambda_{03}\lambda_{11}-2\lambda_{01}\lambda_{13}-\lambda_{03}\lambda_{22}
-4\lambda_{02}\lambda_{23}-\lambda_{00}\lambda_{30}-\lambda_{01}\lambda_{31}-\lambda_{02}\lambda_{32}-2\lambda_{03}\lambda_{33})l\mathcal{A}_{03},\\
\delta S_{\lambda_{10}}
&=&-4|t|\pi
(3\lambda_{10}\lambda_{11}+4\lambda_{12}\lambda_{20}+\lambda_{20}\lambda_{21}+\lambda_{00}\lambda_{01}
+2\lambda_{13}\lambda_{30}+\lambda_{30}\lambda_{31})l\mathcal{A}_{10},\\
\delta S_{\lambda_{11}}
&=&-2|t|\pi
(\lambda_{00}\lambda_{00}+\lambda_{01}\lambda_{01}+\lambda_{02}\lambda_{02}+\lambda_{03}\lambda_{03}
+\lambda_{10}\lambda_{10}+3\lambda_{11}\lambda_{11}+\lambda_{12}\lambda_{12}+\lambda_{13}\lambda_{13}\nonumber\\&&
+\lambda_{20}\lambda_{20}+\lambda_{21}\lambda_{21}+\lambda_{22}\lambda_{22}+\lambda_{23}\lambda_{23}
+\lambda_{30}\lambda_{30}+\lambda_{31}\lambda_{31}+\lambda_{32}\lambda_{32}+\lambda_{33}\lambda_{33}\nonumber\\&&
-2\lambda_{00}\lambda_{11}+8\lambda_{12}\lambda_{21}+2\lambda_{11}\lambda_{22}+2\lambda_{13}\lambda_{31}
+4\lambda_{23}\lambda_{32}+4\lambda_{11}\lambda_{33}-4\lambda_{22}\lambda_{33})l\mathcal{A}_{11},\\
\delta S_{\lambda_{12}}
&=&-4|t|\pi
(-2\lambda_{00}\lambda_{12}+4\lambda_{11}\lambda_{12}+2\lambda_{12}\lambda_{22}-2\lambda_{23}\lambda_{31}
+\lambda_{13}\lambda_{32}+3\lambda_{12}\lambda_{33}+2\lambda_{21}\lambda_{33})l\mathcal{A}_{12},\\
\delta S_{\lambda_{13}}
&=&4|t|\pi
(\lambda_{00}\lambda_{13}-3\lambda_{11}\lambda_{13}-\lambda_{13}\lambda_{22}-4\lambda_{12}\lambda_{23}
-2\lambda_{22}\lambda_{31}+2\lambda_{21}\lambda_{32}-\lambda_{13}\lambda_{33})l\mathcal{A}_{13},\\
\delta S_{\lambda_{20}}
&=&-4|t|\pi
(2\lambda_{10}\lambda_{21}+4\lambda_{20}\lambda_{22}+2\lambda_{23}\lambda_{30})l\mathcal{A}_{20},\\
\delta S_{\lambda_{21}}
&=&4|t|\pi
(-\lambda_{11}\lambda_{21}-5\lambda_{21}\lambda_{22}+\lambda_{00}\lambda_{21}-\lambda_{23}\lambda_{31}
+2\lambda_{13}\lambda_{32}-2\lambda_{12}\lambda_{33}-2\lambda_{21}\lambda_{33})l\mathcal{A}_{21},\\
\delta S_{\lambda_{22}}
&=&-4|t|\pi(\lambda_{12}\lambda_{21}-2\lambda_{00}\lambda_{22}+3\lambda_{11}\lambda_{22}+2\lambda_{22}\lambda_{22}
+2\lambda_{13}\lambda_{31}+\lambda_{23}\lambda_{32}-2\lambda_{11}\lambda_{33}+3\lambda_{22}\lambda_{33})l\mathcal{A}_{22},\\
\delta S_{\lambda_{23}}
&=&4|t|\pi
(-\lambda_{13}\lambda_{21}-2\lambda_{11}\lambda_{23}-5\lambda_{22}\lambda_{23}+\lambda_{00}\lambda_{23}
+2\lambda_{12}\lambda_{31}-2\lambda_{11}\lambda_{32}-\lambda_{23}\lambda_{33})l\mathcal{A}_{23},\\
\delta S_{\lambda_{30}}
&=&-4|t|\pi
(\lambda_{00}\lambda_{03}+\lambda_{20}\lambda_{23}+\lambda_{10}\lambda_{13}+2\lambda_{10}\lambda_{31}
+4\lambda_{20}\lambda_{32}+3\lambda_{30}\lambda_{33})l\mathcal{A}_{30},\\
\delta S_{\lambda_{31}}
&=&4|t|\pi
(-2\lambda_{13}\lambda_{22}+2\lambda_{12}\lambda_{23}-\lambda_{11}\lambda_{31}-\lambda_{22}\lambda_{31}
+\lambda_{00}\lambda_{31}-4\lambda_{21}\lambda_{32}-3\lambda_{31}\lambda_{33})l\mathcal{A}_{31},\\
\delta S_{\lambda_{32}}
&=&4|t|\pi
(2\lambda_{13}\lambda_{21}-\lambda_{12}\lambda_{31}+2\lambda_{00}\lambda_{32}-2\lambda_{22}\lambda_{32}
-2\lambda_{11}\lambda_{23}-3\lambda_{11}\lambda_{32}-4\lambda_{32}\lambda_{33})l\mathcal{A}_{32},\\
\delta S_{\lambda_{33}}
&=&-2|t|\pi
(\lambda_{00}\lambda_{00}+\lambda_{01}\lambda_{01}+\lambda_{02}\lambda_{02}+\lambda_{03}\lambda_{03}
+\lambda_{10}\lambda_{10}+\lambda_{11}\lambda_{11}+\lambda_{12}\lambda_{12}+\lambda_{13}\lambda_{13}\nonumber\\&&
+\lambda_{20}\lambda_{20}+\lambda_{21}\lambda_{21}+\lambda_{22}\lambda_{22}+\lambda_{23}\lambda_{23}
+\lambda_{30}\lambda_{30}+\lambda_{31}\lambda_{31}+\lambda_{32}\lambda_{32}+3\lambda_{33}\lambda_{33}\nonumber\\&&
+4\lambda_{12}\lambda_{21}-4\lambda_{11}\lambda_{22}+2\lambda_{13}\lambda_{31}+8\lambda_{23}\lambda_{32}
-2\lambda_{00}\lambda_{33}+4\lambda_{11}\lambda_{33}+2\lambda_{22}\lambda_{33})l\mathcal{A}_{33},\label{Ap_F33}
\end{eqnarray}
\end{widetext}
where the coefficients $\mathcal{A}_{ij}$ are defined as:
\begin{eqnarray}
\mathcal{A}_{ij}&\equiv&\int_{-\infty}^{+\infty}\frac{d\omega_{1}d\omega_{2}d\omega_{3}}{(2\pi)^{3}}
\int^{b\Lambda}\frac{d^{2}\mathbf{k}_{1}d^{2}\mathbf{k}_{2}d^{2}\mathbf{k}_{3}}{(2\pi)^{6}}\nonumber\\
&&\times\Psi_{\sigma}^{\dagger}(\omega_{1},\mathbf{k}_{1})\sigma_{i}\Psi_{\sigma}(\omega_{2},\mathbf{k}_{2})
\Psi_{\sigma}^{\dagger}(\omega_{3},\mathbf{k}_{3})\sigma_{j}\nonumber\\
&&\times\Psi_{\sigma}(\omega_{1}+\omega_{2}-\omega_{3},\mathbf{k}_{1}+\mathbf{k}_{2}-\mathbf{k}_{3}),
\end{eqnarray}
with $i,j=0,1,2,3$.

\begin{figure}[htbp]
\centering
\subfigure[]{
\includegraphics[width=0.71in]{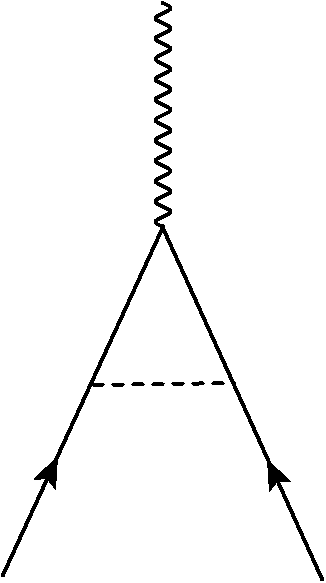}
}\hspace{0.5cm}
\subfigure[]{
\includegraphics[width=0.71in]{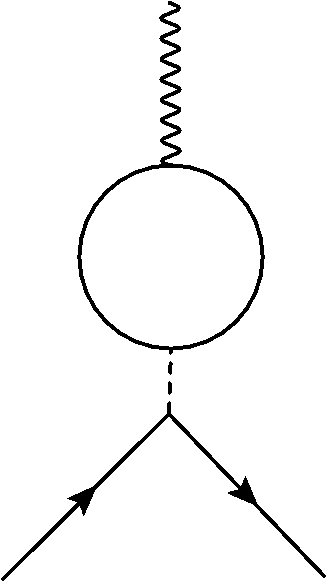}
}\hspace{0.5cm}
\subfigure[]{
\includegraphics[width=0.71in]{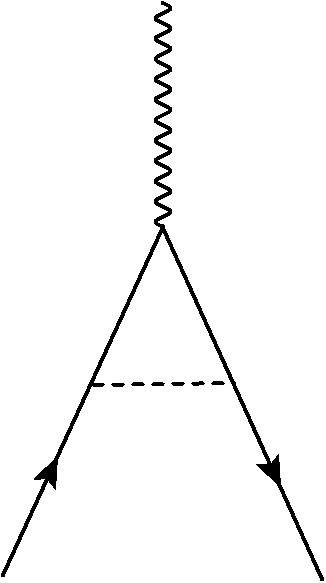}
}
\vspace{-0.1cm}
\caption{One-loop Feynman diagrams contributing to the fermionic bilinear source terms. Panels (a) and (b) correspond to the particle-hole channel, and panel (c) to the particle-particle channel. The solid, dash, and wave lines correspond to the fermionic propagator, fermion-fermion interaction and source term, respectively.}
\label{figure:source-term}
\end{figure}

\section{Flow equations of source terms}\label{Appendix_source_terms}

Given the effective action~(\ref{Eq:Seff}) and the source terms~(\ref{equation:waiyuanxiang}), the underlying
one-loop correction to the source terms is shown in Fig.~\ref{figure:source-term}~\cite{Vafek2014PRB,Wang2017PRB,Wang2017PRB}.
By combining the one-loop calculations presented in Appendix~\ref{Appendix_1L_corrections} with the RG procedure of Sec.~\ref{Sec_RG_analysis}, we are left with the energy-dependent evolution for the source terms of all the fermionic
bilinears as follows,
\begin{widetext}
\begin{eqnarray}
\frac{d\Delta^c_{1}}{dl}
&=&2\Delta^c_{1},\label{Eq_B1}
\\
\frac{d\Delta^c_{2}}{dl}
&=&\Bigl\{2+\frac{1}{2m}\Bigl[\lambda_{00}-7\lambda_{01}-\lambda_{02}-\lambda_{03}+\lambda_{10}+\lambda_{11}\nonumber\\
&&-\lambda_{12}-\lambda_{13}+\lambda_{20}+\lambda_{21}-\lambda_{22}-\lambda_{23}+\lambda_{30}
+\lambda_{31}-\lambda_{32}-\lambda_{33}\Bigr]\Bigr\}\Delta^c_{2},
\\
\frac{d\Delta^c_{3}}{dl}
&=&\Bigl\{2+\frac{1}{2m}\Bigl[\lambda_{00}-\lambda_{01}-7\lambda_{02}-\lambda_{03}+\lambda_{10}-\lambda_{11}\nonumber\\
&&+\lambda_{12}-\lambda_{13}+\lambda_{20}-\lambda_{21}+\lambda_{22}-\lambda_{23}+\lambda_{30}-\lambda_{31}
+\lambda_{32}-\lambda_{33}\Bigr]\Bigr\}\Delta^c_{3},\label{Eq_B3}
\\
\frac{d\Delta^c_{4}}{dl}
&=&\Bigl\{2+\frac{1}{4m}\Bigl[\lambda_{00}-\lambda_{01}-\lambda_{02}-7\lambda_{03}+\lambda_{10}-\lambda_{11}\nonumber\\
&&-\lambda_{12}+\lambda_{13}+\lambda_{20}-\lambda_{21}-\lambda_{22}+\lambda_{23}+\lambda_{30}
-\lambda_{31}-\lambda_{32}+\lambda_{33}\Bigr]\Bigr\}\Delta^c_{4},
\\
\frac{d\Delta^s_{1x}}{dl}
&=&2\Delta^s_{1x},
\\
\frac{d\Delta^s_{1y}}{dl}
&=&2\Delta^s_{1y},
\\
\frac{d\Delta^s_{1z}}{dl}
&=&2\Delta^s_{1z},
\\
\frac{d\Delta^s_{2x}}{dl}
&=&\Big\{2+\frac{1}{4m}\Bigl[\lambda_{00}+\lambda_{01}-\lambda_{02}-\lambda_{03}-7\lambda_{10}+\lambda_{11}\nonumber\\
&&-\lambda_{12}-\lambda_{13}-\lambda_{20}-\lambda_{21}+\lambda_{22}+\lambda_{23}-\lambda_{30}-\lambda_{31}
+\lambda_{32}+\lambda_{33}\Bigr]\Bigr\}\Delta^s_{2x},
\\
\frac{d\Delta^s_{2y}}{dl}
&=&\Big\{2+\frac{1}{4m}\Bigl[\lambda_{00}+\lambda_{01}-\lambda_{02}-\lambda_{03}-\lambda_{10}-\lambda_{11}\nonumber\\
&&+\lambda_{12}+\lambda_{13}-7\lambda_{20}+\lambda_{21}-\lambda_{22}-\lambda_{23}-\lambda_{30}
-\lambda_{31}+\lambda_{32}+\lambda_{33}\Bigr]\Bigr\}\Delta^s_{2y},
\\
\frac{d\Delta^s_{2z}}{dl}
&=&\Big\{2+\frac{1}{4m}\Bigl[\lambda_{00}+\lambda_{01}-\lambda_{02}-\lambda_{03}-\lambda_{10}-\lambda_{11}\nonumber\\
&&+\lambda_{12}+\lambda_{13}-\lambda_{20}-\lambda_{21}+\lambda_{22}+\lambda_{23}-7\lambda_{30}
+\lambda_{31}-\lambda_{32}-\lambda_{33}\Bigr]\Bigr\}\Delta^s_{2z},
\\
\frac{d\Delta^s_{3x}}{dl}
&=&\Big\{2+\frac{1}{2m}\Bigl[\lambda_{00}-\lambda_{01}+\lambda_{02}-\lambda_{03}+\lambda_{10}-\lambda_{11}\nonumber\\
&&-7\lambda_{12}-\lambda_{13}-\lambda_{20}+\lambda_{21}-\lambda_{22}+\lambda_{23}-\lambda_{30}
+\lambda_{31}-\lambda_{32}+\lambda_{33}\Bigr]\Bigr\}\Delta^s_{3x},
\\
\frac{d\Delta^s_{3y}}{dl}
&=&\Big\{2+\frac{1}{2m}\Bigl[\lambda_{00}-\lambda_{01}+\lambda_{02}-\lambda_{03}-\lambda_{10}+\lambda_{11}\nonumber\\
&&-\lambda_{12}+\lambda_{13}+\lambda_{20}-\lambda_{21}-7\lambda_{22}-\lambda_{23}-\lambda_{30}
+\lambda_{31}-\lambda_{32}+\lambda_{33}\Bigr]\Bigr\}\Delta^s_{3y},
\\
\frac{d\Delta^s_{3z}}{dl}
&=&\Big\{2+\frac{1}{2m}\Bigl[\lambda_{00}-\lambda_{01}+\lambda_{02}-\lambda_{03}-\lambda_{10}+\lambda_{11}\nonumber\\
&&-\lambda_{12}+\lambda_{13}-\lambda_{20}+\lambda_{21}-\lambda_{22}+\lambda_{23}
+\lambda_{30}-\lambda_{31}-7\lambda_{32}-\lambda_{33}\Bigr]\Bigr\}\Delta^s_{3z},
\\
\frac{d\Delta^s_{4x}}{dl}
&=&\Big\{2+\frac{1}{4m}\Bigl[\lambda_{00}-\lambda_{01}-\lambda_{02}+\lambda_{03}+\lambda_{10}-\lambda_{11}\nonumber\\
&&-\lambda_{12}-7\lambda_{13}-\lambda_{20}+\lambda_{21}+\lambda_{22}-\lambda_{23}-\lambda_{30}+\lambda_{31}
+\lambda_{32}-\lambda_{33}\Bigr]\Bigr\}\Delta^s_{4x},
\\
\frac{d\Delta^s_{4y}}{dl}
&=&\Big\{2+\frac{1}{4m}\Bigl[\lambda_{00}-\lambda_{01}-\lambda_{02}+\lambda_{03}-\lambda_{10}+\lambda_{11}\nonumber\\
&&+\lambda_{12}-\lambda_{13}+\lambda_{20}-\lambda_{21}-\lambda_{22}-7\lambda_{23}
-\lambda_{30}+\lambda_{31}+\lambda_{32}-\lambda_{33}\Bigr]\Bigr\}\Delta^s_{4y},
\\
\frac{d\Delta^s_{4z}}{dl}
&=&\Big\{2+\frac{1}{4m}\Bigl[\lambda_{00}-\lambda_{01}-\lambda_{02}+\lambda_{03}-\lambda_{10}+\lambda_{11}\nonumber\\
&&+\lambda_{12}-\lambda_{13}-\lambda_{20}+\lambda_{21}+\lambda_{22}-\lambda_{23}
+\lambda_{30}-\lambda_{31}-\lambda_{32}-7\lambda_{33}\Bigr]\Bigr\}\Delta^s_{4z},\label{Eq_B16}
\\
\frac{d\Delta^{\mathrm{PP}}_{1}}{dl}
&=&\Bigl\{2+\frac{1}{2m}\Bigl[\lambda_{00}+\lambda_{01}-\lambda_{02}+\lambda_{03}-\lambda_{10}-\lambda_{11}\nonumber\\
&&+\lambda_{12}-\lambda_{13}-\lambda_{20}-\lambda_{21}+\lambda_{22}-\lambda_{23}
-\lambda_{30}-\lambda_{31}+\lambda_{32}-\lambda_{33}\Bigr]\Bigr\}\Delta^{\mathrm{PP}}_{1},\label{Eq_B17}
\\
\frac{d\Delta^{\mathrm{PP}}_{2}}{dl}
&=&\Bigl\{2+\frac{1}{4m}\Bigl[\lambda_{00}+\lambda_{01}+\lambda_{02}-\lambda_{03}-\lambda_{10}-\lambda_{11}\nonumber\\
&&-\lambda_{12}+\lambda_{13}-\lambda_{20}-\lambda_{21}-\lambda_{22}+\lambda_{23}-\lambda_{30}-\lambda_{31}
-\lambda_{32}+\lambda_{33}\Bigr]\Bigr\}\Delta^{\mathrm{PP}}_{2},
\\
\frac{d\Delta^{\mathrm{PP}}_{3x}}{dl}
&=&2\Delta^{\mathrm{PP}}_{3x},
\\
\frac{d\Delta^{\mathrm{PP}}_{3y}}{dl}
&=&2\Delta^{\mathrm{PP}}_{3y},
\\
\frac{d\Delta^{\mathrm{PP}}_{3z}}{dl}
&=&2\Delta^{\mathrm{PP}}_{3z},
\\
\frac{d\Delta^{\mathrm{PP}}_{4}}{dl}
&=&\Bigl\{2+\frac{1}{4m}\Bigl[\lambda_{00}-\lambda_{01}+\lambda_{02}+\lambda_{03}-\lambda_{10}+\lambda_{11}\nonumber\\
&&-\lambda_{12}-\lambda_{13}-\lambda_{20}+\lambda_{21}-\lambda_{22}-\lambda_{23}
-\lambda_{30}+\lambda_{31}-\lambda_{32}-\lambda_{33}\Bigr]\Bigr\}\Delta^{\mathrm{PP}}_{4}.\label{Eq_B22}
\end{eqnarray}
\end{widetext}



\end{document}